\documentclass{article}
\usepackage{graphicx}
\usepackage{float}
\usepackage{topcapt}

\usepackage{lipsum}
\usepackage{epigraph}

\usepackage{amsmath}
\usepackage{url}
\usepackage{eurosym}

\usepackage[T2A]{fontenc}
\usepackage[utf8]{inputenc}
\usepackage[russian,english,british]{babel}

\usepackage{longtable}
\usepackage{fancyvrb}
\usepackage{tocbibind}

\begin{document}

\title{The Role of Time in Making Risky Decisions and the Function of Choice}
\author{Valerii Salov}
\date{}
\maketitle

\begin{abstract}
The prospects of Kahneman and Tversky, Mega Million and Powerball lotteries, St. Petersburg paradox, premature profits and growing losses criticized by Livermore are reviewed under an angle of view comparing mathematical expectations with awards received. Original prospects have been formulated as a one time opportunity. An award value depends on the number of times the game is played. The random sample mean is discussed as a universal award. The role of time in making a risky decision is important as long as the frequency of games and playing time affect their number. A function of choice mapping properties of two-point random variables to fractions of respondents choosing them is proposed.
\end{abstract}

\section{Introduction}

\setlength{\epigraphwidth}{0.47\textwidth}

\epigraph{\begin{otherlanguage*}{russian} Мы выбираем, нас выбирают. \\ Как это часто не совпадает! \end{otherlanguage*} \\ (We make the choices, we are selected. \\ Often intentions are misdirected!) }
{\textsc{Mikhail Tanich} \\ (VS' translation)}

Decision-making under risk attracts attention of economists for a long time \cite{knight1964} \cite{debreu1951}, \cite{markowitz1952}, \cite{allais1953}, \cite{samuelson1960}, \cite{arrow1971}, \cite{samuelson1977}, \cite{kahneman1979}, \cite{tversky1992}, \cite{kahneman2002}, \cite[Chapter 3]{sharpe2007}. The mathematical expectation of profits and losses in a game or trading is not always the main factor influencing decisions \cite{bernoulli1954}, \cite{buffon1777}, \cite{khinchin1925}, \cite{menger1934}, \cite{kelly1956}, \cite{samuelson1977}, \cite{kahneman1979}, \cite{tversky1992}, \cite{vince1992}, \cite{vince1995}, \cite{kahneman2002}, \cite{sharpe2007}, \cite{peters2011}, \cite{peters2011b}, \cite{varma2013}, \cite[Chapter 4]{salov2007}, \cite{salov2014}. Understanding the nature of award in a game and construction of suitable measures for comparing awards and choosing between games is a challenging task. Here, the \textit{certainty} of \textit{sample mean} profits and losses and \textit{play time} are considered with respect to decision making.

\section{Problem 3 from Kahneman and Tversky}

Based on Maurice Allais's research \cite{allais1953}, Daniel Kahneman and Amos Tversky \cite[p. 266]{kahneman1979} found out that 80 percent of 95 students and university faculty preferred Prospect B in Problem 3: choose between (A) 4,000 Israeli pounds received with 80 percent of chance and (B) 3,000 for sure. 3,000 were the median net monthly income. Studying these results, the author knew how to compute \textit{mathematical expectations} \cite[pp. 57 - 69, Chapter IV Mathematical Expectations]{kolmogorov1974}. It is equal to $1 \times 3,000 + 0 \times 0 = 3,000$ for B and $0.8 \times 4,000 + 0.2 \times 0 = 3,200$ for A. Esteeming the extra 200, he mentally joined the majority declining the gift. Kahneman and Tversky label this phenomenon the \textit{certainty effect}. In Problem 3, 80 percent choosing B is its \textit{experimental measure}. Can the \textit{probability theory} match \textit{theoretically} the experimental \textit{fractions of respondents}?

\paragraph{Two-Point Distribution} represents a \textit{random variable} $\xi$ with two outcomes $A_p$, $A_q$ and probabilities $0 \le p \le 1$, $q = 1 - p$. Its mathematical expectation $E(\xi)$, \textit{variance} $D(\xi)$, \textit{third central moment} $\mu_3(\xi)$, \textit{fourth central moment} $\mu_4(\xi)$, \textit{skewness} $\gamma_1(\xi)$, \textit{excess kurtosis} $\gamma_2(\xi)$, and \textit{entropy} $H(\xi)$ are

\begin{equation}
\label{EqBinaryVariableMean}
E(\xi) = \alpha_1(\xi) = A_p p + A_q q = (A_p - A_q)p + A_q,
\end{equation}
\begin{equation}
\label{EqBinaryVariableVariance}
D(\xi)=\mu_2(\xi)=E([\xi - E(\xi)]^2)=E(\xi^2)-\alpha_1(\xi)^2=(A_p-A_q)^2 p (1 - p),
\end{equation}
\begin{equation}
\mu_3(\xi) = E([\xi - E(\xi)]^3) = (A_p-A_q)^3p(1-p)(1-2p),
\end{equation}
\begin{equation}
\mu_4(\xi) = E([\xi - E(\xi)]^4) = (A_p - A_q)^4p(1-p)(1 - 3p +3p^2),
\end{equation}
\begin{equation}
\label{EqBinaryVariableSkewness}
\gamma_1(\xi)=\frac{\mu_3(\xi)}{D(\xi)^{\frac{3}{2}}}=\frac{E([\xi - E(\xi)]^3)}{D(\xi)^{\frac{3}{2}}}=\frac{A_p-A_q}{|A_p-A_q|} \times \frac{1-2p}{\sqrt{p(1-p)}},
\end{equation}
\begin{equation}
\label{EqBinaryVariableExcessKurtosis}
\gamma_2(\xi) = \frac{\mu_4(\xi)}{D(\xi)^2} - 3=\frac{E([\xi - E(\xi)]^4)}{D(\xi)^2} - 3 = \frac{1 - 6p + 6p^2}{p(1 - p)},
\end{equation}
\begin{equation}
\label{EqBinaryVariableEntropy}
H(\xi) = -p\log_2(p) -(1-p)\log_2(1-p).
\end{equation}
For Problem 3, in A $A_p^{(A)} = 4,000$, $A_q^{(A)} = 0$, $p^{(A)} = 0.8$, $q^{(A)} = 0.2$, $E(\xi^{(A)})=3,200$, \textit{standard deviation} $\sqrt{D(\xi^{(A)})} = 1,600$, $\mu_3(\xi^{(A)})=-6,144,000,000$, $\mu_4(\xi^{(A)})=21,299,200,000,000$, $\gamma_1(\xi^{(A)}) = -\frac{3}{2}$, $\gamma_2(\xi^{(A)})=\frac{1}{4}$, $H(\xi^{(A)})\approx 0.721928$, and in B $A_p^{(B)} = 3,000$, $A_q^{(B)} = 3,000$, $p^{(B)} = 1$, $q^{(B)} = 0$, $E(\xi^{(B)})=3,000$, $\sqrt{D(\xi^{(B)})} = 0$, $\mu_3(\xi^{(B)})=0$, $\mu_4(\xi^{(B)})=0$, $\gamma_1(\xi^{(B)})$ and $\gamma_2(\xi^{(B)})$ are undefined, $H(\xi^{(B)}) = 0$. Computing entropy, we follow \cite[p. 5]{khinchin1953} and set $p\log_2(p)=0$ for $p = 0$. Undoubtedly, declining the greater $E(\xi)$, voters choose a greater \textit{award}. The \textit{"paradox of irrationality"} arises because in B the award coincides with $E(\xi^{(B)})=3,000$ but in A it is not $E(\xi^{(A)})=3,200$. \textit{Indeed, in A the award is random but the mathematical expectation is not. Already because of this the award is not the mathematical expectation.}

\paragraph{Sample Mean.} In order to see better what the award in Prospect A is, the author has "tortured" one of the human beings and formulated four variations of Problem 3, where he might feel comfortable choosing A. A gambler may
\begin{enumerate}
\item[1)]
play a fixed number of times known in advance, and get the mean;
\item[2)]
play unlimitedly, choose the stopping time, and get the known mean;
\item[3)]
play unlimitedly, choose the stopping time, and get the last known value;
\item[4)] gather any number of helpers; all choose A once; gains are summed; the gambler gets the mean.
\end{enumerate}
Each variation wastes gambler's and helpers' time, if all select B. For the author, choosing A, each variation intensifies his feeling to gain more following to the expectation and even more in the third variation. For him, \textit{to go or not to go with the fixed positive mathematical expectation} depends on how many times $N$ the game can be played and the cost of each game, if the expectation does not include the latter. In \cite{salov2014}, $N$ is represented by the \textit{floor function} of the product of the playing frequency $\nu$ and time $t$: $N = \lfloor \nu \times t \rfloor$. The \textit{sample mean} $a_1$ depends on the numbers of outcomes $N_p$ and $N_q$ in $N = N_p + N_q$ trials
\begin{equation}
\label{EqSampleMean}
a_1 = \frac{N_p A_p + N_q A_q}{N} = \frac{N_p A_p + (N - N_p)A_q}{N} = \frac{N_p(A_p - A_q)}{\lfloor \nu t \rfloor} + A_q.
\end{equation}
The $a_1$ is random due to $N_p$. If the fixed $N \ge 1$ games represent \textit{independent identically distributed}, i.i.d., binary variables, then $N_p$ obeys a \textit{binomial distribution}. In Prospect B of Problem 3, $p^{(B)}=1$, $N_p^{(B)} = N = 1$ make $a_1^{(B)} = 3,000$ a \textit{constant award}. In Prospect A of Problem 3 and variations 1, 2, and 4, the $a_1^{(A)}$ is a \textit{random award}. \textit{One cares about all properties of the award but not only one constant} $E(a_1) = E(\xi),$ \textit{if those are responsible for getting nothing.}

Variation 1 fixes $N$ in advance. Variation 2 makes $N$ random and dependent on the observed $a_1$. Variation 3 makes $N$ random, values $N_p>1$ irrelevant, and the award 4,000 likely. Variation 4 switches to the number of gamblers $N$.

For a fixed $N \ge 1$, $E(N_p) = Np$ \cite[p. 175]{gnedenko1988} and
\begin{equation}
\label{EqMeanOfSampleMean}
E(a_1) = E\left( \frac{N_p A_p + (N - N_p)A_q}{N} \right) = (A_p - A_q)p + A_q = E(\xi) = \alpha_1(\xi).
\end{equation}
\begin{equation}
\label{EqVarianceOfSampleMean}
D(a_1) = E([a_1 - E(a_1)]^2) = E(a_1^2) - E^2(a_1) = \frac{(A_p-A_q)^2pq}{N}=\frac{D(\xi)}{\lfloor \nu t \rfloor}.
\end{equation}
The variance decreases with increasing $N$. This formula is in agreement with the \textit{limit theorems} \cite{gnedenko1949} applicable to the mean sum of i.i.d. variables with finite variance $D(\xi) = (A_p-A_q)^2 p q$ in \textit{Bernoulli trials}. An analytic method for computing \textit{beginning moments} $\alpha_k$ of sums of i.i.d. random variables is suggested in \cite{packwood2012}. Then, the \textit{central moments} $\mu_k$ can be expressed via $\alpha_k = E(\xi^k)$ as
\begin{displaymath}
\mu_k = E([\xi - E(\xi)]^k) = E([\xi - \alpha_1]^k) = E\left( \sum_{j = 0}^{j = k} \frac{k!}{j!(k-j)!}\xi^j (-\alpha_1)^{k-j} \right) =
\end{displaymath}
\begin{displaymath}
 = \sum_{j = 0}^{j = k} \frac{k!}{j!(k-j)!} (-\alpha_1)^{k-j} E(\xi^j) = \sum_{j = 0}^{j = k} \frac{k!}{j!(k-j)!} (-\alpha_1)^{k-j} \alpha_j.
\end{displaymath}
This yields the first four $\mu_k$ presented in \cite[pp. 67 - 72]{gnedenko1949}
\begin{displaymath}
\mu_0 = \alpha_0 = 1, \; \mu_1= -\alpha_1 + \alpha_1 = 0, \; \mu_2 = \alpha_2 - \alpha_1^2,
\end{displaymath}
\begin{displaymath}
\mu_3 = \alpha_3 - 3 \alpha_1 \alpha_2 + 2 \alpha_1^3, \; \mu_4 = \alpha_4 - 4 \alpha_1 \alpha_3 + 6 \alpha_1^2 \alpha_2 - 3 \alpha_1^4.
\end{displaymath}
From Equations \ref{EqSampleMean} and \ref{EqMeanOfSampleMean} $\mu_k(a_1)=E([a_1-E(a_1)]^k) = E([\frac{A_p-A_q}{N}]^k[N_p - Np]^k) = (\frac{A_p-A_q}{N})^kE([N_p - Np]^k)$, where the second factor is the $k$th central moment of a binomial distribution. Due to the factor $(\frac{A_p-A_q}{N})^k$ the third standardized moment $\gamma_1$ depends on the sign $-1, 0, 1$ of the difference $A_p - A_q$ but not individual outcomes. The standardized moment $\gamma_2$ does not dependent on the outcomes at all. Formulas for the third and fourth beginning and central moments of a binomial distribution are found in \cite{korn1968}. After accounting the first factor we get
\begin{equation}
\label{EqThirdCentralMomentOfSampleMean}
\mu_3(a_1)=\frac{(A_p-A_q)^3p(1-p)(1-2p)}{N^2}=\frac{\mu_3(\xi)}{N^2}=\frac{\mu_3(\xi)}{\lfloor \nu t\rfloor^2},
\end{equation}
\begin{equation}
\label{EqFourthCentralMomentOfSampleMean}
\mu_4(a_1) = \frac{(A_p-A_q)^4p(1-p)[1 + 3(N-2)p(1-p)]}{N^3},
\end{equation}
\begin{equation}
\label{EqSkewnessOfSampleMean}
\gamma_1(a_1) = \frac{A_p-A_q}{|A_p-A_q|} \times \frac{1-2p}{\sqrt{Np(1-p)}} = \frac{\gamma_1(\xi)}{\sqrt{N}}=\frac{\gamma_1(\xi)}{\sqrt{\lfloor \nu t \rfloor}},
\end{equation}
\begin{equation}
\label{EqExcessKurtosisOfSampleMean}
\gamma_2(a_1) = \frac{1-6p+6p^2}{Np(1-p)}=\frac{\gamma_2(\xi)}{N}=\frac{\gamma_2(\xi)}{\lfloor \nu t \rfloor}.
\end{equation}

The sample mean, its variance, third and fourth central moments, skewness, and excess kurtosis in Equations \ref{EqSampleMean}, \ref{EqVarianceOfSampleMean} - \ref{EqExcessKurtosisOfSampleMean} depend on time. In contrast, the mean of sample mean in Equation \ref{EqMeanOfSampleMean} and its extreme values do not dependent on time. Under other equal conditions, a \textit{big} $N$ achieved in a \textit{short} time without an extra cost makes $a_1$ more symmetric, and certain for the author. It makes $a_1$ less deviating from $E(\xi)$ and "erases" a difference between them. In Problem 3, $N=1$ is the least. \textit{The frequency of games and playing time are important for increasing confidence as long as they determine the actual number of games.} In lotteries, the $\nu$ and $t$ can be compared with the life duration.

\section{Mega Millions}

\begin{figure}
  \centering
  \includegraphics[width=130mm]{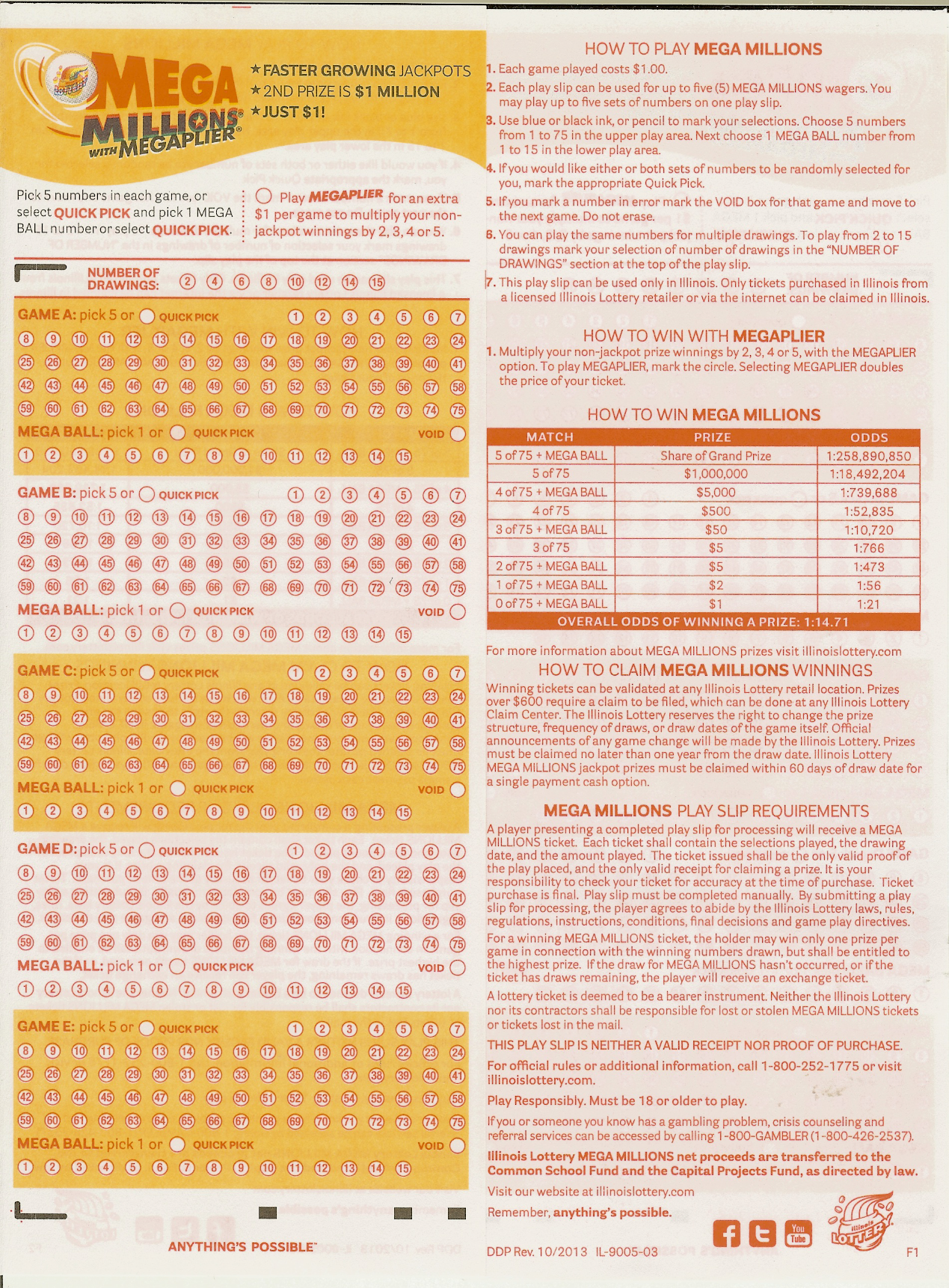}
  \caption[FigLottery]
   {The front (left) and back (right) sides of the Mega Millions with Megaplier play slip in February 2014, Illinois, U.S.A.}
  \label{FigLottery}
\end{figure}

Mathematical expectations are at the heart of lotteries. The history of the \textit{Big Game} began on August 31, 1996. In May 2002 the game was renamed to \textit{Mega Millions}. Since then, the rules have been changing. Currently, the initial jackpot is \$15 million. To win, one has to guess five of 75 numbers in the \textit{Game field} and one of 15 numbers in the \textit{Mega Ball field}, Figure \ref{FigLottery}. A slip allows to play up to five \$1 games. If no tickets win the jackpot, then it grows. Table \ref{TblLottery}, created using data of \url{http://www.megamillions.com/history-of-the-game}, illustrates the jackpot history.

\paragraph{Odds.} Let $F$ is the number of balls in a basket. There are $f$ red and $F-f$ black balls. Ignoring the order, $R$ balls are taken out without replacement with equal chances. The probability to get $r$ red balls among the selected is given by the \textit{hypergeometric distribution} \cite[pp. 55 - 59]{feller1964}
\begin{displaymath}
p = \frac{C(f,r)C(F-f,R-r)}{C(F,R)} = \frac{C(R,r)C(F-R,f-r)}{C(F,f)}.
\end{displaymath}
where $C(N, n) = \frac{N!}{n!(N-n)!}$ is the binomial coefficient. Here, $r \le R$ and $r \le f$.
\begin{center}
\begin{longtable}{|c|r|r|c|r|r|c|r|r|}
\caption[History of Mega Millions]{History of Mega Millions Jackpot. Dates of drawings, jackpot amounts in million U.S.A. dollars, and numbers of tickets winning the jackpot.} \label{TblLottery} \\
 \hline
 \multicolumn{1}{|c|}{Date} &
 \multicolumn{1}{c|}{Amount} &
 \multicolumn{1}{c|}{\#} &
 \multicolumn{1}{|c|}{Date} &
 \multicolumn{1}{c|}{Amount} &
 \multicolumn{1}{c|}{\#} &
 \multicolumn{1}{|c|}{Date} &
 \multicolumn{1}{c|}{Amount} &
 \multicolumn{1}{c|}{\#}\\
 \hline 
 \endfirsthead
 \multicolumn{9}{c}%
 {\tablename\ \thetable{} -- continued from previous page} \\
 \hline
 \multicolumn{1}{|c|}{Date} &
 \multicolumn{1}{c|}{Amount} &
 \multicolumn{1}{c|}{\#} &
 \multicolumn{1}{|c|}{Date} &
 \multicolumn{1}{c|}{Amount} &
 \multicolumn{1}{c|}{\#} &
 \multicolumn{1}{|c|}{Date} &
 \multicolumn{1}{c|}{Amount} &
 \multicolumn{1}{c|}{\#}\\
 \hline 
 \endhead
 \hline \multicolumn{9}{|r|}{{Continued on next page}} \\ \hline
 \endfoot
 \hline
 \endlastfoot
1/3/2014 & 61 & 1 & 3/5/2010 & 134 & 1 & 6/16/2006 & 35 & 1\\
12/17/2013 & $648^*$ & 2 & 1/29/2010 & 144 & 1 & 6/2/2006 & 47 & 1\\
10/1/2013 & 189 & 1 & 12/22/2009 & 165 & 1 & 5/16/2006 & 94 & 1\\
7/26/2013 & 19 & 1 & 11/10/2009 & 77 & 2 & 4/18/2006 & 265 & 1\\
7/16/2013 & 20 & 1 & 10/16/2009 & 200 & 1 & 2/28/2006 & 270 & 1\\
7/5/2013 & 80 & 1 & 9/1/2009 & 12 & 1 & 1/6/2006 & 15 & 1\\
5/31/2013 & 30 & 1 & 8/28/2009 & 336 & 2 & 12/30/2005 & 88 & 1\\
5/17/2013 & 198 & 2 & 7/7/2009 & 133 & 1 & 11/29/2005 & 35 & 2\\
3/12/2013 & 41 & 1 & 5/29/2009 & 35 & 1 & 11/15/2005 & 315 & 1\\
2/19/2013 & 26 & 1 & 5/15/2009 & 38 & 2 & 9/16/2005 & 258 & 1\\
2/5/2013 & 19 & 1 & 5/1/2009 & 227 & 3 & 7/22/2005 & 170 & 1\\
1/25/2013 & 89 & 1 & 3/13/2009 & 26 & 1 & 6/3/2005 & 106 & 1\\
12/14/2012 & 35 & 1 & 3/3/2009 & 216 & 1 & 4/22/2005 & 208 & 1\\
11/27/2012 & 50 & 1 & 1/13/2009 & 22 & 1 & 3/1/2005 & 115 & 1\\
11/2/2012 & 33 & 1 & 1/2/2009 & 47 & 1 & 1/18/2005 & 131 & 1\\
10/16/2012 & 61 & 2 & 12/12/2008 & 207 & 1 & 12/3/2004 & 25 & 1\\
9/18/2012 & 14 & 1 & 10/21/2008 & 42 & 1 & 11/19/2004 & 149 & 1\\
9/11/2012 & 120 & 1 & 10/3/2008 & 42 & 1 & 10/1/2004 & 106 & 1\\
7/27/2012 & 52 & 1 & 9/16/2008 & 15 & 1 & 8/20/2004 & 52 & 1\\
7/3/2012 & 85 & 1 & 9/9/2008 & 24 & 1 & 7/27/2004 & 10 & 1\\
5/29/2012 & 32 & 1 & 8/29/2008 & 133 & 1 & 7/23/2004 & 47 & 1\\
5/15/2012 & 25 & 1 & 7/22/2008 & 126 & 1 & 7/2/2004 & $294^*$ & 1\\
5/4/2012 & 118 & 1 & 6/13/2008 & 57 & 1 & 5/7/2004 & 67 & 1\\
3/30/2012 & $656^*$ & 3 & 5/23/2008 & 17 & 2 & 4/9/2004 & 109 & 1\\
1/24/2012 & 72 & 1 & 5/16/2008 & 196 & 1 & 3/2/2004 & 21 & 1\\
12/27/2011 & 208 & 1 & 4/1/2008 & 136 & 1 & 2/20/2004 & 239 & 1\\
11/1/2011 & 78 & 1 & 2/22/2008 & 275 & 1 & 12/30/2003 & 162 & 1\\
9/30/2011 & 113 & 2 & 1/1/2008 & 33 & 1 & 11/11/2003 & 70 & 3\\
8/19/2011 & 32 & 1 & 12/18/2007 & 163 & 2 & 10/7/2003 & 12 & 1\\
8/5/2011 & 99 & 1 & 11/2/2007 & 75 & 1 & 9/30/2003 & 150 & 1\\
7/1/2011 & 107 & 1 & 10/5/2007 & 27 & 1 & 8/8/2003 & 50 & 1\\
5/27/2011 & 35 & 1 & 9/25/2007 & 12 & 1 & 7/11/2003 & 34 & 1\\
5/13/2011 & 27 & 1 & 9/21/2007 & 61 & 1 & 6/20/2003 & 183 & 1\\
5/3/2011 & 51 & 1 & 8/31/2007 & 330 & 4 & 4/25/2003 & 46 & 1\\
4/15/2011 & 72 & 1 & 7/6/2007 & 128 & 1 & 3/28/2003 & 20 & 1\\
3/25/2011 & 319 & 1 & 5/29/2007 & 44 & 1 & 3/14/2003 & 43 & 1\\
2/1/2011 & 93 & 2 & 5/11/2007 & 113 & 1 & 2/18/2003 & 12 & 1\\
1/4/2011 & $380^*$ & 2 & 4/6/2007 & 105 & 1 & 2/11/2003 & 128 & 1\\
11/9/2010 & 25 & 1 & 3/6/2007 & $390^*$ & 2 & 12/24/2002 & 68 & 1\\
10/29/2010 & 141 & 1 & 1/9/2007 & 125 & 1 & 11/19/2002 & 16 & 1\\
9/17/2010 & 54 & 1 & 12/1/2006 & 40 & 1 & 11/8/2002 & 93 & 1\\
8/27/2010 & 135 & 1 & 11/14/2006 & 75 & 1 & 9/27/2002 & 37 & 1\\
7/16/2010 & 64 & 1 & 10/17/2006 & 55 & 1 & 9/6/2002 & 17 & 1\\
6/22/2010 & 26 & 1 & 9/26/2006 & 15 & 1 & 8/27/2002 & 108 & 1\\
6/11/2010 & 36 & 1 & 9/19/2006 & 12 & 1 & 7/16/2002 & 165 & 1\\
5/28/2010 & 12 & 1 & 9/15/2006 & 163 & 1 & 5/24/2002 & 12 & 1\\
5/25/2010 & 64 & 1 & 8/1/2006 & 31 & 1 & 5/17/2002 & 28 & 1\\
5/4/2010 & 266 & 1 & 7/18/2006 & 49 & 1 & & & \\
3/12/2010 & 20 & 1 & 6/27/2006 & 24 & 1 & & &
\end{longtable}
\end{center}
If there are two baskets represented by $F_1$, $f_1$, $F_1-f_1$ and $F_2$, $f_2$, $F_2-f_2$ and $R_1$ and $R_2$ balls are taken out of each, then the probability to get $r_1$ and $r_2$ red balls distinguished between baskets is the product $p_1p_2$. For $m$ "independent" baskets the reciprocal $\prod_{i=1}^{i=m} \frac{1}{p_i}$ is equal to
\begin{displaymath}
Q = \prod_{i=1}^{i=m} \frac{C(F_i,R_i)}{C(f_i,r_i)C(F_i-f_i,R_i-r_i)} = \prod_{i=1}^{i=m} \frac{C(F_i,f_i)}{C(R_i,r_i)C(F_i-R_i,f_i-r_i)}.
\end{displaymath}
For a particular case $f_i = R_i$ the formula
\begin{equation}
\label{EqLottery}
Q(r_1, \; \dots, \; r_m) = \prod_{i=1}^{i=m} \frac{C(F_i,f_i)}{C(f_i,r_i)C(F_i-f_i,f_i-r_i)}.
\end{equation}
describes reciprocal odds in the current Mega Millions and \textit{Power Ball} lotteries. The red and black balls are winning and losing numbers. The two baskets are the Game and Mega Ball and Game and \textit{Powerball} fields on slips. The word \textit{field} justifies the denominations $F_i$ and $f_i$.

For Mega Millions $m=2$, $F_1=75$, $f_1=5$, $F_2=15$, $f_2=1$. The jackpot's $Q(5,1) = \frac{C(75,5)}{C(5,5)C(70,0)}\frac{C(15,1)}{C(1,1)C(14,0)}=17259390 \times 15 = 258890850$. This coincides with the entry in "How to Win Mega Millions", Figure \ref{FigLottery}. The values $Q(5,0) = 18492203.571$, $Q(4,1) = 739688.143$, $Q(4,0) = 52834.867$, $Q(3,1) = 10720.118$, $Q(3,0) = 765.723$, $Q(2,1) = 472.946$, $Q(1,1) = 56.471$, $Q(0,1) = 21.391$ match remaining entries. The sum of the nine reciprocals of $Q$ is equal to $0.0679916034$. This is the probability of winning something per ticket with one chance in 14.707698.

\paragraph{Numbers of tickets.} For drawing on January 31, 2014, 1,440,661 winning tickets were announced. The jackpot was not won. If each ticket would be equally probable, then their total number would be $\approx 1,440,661 \times 14.707698 = 21,188,807$ including $21,188,807 - 1,440,661 = 19,748,146$ losing tickets.

\paragraph{Expectations.} All awards except the \textit{effective jackpot} $J^*$, the odds given by Equation \ref{EqLottery}, and ticket price $P = \$1$ are fixed. The mathematical expectation of the \textit{profit and loss}, $P\&L$, of the game without \textit{Megaplier} and tax is equal to (see Figure \ref{FigLottery})
\begin{displaymath}
E(P\&L) = \frac{J^*}{Q(5,1)}+\frac{1,000,000}{Q(5,0)}+\frac{5,000}{Q(4,1)}+\frac{500}{Q(4,0)}+\frac{50}{Q(3,1)}+\frac{5}{Q(3,0)}+
\end{displaymath}
\begin{displaymath}
+\frac{5}{Q(2,1)}+\frac{2}{Q(1,1)}+\frac{1}{Q(0,1)}-P = \frac{J^*}{258,890,850} - 0.82576841166846955.
\end{displaymath}
The jackpot and large prizes \$1,000,000 and \$5,000 are taxable. With 40 percent tax the last two numbers decrease to \$600,000 and \$3,000 yielding
\begin{displaymath}
E(P\&L) = \frac{J^*}{258,890,850} - 0.85010299127991584.
\end{displaymath}
It would be positive for $\$220,083,886 < J^*$.

For Powerball $m=2$, $F_1=59$, $f_1=5$, $F_2=35$, $f_2=1$, $P=2$. Using Equation \ref{EqLottery} we get $Q(5,1) = 175223510$, $Q(5,0) = 5153632.647$, $Q(4,1) = 648975.963$, $Q(4,0) = 19087.528$, $Q(3,1) = 12244.829$, $Q(3,0) = 360.142$, $Q(2,1) = 706.432$, $Q(1,1) = 110.813$, $Q(0,1) = 55.406$. Figure of a Powerball slip is omitted. The expectation without  \textit{Powerplay} and tax is
\begin{displaymath}
E(P\&L) = \frac{J^*}{Q(5,1)}+\frac{1,000,000}{Q(5,0)}+\frac{10,000}{Q(4,1)}+\frac{100}{Q(4,0)}+\frac{100}{Q(3,1)}+\frac{7}{Q(3,0)}+
\end{displaymath}
\begin{displaymath}
+\frac{7}{Q(2,1)}+\frac{4}{Q(1,1)}+\frac{4}{Q(0,1)}-2 = \frac{J^*}{175,223,510} - 1.6395111592046067.
\end{displaymath}
Decreasing \$1,000,000 and \$10,000 to \$600,000 and \$6,000 after 40\% tax yields
\begin{displaymath}
E(P\&L) = \frac{J^*}{175,223,510} - 1.7232898713192083.
\end{displaymath}
It would be positive for $\$301,960,900 < J^*$.

The effective jackpot $J^*$ should differ from the quoted $J$. The jackpot in Table \ref{TblLottery} assumes annuity payments. It has to be converted to a cash payout for declining annuity. The latter splits between all tickets winning jackpot and should be adjusted. Finally, the adjusted cash payout is a subject for tax.

The first step based on the \textit{ordinary annuity} formula \cite[p. 13]{fabozzi1996} yields $J_1^*$
\begin{displaymath}
J_1^* = \frac{years \times rate}{(1 + rate)^{years} - 1}J.
\end{displaymath}
For 30 years and rates of 2, 3, or 5 percent, the coefficient in front of $J$ is equal to 0.7395, 0.6306, or 0.4515.

\begin{figure}[!h]
  \centering
  \includegraphics[width=130mm]{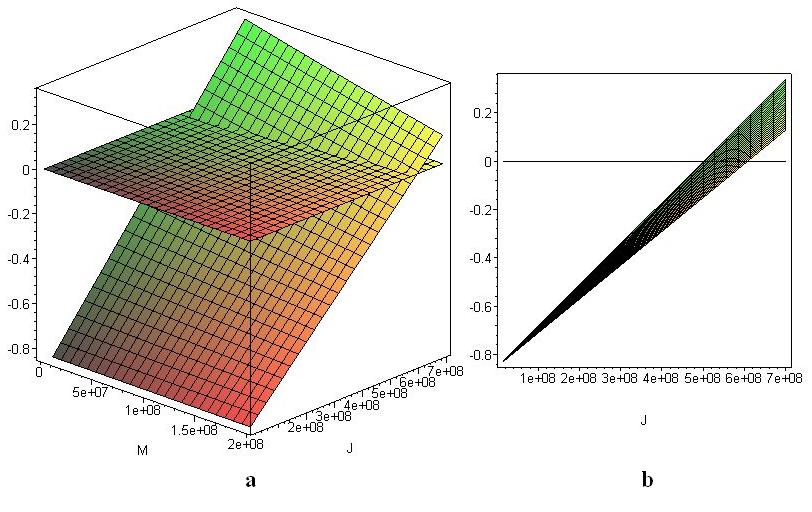}
  \caption[FigLotteryME]
   {Mathematical expectation of Mega Millions lottery as a function of the announced jackpot $J$ and purchased number of tickets $M$, $E(P\&L) = 1.71 \times 10^{-9} \times coeff_{2013}(M) \times J - 0.85$, given by Equation \ref{EqLotteryME} with $years=30$ and $rate=0.02$. Two projections a and b and the plane $0 \times M + 0 \times J$ help to see better the range of parameters making the expectation positive. Plots are done using Maple 10 from Maplesoft.}
  \label{FigLotteryME}
\end{figure}

The second step has to take into consideration the number of purchased tickets $M$. If each ticket wins jackpot equally likely with probability $1/Q(5,1)$ and they are filled independently, then the probability of $K$ winning coincidences is binomial $C(M,K)(\frac{1}{Q(5,1)})^K (1 - \frac{1}{Q(5,1)})^{M-K}$. Each winner gets $\frac{J_1^*}{K}$. For $K=0$ the payoff is zero. The mathematical expectation of the non-zero payoffs is
\begin{displaymath}
J_1^* \sum_{K=1}^{M} \frac{C(M,K)}{K} \left( \frac{1}{Q(5,1)} \right)^K \left(1 - \frac{1}{Q(5,1)} \right)^{M-K}
\end{displaymath}
The probability that it is won is equal to
\begin{displaymath}
\sum_{K=1}^{M} C(M,K)\left(\frac{1}{Q(5,1)}\right)^K \left(1 - \frac{1}{Q(5,1)}\right)^{M-K} = 1 - \left(1 - \frac{1}{Q(5,1)} \right)^M 
\end{displaymath}
For $M$ equal to 10,000,000 and 100,000,000 it is 0.03788983372734210 and 0.3204083454349182. The weighted jackpot yielding the payoff expectation is
\begin{displaymath}
J_2^* = \frac{\sum_{K=1}^{M} \frac{C(M,K)}{K} \left( \frac{1}{Q(5,1)} \right)^K \left(1 - \frac{1}{Q(5,1)} \right)^{M-K}}{1 - \left(1 - \frac{1}{Q(5,1)} \right)^M}J_1^* = coeff(M) \times J_1^*.
\end{displaymath}
The third step is to apply the tax percent and get $J^* = (1 - tax) \times J_2^*$. Finally, the mathematical expectation in Mega Millions with $tax = 0.4$ is
\begin{equation}
\label{EqLotteryME}
\begin{split}
E(P\&L) = \\
& \frac{0.6 \times years \times rate \times \sum_{K=1}^{M} \frac{C(M,K)}{K} \left( \frac{1}{Q(5,1)} \right)^K \left(1 - \frac{1}{Q(5,1)} \right)^{M-K}}
{258,890,850 \times  \left((1 + rate)^{years} - 1\right) \times \left(1 - \left(1 - \frac{1}{Q(5,1)} \right)^M\right)} J \\
& - 0.85010299127991584.
\end{split}
\end{equation}
Under other equal conditions $E(P\&L)$ is the function of the announced jackpot $J$ and the number of purchased tickets $M$, Figure \ref{FigLotteryME}. The former is known prior a drawing. The latter can be estimated multiplying by three the difference between the current and previous, if it was not won, cash jackpot amounts $J_1^*$. This assumes that one third of the collected money was used for increasing the cash jackpot and the current number of tickets is not less.

With the initial jackpot \$15,000,000 the mathematical expectation of profits and losses in Mega Millions is negative. Equation \ref{EqLotteryME} and Figure \ref{FigLotteryME} indicate that it can be positive, even, after declining the annuity option, splitting jackpot between tickets winning it, and taxes. Since 2002 until 2005 it was needed to guess 5 of 52 and 1 of 52 numbers. Since 2005 until 2013 - 5 of 56 and 1 of 46 numbers. The eight prizes after jackpot were \$250,000, \$10,000, \$150, \$150, \$7, \$10, \$3, and \$2. Equations for the mathematical expectations of profit and loss after 2002, 2005, and 2013 are $3.28 \times 10^{-9} \times coeff_{2002}(M) \times J - 0.82$, $2.53 \times 10^{-9} \times coeff_{2005}(M) \times J - 0.85$, and $1.71 \times 10^{-9} \times coeff_{2013}(M) \times J - 0.85$. With $coeff(M)=0.9$, the expectations are positive for $J_{2002} > \$278$ million, $J_{2005} > \$373$ million, and $J_{2013} > \$552$ million. Table \ref{TblLottery} marks by stars five such jackpot candidates occurred in 12 years. Let us assume that a gambler decides to play only, when the expectation becomes positive. \textit{Does it mean that the game under such a favorable condition is for certain?}

\paragraph{Game for Sure?} Positive $E(P\&L)$ does not make the "favorable" drawings a \textit{game for sure}. During 80 years of an adult life a gambler may participate in such infrequent events $\frac{80 \times 5}{12} \approx 33$ times. A gambler betting twice every week can make $80 * 52 * 2 \approx 8,320$ attempts. Both numbers are insignificant comparing with the odds 1 : 258,890,850. The discrete nature of gains and losses implies either a chain of -\$1 (ignore smaller prizes) or one win making previous losses unimportant. \textit{There are no drawings returning the favorable expected values counted by a few dozen of cents. The potential extreme gain and actual low cost are more important than the mathematical expectation for a decision to play}. 

Let one in a favorable case with \$552,000,000 jackpot agrees to get annuity, pays for the tickets \$258,890,850, and technically fills them by unique combinations of numbers. This ensures winning the jackpot. We have seen that the probability of multiple wins of the jackpot does not reduce much the mean number of winning tickets. The coefficient 0.8 - 0.9 increases 1 to $1.1 - 1.25$. Tax 40\% takes more. But this is the mean number of winning tickets. In reality it is either 1, 2, or 3 ... While the tax is fixed (not for annuity payments), the number of winning tickets is random. \textit{Just only two else claim a winning ticket, the favorable attempt becomes a financial disaster because of the sharing. Again, not only the mathematical expectation of the number of winning tickets but a probability that the actual number will deviate from it look important}.

\section{Problem 3' from Kahneman and Tversky}

Problem 3 was changed to 3' \cite[268]{kahneman1979}. Now, Prospect A' is losing 4,000 with the probability 0.8 and Prospect B' is losing 3,000 for sure. In A' the mathematical expectation $0.2 \times 0 + 0.8 \times -4,000 = -3,200$ is less than in B' -3,000. However,  92 percent of 95 respondents choose A'. Kahneman and Tversky label this pattern, switching the signs of outcomes, the \textit{reflection effect}. Respondents attempt to avoid a loss with the small probability 0.2. Again, the author "feels" that losing the sample mean $a_1 = \frac{N_2 \times (-4,000)}{N}$ characterized by the standard deviation $\sqrt{D(a_1)} = \sqrt{\frac{(-4,000)^2 \times 0.8 \times 0.2}{N}} = \frac{1,600}{\sqrt{N}}$ in a \textit{big  guaranteed} number of games $N = \lfloor \nu \times t \rfloor$ without extra cost makes A' worse than B'.

Properties of variables for Problem 3' are $A_p^{(A')} = 0$, $A_q^{(A')} = -4,000$, $p^{(A')} = 0.2$, $q^{(A')} = 0.8$, $E(\xi^{(A')})=-3,200$, \textit{standard deviation} $\sqrt{D(\xi^{(A')})} = 1,600$, $\mu_3(\xi^{(A')})=6,144,000,000$, $\mu_4(\xi^{(A')})=21,299,200,000,000$, $\gamma_1(\xi^{(A')}) = \frac{3}{2}$, $\gamma_2(\xi^{(A')})=\frac{1}{4}$, $H(\xi^{(A')})\approx 0.721928$, and $A_p^{(B')} = -3,000$, $A_q^{(B')} = -3,000$, $p^{(B')} = 1$, $q^{(B')} = 0$, $E(\xi^{(B')})=-3,000$, $\sqrt{D(\xi^{(B')})} = 0$, $\mu_3(\xi^{(B')})=0$, $\mu_4(\xi^{(B')})=0$, $\gamma_1(\xi^{(B')})$ and $\gamma_2(\xi^{(B')})$ are undefined, $H(\xi^{(B')}) = 0$. Mathematical expectations, third central moments, skewness of the random variables in Problems 3A and 3'A are equal by absolute value and have opposite sign. Mathematical expectations of the deterministic variables in Problems 3B and 3'B are equal by absolute value and have opposite sign.

\section{Livermore about the hope and fear}

The author has "discovered" \cite[p. 5]{salov2013} that the legendary speculator Jesse Livermore \cite{livermore1940}, \cite{lefevre1923}, \cite{thomson1983}, \cite{smitten2001} guessed the certainty and reflection effects \cite{kahneman1979}.

Indeed, \cite[pp. 11 - 12]{livermore1940}: \textit{"... You buy a stock at \$30.00. The next day it has a quick run-up to \$32.00 or \$32.50. You immediately become fearful that if you don't take the profit, the next day you may see it fade away - so out you go with a small profit, when that is very time you should entertain all the hope in the world."} This is a certainty effect.

Further, \cite[pp. 12 - 13]{livermore1940}: \textit{"On the other hand, suppose you buy a stock at \$30.00, and the next day it goes to \$28.00, showing a two-point loss. You would not be fearful that the next day would possibly see a three-point loss or more. No, you would regard it merely as a temporary reaction, feeling certain that the next day it would recover its loss. ... That is when you should be fearful, because if you do not get out, you might be forced to take a much greater loss later on."} This is a reflection effect.

Livermore believes that a human being "injects a hope and fear into the business of speculation" and "is apt to get the two confused and in reverse positions". His 40 years trading wisdom is \cite[p. 13]{livermore1940} \textit{"Profits always take care of themselves, but losses never do"}. His intuition was trained by winning and losing several fortunes. He played the game many times feeling but not measuring the odds behind his advices. Empirical distributions of prices, their increments, and waiting times between transactions vary \cite{salov2013}. This strengthens uncertainty of profits and losses presenting a trading opportunity as the last one. Under such conditions following his advices is difficult psychologically and increasing $N = \lfloor \nu \times t \rfloor$ does not guarantee a fast convergence of $a_1$.

\section{The St. Petersburg paradox}

The author considered \textit{random sample means} as universal awards in Prospects and the St. Petersburg game independently on Khinchin \cite{khinchin1925} and only after that found his "forgotten" paper \cite{salov2014}. Khinchin concentrates on mathematical properties of random geometric and arithmetic means, which can explain the "paradox". He reviews psychology \cite[p. 330]{khinchin1925}, \cite[p. 7]{salov2014}):

\textit{"Let us notice only that in this case, of course, no speech may go about any mathematical paradox but at most about that the mathematical expectation is not always adequate to those worldly-psychological representations, which it is commonly connected to. In the case of the Petersburg game, it is often pointed to that Petr in his expectation of winning, naturally, orients not on the mathematical expectation of winning in a particular game, which is difficult to account psychologically, but on some average winning during big number of games. Such understanding of psychological prerequisites of the "paradox" puts in front of us a certain mathematical task, which can be formulated as follows: Find such an estimate of the mean winning of Petr during a big number of games, that its probability would go to unit with infinite increasing the number of games. However, it makes sense to say, that the task will get a quite determined sense only after a certain notion of the mean winning will be exactly defined. In the current note, we shall consider in details the set problem in two of the most simple (and also the most important) cases, namely in assumption that the mean winning is defined as the geometric and arithmetic mean of particular games"}.

Khinchin is indifferent, if $N$ is achievable. If "yes", then his two theorems work. For the author, the sample mean $a_1$ is an award in Prospects with $N = 1$. The case $N = \lfloor \nu \times t \rfloor$ is generic. The $a_1$ is a universal random award in the St. Petersburg game, Prospects, and variations 1, 2, and 4. For $N = 1$ it coincides with underlying random variables.

In a private conversation with Timur Misirpashaev the author discussed \textit{credit nuances} of the St. Petersburg game and believes that the topic "correlates" with Samuelson's \textit{bankruptcy} consideration \cite{samuelson1977} and the following Buffon's comment \cite[p. 82]{buffon1777}: \textit{"... All money on the Earth is not enough to accomplish this [VS: to pay the win], if the game stops on 40th trial, because it will require 1024 times more money than there exists in the entire kingdom of France"}. During the discussion, the author has proposed to pay \$3 for the right to play one time and \$18 to play 10 times, under a condition that in both cases the third party reserves the deposit \$128 to pay a win, if such will be drawn. This proposal was done after experimenting with the C++ program khinchin.cpp \cite[Appendix]{salov2014}.

Making decisions for the four variations and Prospects depends on probabilistic properties of $a_1$. The St. Petersburg game adds the credit uncertainty of paying the win. The creditworthiness is not in scope of this article.

\section{Prokhorov's estimates}

\begin{figure}
  \centering
  \includegraphics[width=90mm]{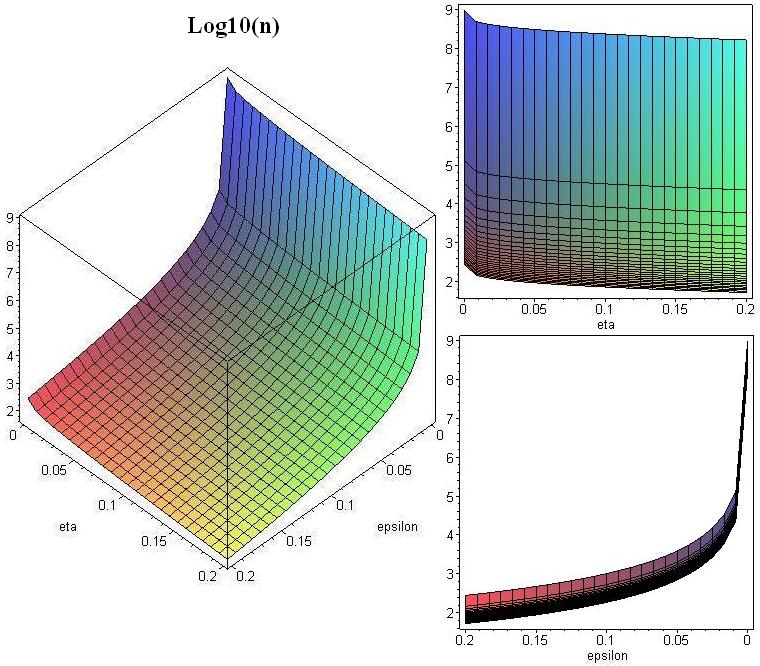}
  \caption[FigProkhorov]
   {3D dependence of decimal logarithm of the rightmost expression of Inequality \ref{EqProkhorov} on $\eta$ and $\epsilon$ and two projections. Plots are done using Maple 10 from Maplesoft}
  \label{FigProkhorov}
\end{figure}

If one choses between random variables $a_1^A$ and $a_1^B$, then a measure affecting the decision should account their properties. The sums and mean sums of i.i.d. variables obey the Laplace theorem, the law of big numbers, and the law of the iterated logarithm \cite{khinchin1932}, \cite{kolmogorov1929}, \cite{gnedenko1949}, \cite{kolmogorov1974}, \cite{prokhorov1983}, \cite{gnedenko1988}. Yurii Vasilevich Prokhorov presents estimates for the law of big numbers \cite[p. 281]{prokhorov1983}
\begin{equation}
\label{EqProkhorov}
\forall \epsilon > 0, \eta > 0, \; P\{|\frac{\mu_n}{n} - p| \le \epsilon \} > 1 - \eta \textrm{, if } n \ge n_0 > \frac{1 + \epsilon}{\epsilon^2} \log \frac{1}{\eta} + \frac{1}{\epsilon},
\end{equation}
where $\mu_n$ is the number of successes in $n$ Bernoulli $(p, 1 - p)$ trials and $P$ is the probability of inequality. Figure \ref{FigProkhorov} plots decimal logarithm of the right expression.
In Problem 3, $a_1^B = E(a_1^B) = E(\xi^B) = 3,000$ is constant and $a_1^A$ is random with $N = 1$. In variations 1, 2, and 4 with increasing $N$, $a_1^A \rightarrow E(a_1^A) = E(\xi^A) = 3,200$. If $a_1^A$ does not drop below $a_1^B = 3,000$, then it becomes attractive. Such a drop occurs, if $\frac{\mu_n}{n} \le \frac{A_1^B}{A_1^A} = \frac{3,000}{4,000} = \frac{3}{4}$. Thus, the threshold value is $\epsilon = |\frac{3}{4} - \frac{4}{5}| = \frac{1}{20} = 0.05$ and $N > 420\log\frac{1}{\eta}+20$. For $\eta = 0.01$ and $\eta = 0.001$, we get $P > 0.99$ and $P > 0.999$, if $N > 1,955$ and $N > 2,922$. It is assumed that $N = \lfloor \nu \times t \rfloor$ games can be played and there is no additional cost comparing with playing one game.

\section{Entropy of Sample Mean}

In probability theory, a \textit{full system of events} $A_1$, $A_2$, ..., $A_n$ is a group of events, where in each trial one and only one of them comes. The events, specified together with their probabilities $p1$, $p_2$, ..., $p_n$ ($p_k \ge 0$, $\sum_{k=1}^{k=n}p_k=1$), are named in \cite{khinchin1953} a \textit{finite schema}
\begin{displaymath}
\left( \begin{array}{cccc}
A_1 & A_2 & ... & A_n \\
p_1 & p_2 & ... & p_n
\end{array}\right).
\end{displaymath}
The Shannon's entropy of the finite schema is the function
\begin{displaymath}
H(p_1, p_2, ..., p_n) = -\sum_{k=1}^{k=n}p_k\log_2(p_k),
\end{displaymath}
where $p_k\log_2(p_k)=0$ for $p_k=0$. It is a reasonable measure of uncertainty of the finite schema. For a two-point random variable we get $H$ from Equation \ref{EqBinaryVariableEntropy}.

For two \textit{independent} finite scheme
\begin{displaymath}
\left( \begin{array}{cccc}
A_1 & A_2 & ... & A_n \\
p_1 & p_2 & ... & p_n
\end{array}\right), \;
\left( \begin{array}{cccc}
B_1 & B_2 & ... & B_m \\
q_1 & q_2 & ... & q_m
\end{array}\right)\end{displaymath}
the probability $\pi_{kl}$ of the joint events $A_k$ and $A_l$ is equal to $p_kq_l$ ($1 \le k \le n$, $1 \le l \le m$). The set of the joint events $A_kB_l$ form a new schema $AB$ - the \textit{union of scheme} $A$ and $B$. In this case, $H(AB)=H(A)+H(B)$. If for two scheme, representing two i.i.d. two-point variables, the four events $(A_p, A_p)$, $(A_p, A_q)$, $(A_q, A_p)$, $(A_q, A_q)$ are distinguished, then the entropy is equal to
\begin{displaymath}
\begin{split}
-p^2\log_2(p^2)-p(1-p)\log_2(p(1-p))-(1-p)p\log_2((1-p)p)-\\
(1-p)^2\log_2((1-p)^2)=2(-p\log_2(p)-(1-p)\log_2(1-p))=2H.
\end{split}
\end{displaymath}
For the sum of $N$ i.i.d. two-point variables, where $2^N$ events are distinguished, the entropy grows linearly with $N$ and is equal to $N \times H$. 

However, from profit or loss point of view, the events $(A_p, A_q)$ and $(A_q, A_p)$ are identical yielding the schema for sample mean money outcomes
\begin{displaymath}
\left( \begin{array}{cccc}
A_p & A_p + A_q & A_q \\
p^2 & 2p(1-p) & (1-p)^2
\end{array}\right).
\end{displaymath}
Its entropy is less than or equal to $2H$
\begin{displaymath}
-p^2\log_2(p^2)-2p(1-p)\log_2(2p(1-p))-(1-p)^2\log_2((1-p)^2)= 2H + 2p^2-2p,
\end{displaymath}
because $p^2-p \le 0$. From Equation \ref{EqSampleMean}, for $N \ge 1$ we deal with a binomial distribution of $N + 1$ outcomes. The schema for the sum of $N$ terms is
\begin{displaymath}
\left( \begin{array}{cccccc}
NA_q & (N-1)A_q+A_p & ... & NA_q+(A_p-A_q)N_p & ... & NA_p \\
(1-p)^N & Np(1-p)^{N-1} & ... & \frac{N!}{N_p!(N-N_p)!}p^{N_p}(1-p)^{N-N_p} & ... & p^N
\end{array}\right).
\end{displaymath}
The schema for the mean sum of $N$ terms is
\begin{displaymath}
\left( \begin{array}{cccccc}
A_q & \frac{(N-1)A_q+A_p}{N} & ... & A_q+(A_p-A_q)\frac{N_p}{N} & ... & A_p \\
(1-p)^N & Np(1-p)^{N-1} & ... & \frac{N!}{N_p!(N-N_p)!}p^{N_p}(1-p)^{N-N_p} & ... & p^N
\end{array}\right).
\end{displaymath}
For both scheme, where random $N_p=0, 1, \; \dots, \; N$,
\begin{displaymath}
H(p, N) = -\sum_{N_p=0}^{N_p=N}   \frac{N!}{N_p!(N-N_p)!}p^{N_p}(1-p)^{N-N_p} \log_2 \left (\frac{N!}{N_p!(N-N_p)!}p^{N_p}(1-p)^{N-N_p} \right ).
\end{displaymath}
$H(p, N)$ is positive for $0 < p < 1$ and $1 \le N$. The random sum $\xi_{sum}$ and sample mean $a_1$ are linear functions of the binomial random variable $N_p$: $\xi_{sum}=slope_{sum} \times N_p + intercept_{sum} = (A_p-A_q)N_p+A_q$ and $a_1=\frac{A_p-A_q}{N}N_p+A_q$.

Let us notice, that for a normal distribution with mean $\alpha_1$ and variance $\mu_2$ the \textit{differential} (or continuous) entropy is equal to
\begin{displaymath}
\begin{split}
H(\alpha_1, \mu_2)=-\int_{-\infty}^{\infty}\frac{1}{\sqrt{2\pi\mu_2}}e^{-\frac{(x-\alpha_1)^2}{2\mu_2}}\log_2 \left (\frac{1}{\sqrt{2\pi\mu_2}}e^{-\frac{(x-\alpha_1)^2}{2\mu_2}} \right )dx=\\
-\int_{-\infty}^{\infty}\frac{1}{\sqrt{2\pi\mu_2}}e^{-\frac{(x-\alpha_1)^2}{2\mu_2}} \left [\log_2(\frac{1}{\sqrt{2\pi\mu_2}}) - \frac{(x-\alpha_1)^2}{2\mu_2}\log_2(e) \right ]dx=\\
\log_2(\sqrt{2\pi\mu_2}) + \frac{\log_2(e)}{2} = \frac{\log_2(2\pi e \mu_2)}{2}.
\end{split}
\end{displaymath}
$H(\alpha_1, \mu_2)$ does not depend on $\alpha_1$ and can be negative for tiny variance $\mu_2$. From limit theorems, for $N \rightarrow \infty$ binomial distribution approaches normal with mean $Np$ and variance $Np(1-p)$. There is temptation to approximate the binomial entropy by $\frac{\log_2(2\pi e Np(1-p))}{2}$, which tends to infinity for $0 < p < 1$ and $N \rightarrow \infty$. Similarly, a sample mean distribution approaches normal with mean $(A_p-A_q)p+A_q$ and variance $\frac{(A_p-A_q)^2p(1-p)}{N}$. However, substituting variance into the normal distribution entropy expression yields $\frac{\log_2(2\pi e (A_p-A_q)^2p(1-p)))}{2}-\frac{\log_2(N)}{2}$, which becomes negative for $N \ge 2\pi e (A_p-A_q)^2p(1-p)$. This contradicts to non-negative value $H(p, N)$ and emphasizes the difference between discrete and differential entropy. Let us denote binomial probability $P_{N_{p}}=\frac{N!}{N_p!(N-N_p)!}p^{N_p}(1-p)^{N-N_p}$ and recollect that $\sum_{N_p=0}^{N_p=N} P_{N_{p}}=1$, $\sum_{N_p=0}^{N_p=N} N_{p}P_{N_{p}}=Np$, $P_0=(1-p)^N$, $P_1=Np(1-p)^{N-1}$, $P_{N-1}=Np^{N-1}(1-p)$, and $P_N=p^N$. Then,
\begin{displaymath}
\begin{split}
H(p, N)= -\sum_{N_p=0}^{N_p=N} P_{N_{p}} \log_2(P_{N_{p}}) = -\sum_{N_p=0}^{N_p=N} P_{N_{p}} \log_2 \left(\frac{N!}{N_p!(N-N_p)!}\right)\\
-\sum_{N_p=0}^{N_p=N} P_{N_{p}} N_p \log_2(p) - \sum_{N_p=0}^{N_p=N} P_{N_{p}} (N-N_p) \log_2(1-p)=\\
-\sum_{N_p=0}^{N_p=N} P_{N_{p}} \log_2 \left(\frac{N!}{N_p!(N-N_p)!}\right) - Np\log_2(p) - N(1-p)\log_2(1-p)=\\
N \times H -\sum_{N_p=0}^{N_p=N} P_{N_{p}} \log_2 \left(\frac{N!}{N_p!(N-N_p)!}\right).
\end{split}
\end{displaymath}
Since $1 \le \frac{N!}{N_p!(N-N_p)!}$, $H(p, N) \le N \times H$. Next,
\begin{displaymath}
\begin{split}
\sum_{N_p=0}^{N_p=N} P_{N_{p}} \log_2 \left(\frac{N!}{N_p!(N-N_p)!}\right)=(P_1+P_{N-1})\log_2(N)+\\
\sum_{N_p=2}^{N_p=N-2} P_{N_{p}} (\log_2(N!)-\log_2(N_p!)-\log_2((N-N_p)!)) \approx (P_1+P_{N-1})\log_2(N)+\\
\sum_{N_p=2}^{N_p=N-2} P_{N_{p}} ((N+\frac{1}{2})\log_2(N)-(N_p+\frac{1}{2})\log_2(N_p)-(N-N_p+\frac{1}{2})\log_2(N-N_p))-\\
\frac{1}{2}\log_2(2\pi)(1-P_0-P_1-P_{N-1}-P_N)=(P_1+P_{N-1})\log_2(N)+\\
((N+\frac{1}{2})\log_2(N)-\frac{1}{2}\log_2(2\pi))(1-P_0-P_1-P_{N-1}-P_N)-\\
\sum_{N_p=2}^{N_p=N-2} P_{N_{p}} ((N_p+\frac{1}{2})\log_2(N_p)+(N-N_p+\frac{1}{2})\log_2(N-N_p)).
\end{split}
\end{displaymath}
The Stirling's approximation $\ln(n!) \approx n\ln(n)-n+\frac{1}{2}\ln(2\pi n)=(n+\frac{1}{2})\ln(\frac{n}{e})+\frac{1}{2}\ln(2\pi e)$ and $\log_2(n!) \approx (n+\frac{1}{2})\log_2(\frac{n}{e})+\frac{1}{2}\log_2(2\pi e)$ has been applied. We get
\begin{displaymath}
\begin{split}
H(p,N) \approx N H - (P_1+P_{N-1})\log_2(N) -\\
((N+\frac{1}{2})\log_2(N)-\frac{1}{2}\log_2(2\pi))(1-P_0-P_1-P_{N-1}-P_N)+\\
\sum_{N_p=2}^{N_p=N-2} P_{N_{p}} ((N_p+\frac{1}{2})\log_2(N_p)+(N-N_p+\frac{1}{2})\log_2(N-N_p)),
\end{split}
\end{displaymath}
where evaluation of large factorials is avoided.
While the number of outcomes $N+1$ increases linearly with $N$, the change of corresponding binomial probabilities causes non-linear and slow growth of $H(p,N)$. The role of time here is in $N=\lfloor \nu t \rfloor$.  The $H(p,N)$ is common for the sample sum and mean sum. The range of the former $[NA_q,NA_p]$ is getting wider. The range of the latter $[A_q, A_p]$ remains intact with uniform increasing density of outcomes. The common for both sums discrete entropy, being a function of only probabilities, does not reflect these nuances. \textit{Increasing the discrete Shannon's entropy widely associated with uncertainty does not tell that the mean sum gets lower variance. However, it is the latter decreasing parameter causes author's impression that certainty of an award increases}.

\paragraph{Discrete vs. differential entropy.} In contrast with discrete entropy of a binomial distribution, the differential entropy of a normal one, approached by both sums for $N \rightarrow \infty$, better corresponds to a common point of view that a higher entropy associates with higher uncertainty. This happens because the expression for $H(\alpha_1, \mu_2)$ contains $\mu_2$ under logarithm. For the sample mean sum the entropy at the Gaussian limit grows by absolute value but with negative sign - decreases. The negative sign creates inconvenience. Accordingly, we shall use entropy for studying fractions of responds carefully and restrictedly. 

\section{Fractions of respondents}

Let there exists a function that maps a \textit{random} prospect to \textit{deterministic} money. If prospects and equivalents could be exchanged at any time without cost, then respondents choosing between two prospects, would \textit{rationally} select the greater value. The fractions would be zero and one and for equal values any. Divergence of opinion means that the function that eliminates uncertainty was unknown. In contrast, a \textit{risk neutral portfolio} of a stock and call options on it illustrates a possibility to eliminate randomness \cite{black1973}. \textit{We shall look for a different function that maps properties of random prospects to fractions of respondents,} which can be verified using questionnaires like in \cite{kahneman1979} and \cite{tversky1992}. 

Selecting a profit opportunity in Problem 3, majority \textit{excludes a chance to get nothing}. Choosing between losing scenarios in Problem 3', respondents want to \textit{avoid a loss completely}. If in Prospects 3B and 3A the sure, $p = 1$, and uncertain, $p = 0.8$, gains would be 1 but not 3,000 and a Mega Millions jackpot but not 4,000, then the author would favor 3A. Set the uncertain, $q = 0.8$, loss in Prospect 3'A substantially greater by absolute value than -4,000 (home, Mega Millions jackpot, ... \textit{life}) and 3'B with the sure, -3,000, loss looks as the lesser of evils. The fractions of voters depend on \textit{known properties} of awards or losses and poorly known nature of human beings. \textit{Can fractions of respondents be derived from properties of prospects?}

Several prospects \cite{kahneman1979}, \cite{tversky1992} are based on two outcomes $A_p$ and $A_q$ and corresponding probabilities $p$ and $q = 1 - p$. These are two-point variables, which the author considers as random sample means $a_1$, mean sums of the random i.i.d. two-point variables, where the number of summands $N = 1$. This generalization includes cases with $N = \lfloor \nu \times t \rfloor > 1$. The four \textit{independent} quantities $\{A_p, A_q, p, N\}$ fully characterize $a_1$. Without losing generality $A_q \le A_p$.

Prospect 3B is the deterministic award $a_1$ with $\{A_p = 3,000, A_q = 3,000, p = 1, N = 1\}$. Prospect 3A is the random award $a_1$ with $\{A_p = 4,000, A_q = 0, p = 0.8, N = 1\}$. Prospect 3'B is the deterministic loss with $\{A_p = -3,000, A_q = -3,000, p = 1, N = 1\}$. Prospect 3'A is the random loss with $\{A_p = 0, A_q = -4,000, p = 0.2, N = 1\}$. The two-point variable is deterministic, if $A_p = A_q$ and/or $p=0$ or $1$. Empirical fractions of respondents choosing between $a_1^{(1)}$ and $a_1^{(2)}$ summing to $1 = F^{(1)} + F^{(2)}$ should be approximated by model functions $f^{(1)}=f^{(1)}(a_1^{(1)}, a_1^{(2)}) = f^{(1)}(\{A_p^{(1)}, A_q^{(1)},p^{(1)},N^{(1)}\}, \{A_p^{(2)}, A_q^{(2)},p^{(2)},N^{(2)}\})$ and $f^{(2)} = 1 - f^{(1)}$. Ideal is a single function $f(a_1^{(1)}, a_1^{(2)})$ returning a fraction of respondents depending on the order of arguments. Then, $0 \le f(x, y) \le 1$ is a solution of the functional equation $f(x, y) + f(y, x) = 1$
\begin{equation}
\label{EqFunctional}
\begin{split}
f(\{A_p^{(1)}, A_q^{(1)},p^{(1)},N^{(1)}\}, \{A_p^{(2)}, A_q^{(2)},p^{(2)},N^{(2)}\}) + \\
+ f(\{A_p^{(2)}, A_q^{(2)},p^{(2)},N^{(2)}\}, \{A_p^{(1)}, A_q^{(1)},p^{(1)},N^{(1)}\}) = 1.
\end{split}
\end{equation}

\section{Predictable properties of $f^{(1)}$ and $f^{(2)}$}

For some arguments, fractions of respondents make \textit{rational sense}. Each discrete random variable with finite number of finite outcomes has a minimum and maximum value. Intervals determined by the extreme values of two variables can be disjoined, adjacent, or overlapping. An interval and variable is a point and constant, if $A_{min} = A_{max}$. The symbols $A_{min}^{(1)}$, $A_{max}^{(1)}$, $A_{min}^{(2)}$, $A_{max}^{(2)}$ form $4! = 24$ permutations. Each has three places to be filled by $=$ or $<$ like $A_{min}^{(1)} = A_{max}^{(1)} < A_{min}^{(2)} < A_{max}^{(2)}$. This yields eight variations for each permutation and creates $24 \times 8 = 192$ combinations.

\paragraph{Canonical Combinations.} Some combinations are equivalent: $A_{min}^{(1)} = A_{max}^{(1)} < A_{min}^{(2)} < A_{max}^{(2)}$ and $A_{max}^{(1)} = A_{min}^{(1)} < A_{min}^{(2)} < A_{max}^{(2)}$. Symbols, \textit{connected by equality signs}, were sorted in the order $A_{min}^{(1)} < A_{max}^{(1)} < A_{min}^{(2)} < A_{max}^{(2)}$. This translates $A_{max}^{(1)} = A_{min}^{(1)} < A_{min}^{(2)} < A_{max}^{(2)}$ to $A_{min}^{(1)} = A_{max}^{(1)} < A_{min}^{(2)} < A_{max}^{(2)}$ and $A_{max}^{(1)} = A_{min}^{(2)} = A_{min}^{(1)} < A_{max}^{(2)}$ to $A_{min}^{(1)} = A_{max}^{(1)} = A_{min}^{(2)} < A_{max}^{(2)}$. The images were named \textit{canonical combinations} and only one repetition was selected. The author has written a C++ program and found 75 unique canonical combinations with 26 satisfying numerical conventions $A_{min}^{(1)} \le A_{max}^{(1)}$ and $A_{min}^{(2)} \le A_{max}^{(2)}$, Table \ref{TblTheoryF1}.

In 1, 2, 9, 10, 17, 18, 25, 26 the worst outcome of one prospect is better than the best outcome of another. This advantage depends on neither probabilities nor positive numbers of trials. \textit{Rational respondents will predictably chose a prospect with greater outcomes, if the intervals are disjoined}.

In 3, 4, 11, 16, 23, 24 one prospect is not worse than another and the fractions of respondents are either 1 or 0. For these combinations the fractions can be undefined $\uparrow$, if probabilities make both variables equal constants. In $A_q^{(1)} = A_p^{(1)} = A_q^{(2)} < A_p^{(2)}$ a two-point prospect (2) is better than (1), if $p^{(2)} > 0$, and undefined otherwise. \textit{The fractions} 0, 1, $\uparrow$ \textit{for adjacent intervals are predictable}.

In 12 both prospects are equal constants: fractions are predictably undefined.

In 5, 6, 7, 8, 13, 14, 15, 19, 20, 21, 22 the intervals overlap and fractions are not obvious.

\textit{If the intervals of two variables determined by the maximum and minimum outcomes are disjoined, then the fractions of respondents are deduced from a common rational sense, where probabilities and positive numbers of trials are irrelevant for a decision making. If the intervals are adjacent, then the fractions are deduced from rational sense or undefined for equal deterministic variables. If the intervals overlap, then probabilities and numbers of trials matter}.

\begin{center}
\begin{longtable}{|r|c|c|c|c|}
\caption[TheoryF1]{Canonical combinations, $f^{(1)}$, $f^{(2)}$ for $N^{(1)}>0$, $N^{(2)}>0$.} \label{TblTheoryF1} \\
 \hline
 \multicolumn{1}{|c|}{\#} &
 \multicolumn{1}{c|}{Condition} &
 \multicolumn{1}{c|}{$f^{(1)}$} &
 \multicolumn{1}{c|}{$f^{(2)}$} &
 \multicolumn{1}{c|}{Comment}\\
 \hline 
 \endfirsthead
 \multicolumn{5}{c}%
 {\tablename\ \thetable{} -- continued from previous page} \\
 \hline
 \multicolumn{1}{|c|}{\#} &
 \multicolumn{1}{c|}{Condition} &
 \multicolumn{1}{c|}{$f^{(1)}$} &
 \multicolumn{1}{c|}{$f^{(2)}$} &
 \multicolumn{1}{c|}{Comment}\\
 \hline 
 \endhead
 \hline \multicolumn{5}{|r|}{{Continued on next page}} \\ \hline
 \endfoot
 \hline
 \endlastfoot
 1 & $A_{min}^{(1)} < A_{max}^{(1)} < A_{min}^{(2)} < A_{max}^{(2)}$ & 0 &1 & (1) < (2)\\
 2 & $A_{min}^{(1)} < A_{max}^{(1)} < A_{min}^{(2)} = A_{max}^{(2)}$ & 0 &1 & (1) < (2); constant $a_1^{(2)}$\\
 3 & $A_{min}^{(1)} < A_{max}^{(1)} = A_{min}^{(2)} < A_{max}^{(2)}$ & $0  \vee \uparrow$ & $1  \vee \uparrow$ & $(1) \le (2)$; adjacent\\
 4 & $A_{min}^{(1)} < A_{max}^{(1)} = A_{min}^{(2)} = A_{max}^{(2)}$ & $0  \vee \uparrow$ & $1  \vee \uparrow$ & $(1) \le (2)$; constant $a_1^{(2)}$; adjacent\\
 5 & $A_{min}^{(1)} < A_{min}^{(2)} < A_{max}^{(1)} < A_{max}^{(2)}$ & ? & ? & overlapping\\
 6 & $A_{min}^{(1)} < A_{min}^{(2)} < A_{max}^{(1)} = A_{max}^{(2)}$ & ? & ? & overlapping\\
 7 & $A_{min}^{(1)} < A_{min}^{(2)} < A_{max}^{(2)} < A_{max}^{(1)}$ & ? & ? & overlapping\\
 8 & $A_{min}^{(1)} < A_{min}^{(2)} = A_{max}^{(2)} < A_{max}^{(1)}$ & ? & ? & constant $a_1^{(2)}$; overlapping\\
 9 & $A_{min}^{(1)} = A_{max}^{(1)} < A_{min}^{(2)} < A_{max}^{(2)}$ & 0 & 1 & (1) < (2); constant $a_1^{(1)}$\\
10 & $A_{min}^{(1)} = A_{max}^{(1)} < A_{min}^{(2)} = A_{max}^{(2)}$ & 0 & 1 & (1) < (2); constant $a_1^{(1)}$, $a_1^{(2)}$\\
11 & $A_{min}^{(1)} = A_{max}^{(1)} = A_{min}^{(2)} < A_{max}^{(2)}$ & $0  \vee \uparrow$ & $1  \vee \uparrow$ & $(1) \le (2)$; constant $a_1^{(1)}$; adjacent\\
12 & $A_{min}^{(1)} = A_{max}^{(1)} = A_{min}^{(2)} = A_{max}^{(2)}$ & $\uparrow$ & $\uparrow$ & constant $a_1^{(1)} = a_1^{(2)}$\\
13 & $A_{min}^{(1)} = A_{min}^{(2)} < A_{max}^{(1)} < A_{max}^{(2)}$ & ? & ? & overlapping\\
14 & $A_{min}^{(1)} = A_{min}^{(2)} < A_{max}^{(1)} = A_{max}^{(2)}$ & ? & ? & overlapping\\
15 & $A_{min}^{(1)} = A_{min}^{(2)} < A_{max}^{(2)} < A_{max}^{(1)}$ & ? & ? & overlapping\\
16 & $A_{min}^{(1)} = A_{min}^{(2)} = A_{max}^{(2)} < A_{max}^{(1)}$ & $1  \vee \uparrow$ & $0  \vee \uparrow$ & $(2) \le (1)$; constant $a_1^{(2)}$; adjacent\\
17 & $A_{min}^{(2)} < A_{max}^{(2)} < A_{min}^{(1)} < A_{max}^{(1)}$ & 1 & 0 &  (2) < (1)\\
18 & $A_{min}^{(2)} < A_{max}^{(2)} < A_{min}^{(1)} = A_{max}^{(1)}$ & 1 & 0 &  (2) < (1); constant $a_1^{(1)}$\\
19 & $A_{min}^{(2)} < A_{min}^{(1)} < A_{max}^{(1)} < A_{max}^{(2)}$ & ? & ? & overlapping\\
20 & $A_{min}^{(2)} < A_{min}^{(1)} < A_{max}^{(1)} = A_{max}^{(2)}$ & ? & ? & overlapping\\
21 & $A_{min}^{(2)} < A_{min}^{(1)} < A_{max}^{(2)} < A_{max}^{(1)}$ & ? & ? & overlapping\\
22 & $A_{min}^{(2)} < A_{min}^{(1)} = A_{max}^{(1)} < A_{max}^{(2)}$ & ? & ? & constant $a_1^{(1)}$; overlapping\\
23 & $A_{min}^{(2)} < A_{min}^{(1)} = A_{max}^{(1)} = A_{max}^{(2)}$ & $1  \vee \uparrow$ & $0  \vee \uparrow$ & $(2) \le (1)$; constant $a_1^{(1)}$; adjacent\\
24 & $A_{min}^{(2)} < A_{min}^{(1)} = A_{max}^{(2)} < A_{max}^{(1)}$ & $1  \vee \uparrow$ & $0  \vee \uparrow$ & $(2) \le (1)$; adjacent\\
25 & $A_{min}^{(2)} = A_{max}^{(2)} < A_{min}^{(1)} < A_{max}^{(1)}$ & 1 & 0 & (2) < (1); constant $a_1^{(2)}$\\
26 & $A_{min}^{(2)} = A_{max}^{(2)} < A_{min}^{(1)} = A_{max}^{(1)}$ & 1 & 0 & (2) < (1); constant $a_1^{(2)} < a_1^{(1)}$
\end{longtable}
\end{center}

Swapping indexes 1 and 2 does not change validity of a canonical combination and may require conventional sorting. Following Equation \ref{EqFunctional} this swaps values of fractions of respondents. Swapping transforms $1 \leftrightarrow 17$, $2 \leftrightarrow 18$, $3 \leftrightarrow 24$, $4 \leftrightarrow 23$, $5 \leftrightarrow 21$, $6 \leftrightarrow 20$, $7 \leftrightarrow 19$, $8 \leftrightarrow 22$, $9 \leftrightarrow 25$, $10 \leftrightarrow 26$, $11 \leftrightarrow 16$, $13 \leftrightarrow 15$. Due to sorting $12 \leftrightarrow 12$, and $14 \leftrightarrow 14$. 12 and 14 represent identical intervals, where 12 is a point. Due to the symmetry we have to find $f(a_1^{(1)}, a_1^{(2)})$ only for 1 - 14. The cases 1, 2, 3, 4, 9, 10, 11, 12 are either predictable or undefined, axiomatic. \textit{The cases} 5, 6, 7, 8, 13, 14 \textit{require attention}.

\paragraph{For two-point variables,} $A_{min}^{(1)}=A_q^{(1)} \le A_{max}^{(1)}=A_p^{(1)}$ and $A_{min}^{(2)}=A_q^{(2)} \le A_{max}^{(2)}=A_p^{(2)}$. Probabilities $p^{(1)}$ and $p^{(2)}$ are independent. Their combinations are in Table \ref{TblProbabilitiesF1}. A variable is constant, if $p=1$ or 0. \textit{If both variables are constant because of probabilities and/or equal outcomes, then the choice is predictable}.
\begin{center}
\begin{longtable}{|r|c|c|}
\caption[TheoryF1]{Combinations of probabilities of two-point distributions.} \label{TblProbabilitiesF1} \\
 \hline
 \multicolumn{1}{|c|}{\#} &
 \multicolumn{1}{c|}{Condition} &
 \multicolumn{1}{c|}{Comment}\\
 \hline 
 \endfirsthead
 \multicolumn{3}{c}%
 {\tablename\ \thetable{} -- continued from previous page} \\
 \hline
 \multicolumn{1}{|c|}{\#} &
 \multicolumn{1}{c|}{Condition} &
 \multicolumn{1}{c|}{Comment}\\
 \hline 
 \endhead
 \hline \multicolumn{3}{|r|}{{Continued on next page}} \\ \hline
 \endfoot
 \hline
 \endlastfoot
 1 & $p^{(1)}=0$, $p^{(2)}=0$ & constant $a_1^{(1)}=A_q^{(1)}$, $a_1^{(2)}=A_q^{(2)}$, $H(a_1^{(1)})=H(a_1^{(2)})=0$\\
 2 & $p^{(1)}=0 < p^{(2)}<1$ & constant $a_1^{(1)}=A_q^{(1)}$, $H(a_1^{(1)})=0$\\
 3 & $p^{(1)}=0$, $p^{(2)}=1$ & constant $a_1^{(1)}=A_q^{(1)}$, $a_1^{(2)}=A_p^{(2)}$, $H(a_1^{(1)})=H(a_1^{(2)})=0$\\
 4 & $0 < p^{(1)} <1$, $p^{(2)}=0$ & constant $a_1^{(2)}=A_q^{(2)}$, $H(a_1^{(2)})=0$\\
 5 & $0 < p^{(1)} = p^{(2)}=p < 1$ & $H(\xi_1)=H(\xi_2)=-p\log_2(p)-(1-p)\log_2(1-p)$\\
 6 & $0 < p^{(1)} <1$, $p^{(2)}=1$ & constant $a_1^{(2)}=A_p^{(2)}$, $H(a_1^{(2)})=0$\\
 7 & $p^{(1)}=1$, $p^{(2)}=0$ & constant $a_1^{(1)}=A_p^{(1)}$, $a_1^{(2)}=A_q^{(2)}$, $H(a_1^{(1)})=H(a_1^{(2)})=0$\\
 8 & $p^{(1)}=1$, $0< p^{(2)}<1$ & constant $a_1^{(1)}=A_p^{(1)}$, $H(a_1^{(1)})=0$\\
 9 & $p^{(1)}=1$, $p^{(2)}=1$ & constant $a_1^{(1)}=A_p^{(1)}$, $a_1^{(2)}=A_p^{(2)}$, $H(a_1^{(1)})=H(a_1^{(2)})=0$\\
10 & $0 < p^{(1)} < p^{(2)} < 1$ &\\
11 & $0 < p^{(2)} < p^{(1)} < 1$ &\\
\end{longtable}
\end{center}
Combinations 1, 3, 7, 9 from Table \ref{TblProbabilitiesF1} make both prospects deterministic. Combinations 2, 4, 6, 8 make one prospect deterministic. The alternative prospect can become deterministic due to outcomes. Example is $A_{min}^{(1)} < A_{min}^{(2)} = A_{max}^{(2)} < A_{max}^{(1)}$, Table \ref{TblTheoryF1}(8), combined with $p^{(1)}=0 < p^{(2)}<1$, Table \ref{TblProbabilitiesF1}(2): Prospect (2) is constant because $a_1^{(2)}=A_{min}^{(2)} = A_{max}^{(2)}=A_q^{(2)} = A_p^{(2)}$ and Prospect (1) is constant because $p^{(1)}=0$ and $a_1^{(1)}=A_{min}^{(1)}=A_q^{(1)}$. This is true $\forall \; 0<N^{(1)}, \; 0<N^{(2)}$. Therefore, $f^{(1)}=0$, $f^{(2)}=1$.

Cartesian product Table \ref{TblTheoryF1}(5, 6, 7, 8, 13, 14) $\times$ Table \ref{TblProbabilitiesF1}(2, 4, 5, 6, 8, 10, 11) = 42 requires further attention, Table \ref{TblCartesianF1}.
\begin{center}
\begin{longtable}{|r|r|c|r|c|c|c|c|}
\caption[TheoryF1]{Cartesian product of selective conditions for outcomes $A_{min}^{(1)}=A_q^{(1)}$, $A_{max}^{(1)}=A_p^{(1)}$, $A_{min}^{(2)}=A_q^{(2)}$, $A_{max}^{(2)}=A_p^{(2)}$, Table \ref{TblTheoryF1}, and probabilities, Table \ref{TblProbabilitiesF1}, and $f^{(1)}$, $f^{(2)}$ for $N^{(1)}>0$, $N^{(2)}>0$.} \label{TblCartesianF1} \\
 \hline
 \multicolumn{1}{|c|}{\#} &
 \multicolumn{1}{c|}{\ref{TblTheoryF1}} &
 \multicolumn{1}{c|}{Condition} &
 \multicolumn{1}{c|}{\ref{TblProbabilitiesF1}} &
 \multicolumn{1}{c|}{Condition} &
 \multicolumn{1}{c|}{$f^{(1)}$} &
 \multicolumn{1}{c|}{$f^{(2)}$} &
 \multicolumn{1}{c|}{Comment}\\
 \hline 
 \endfirsthead
 \multicolumn{8}{c}%
 {\tablename\ \thetable{} -- continued from previous page} \\
 \hline
 \multicolumn{1}{|c|}{\#} &
 \multicolumn{1}{c|}{\ref{TblTheoryF1}} &
 \multicolumn{1}{c|}{Condition} &
 \multicolumn{1}{c|}{\ref{TblProbabilitiesF1}} &
 \multicolumn{1}{c|}{Condition} &
 \multicolumn{1}{c|}{$f^{(1)}$} &
 \multicolumn{1}{c|}{$f^{(2)}$} &
 \multicolumn{1}{c|}{Comment}\\
 \hline 
 \endhead
 \hline \multicolumn{8}{|r|}{{Continued on next page}} \\ \hline
 \endfoot
 \hline
 \endlastfoot
 1 & 5 & $A_q^{(1)} < A_q^{(2)} < A_p^{(1)} < A_p^{(2)}$ & 2 & $p^{(1)}=0 < p^{(2)}<1$ &  0 &1 & (1) < (2)\\
 2 & 5 & $A_q^{(1)} < A_q^{(2)} < A_p^{(1)} < A_p^{(2)}$ & 4 & $0 < p^{(1)} <1$, $p^{(2)}=0$ &  ? & ? &\\
 3 & 5 & $A_q^{(1)} < A_q^{(2)} < A_p^{(1)} < A_p^{(2)}$ & 5 & $0 < p^{(1)} = p^{(2)}=p < 1$ &  0 & 1 &\\
 4 & 5 & $A_q^{(1)} < A_q^{(2)} < A_p^{(1)} < A_p^{(2)}$ & 6 & $0 < p^{(1)} <1$, $p^{(2)}=1$ &  0 & 1 & (1) < (2)\\
 5 & 5 & $A_q^{(1)} < A_q^{(2)} < A_p^{(1)} < A_p^{(2)}$ & 8 & $p^{(1)}=1$, $0< p^{(2)}<1$ &  ? & ? &\\
 6 & 5 & $A_q^{(1)} < A_q^{(2)} < A_p^{(1)} < A_p^{(2)}$ & 10 & $0 < p^{(1)} < p^{(2)} < 1$ &  ? & ? &\\
 7 & 5 & $A_q^{(1)} < A_q^{(2)} < A_p^{(1)} < A_p^{(2)}$ & 11 & $0 < p^{(2)} < p^{(1)} < 1$ &  ? & ? &\\
 \hline
 8 & 6 & $A_q^{(1)} < A_q^{(2)} < A_p^{(1)} = A_p^{(2)}$ & 2 & $p^{(1)}=0 < p^{(2)}<1$ &  0 &1 & (1) < (2)\\
 9 & 6 & $A_q^{(1)} < A_q^{(2)} < A_p^{(1)} = A_p^{(2)}$ & 4 & $0 < p^{(1)} <1$, $p^{(2)}=0$ &  ? & ? &\\
10 & 6 & $A_q^{(1)} < A_q^{(2)} < A_p^{(1)} = A_p^{(2)}$ & 5 & $0 < p^{(1)} = p^{(2)}=p < 1$ &  0 & 1 &\\
11 & 6 & $A_q^{(1)} < A_q^{(2)} < A_p^{(1)} = A_p^{(2)}$ & 6 & $0 < p^{(1)} <1$, $p^{(2)}=1$ &  0 & 1 & $(1) \le (2)$\\
12 & 6 & $A_q^{(1)} < A_q^{(2)} < A_p^{(1)} = A_p^{(2)}$ & 8 & $p^{(1)}=1$, $0< p^{(2)}<1$ &  1 & 0 &$(2) \le (1)$\\
13 & 6 & $A_q^{(1)} < A_q^{(2)} < A_p^{(1)} = A_p^{(2)}$ & 10 & $0 < p^{(1)} < p^{(2)} < 1$ &  0 & 1 &\\
14 & 6 & $A_q^{(1)} < A_q^{(2)} < A_p^{(1)} = A_p^{(2)}$ & 11 & $0 < p^{(2)} < p^{(1)} < 1$ &  ? & ? &\\
 \hline
15 & 7 & $A_q^{(1)} < A_q^{(2)} < A_p^{(2)} < A_p^{(1)}$ & 2 & $p^{(1)}=0 < p^{(2)}<1$ &  0 &1 & (1) < (2)\\
16 & 7 & $A_q^{(1)} < A_q^{(2)} < A_p^{(2)} < A_p^{(1)}$ & 4 & $0 < p^{(1)} <1$, $p^{(2)}=0$ &  ? & ? &\\
17 & 7 & $A_q^{(1)} < A_q^{(2)} < A_p^{(2)} < A_p^{(1)}$ & 5 & $0 < p^{(1)} = p^{(2)}=p < 1$ &  ? & ? &\\
18 & 7 & $A_q^{(1)} < A_q^{(2)} < A_p^{(2)} < A_p^{(1)}$ & 6 & $0 < p^{(1)} <1$, $p^{(2)}=1$ &  ? & ? &\\
19 & 7 & $A_q^{(1)} < A_q^{(2)} < A_p^{(2)} < A_p^{(1)}$ & 8 & $p^{(1)}=1$, $0< p^{(2)}<1$ &  1 & 0 &$(2) < (1)$\\
20 & 7 & $A_q^{(1)} < A_q^{(2)} < A_p^{(2)} < A_p^{(1)}$ & 10 & $0 < p^{(1)} < p^{(2)} < 1$ &  ? & ? &\\
21 & 7 & $A_q^{(1)} < A_q^{(2)} < A_p^{(2)} < A_p^{(1)}$ & 11 & $0 < p^{(2)} < p^{(1)} < 1$ &  ? & ? &\\
 \hline
22 & 8 & $A_q^{(1)} < A_q^{(2)} = A_p^{(2)} < A_p^{(1)}$ & 2 & $p^{(1)}=0 < p^{(2)}<1$ &  0 &1 & (1) < (2)\\
23 & 8 & $A_q^{(1)} < A_q^{(2)} = A_p^{(2)} < A_p^{(1)}$ & 4 & $0 < p^{(1)} <1$, $p^{(2)}=0$ &  ? & ? &\\
24 & 8 & $A_q^{(1)} < A_q^{(2)} = A_p^{(2)} < A_p^{(1)}$ & 5 & $0 < p^{(1)} = p^{(2)}=p < 1$ &  ? & ? &\\
25 & 8 & $A_q^{(1)} < A_q^{(2)} = A_p^{(2)} < A_p^{(1)}$ & 6 & $0 < p^{(1)} <1$, $p^{(2)}=1$ &  ? & ? &\\
26 & 8 & $A_q^{(1)} < A_q^{(2)} = A_p^{(2)} < A_p^{(1)}$ & 8 & $p^{(1)}=1$, $0< p^{(2)}<1$ &  1 & 0 &$(2) < (1)$\\
27 & 8 & $A_q^{(1)} < A_q^{(2)} = A_p^{(2)} < A_p^{(1)}$ & 10 & $0 < p^{(1)} < p^{(2)} < 1$ &  ? & ? &\\
28 & 8 & $A_q^{(1)} < A_q^{(2)} = A_p^{(2)} < A_p^{(1)}$ & 11 & $0 < p^{(2)} < p^{(1)} < 1$ &  ? & ? &\\
 \hline
29 & 13 & $A_q^{(1)} = A_q^{(2)} < A_p^{(1)} < A_p^{(2)}$ & 2 & $p^{(1)}=0 < p^{(2)}<1$ &  0 &1 & $(1) \le (2)$\\
30 & 13 & $A_q^{(1)} = A_q^{(2)} < A_p^{(1)} < A_p^{(2)}$ & 4 & $0 < p^{(1)} <1$, $p^{(2)}=0$ &  1 & 0 & $(2) \le (1)$\\
31 & 13 & $A_q^{(1)} = A_q^{(2)} < A_p^{(1)} < A_p^{(2)}$ & 5 & $0 < p^{(1)} = p^{(2)}=p < 1$ &  0 & 1 & $(1) \le (2)$\\
32 & 13 & $A_q^{(1)} = A_q^{(2)} < A_p^{(1)} < A_p^{(2)}$ & 6 & $0 < p^{(1)} <1$, $p^{(2)}=1$ &  0 & 1 &(1) < (2)\\
33 & 13 & $A_q^{(1)} = A_q^{(2)} < A_p^{(1)} < A_p^{(2)}$ & 8 & $p^{(1)}=1$, $0< p^{(2)}<1$ &  ? & ? &\\
34 & 13 & $A_q^{(1)} = A_q^{(2)} < A_p^{(1)} < A_p^{(2)}$ & 10 & $0 < p^{(1)} < p^{(2)} < 1$ &  0 & 1 &\\
35 & 13 & $A_q^{(1)} = A_q^{(2)} < A_p^{(1)} < A_p^{(2)}$ & 11 & $0 < p^{(2)} < p^{(1)} < 1$ &  ? & ? &\\
 \hline
36 & 14 & $A_q^{(1)} = A_q^{(2)} < A_p^{(1)} = A_p^{(2)}$ & 2 & $p^{(1)}=0 < p^{(2)}<1$ &  0 &1 & $(1) \le (2)$\\
37 & 14 & $A_q^{(1)} = A_q^{(2)} < A_p^{(1)} = A_p^{(2)}$ & 4 & $0 < p^{(1)} <1$, $p^{(2)}=0$ &  1 & 0 & $(2) \le (1)$\\
38 & 14 & $A_q^{(1)} = A_q^{(2)} < A_p^{(1)} = A_p^{(2)}$ & 5 & $0 < p^{(1)} = p^{(2)}=p < 1$ &  $\uparrow$ & $\uparrow$ &(1) = (2)\\
39 & 14 & $A_q^{(1)} = A_q^{(2)} < A_p^{(1)} = A_p^{(2)}$ & 6 & $0 < p^{(1)} <1$, $p^{(2)}=1$ &  0 & 1 & $(1) \le (2)$\\
40 & 14 & $A_q^{(1)} = A_q^{(2)} < A_p^{(1)} = A_p^{(2)}$ & 8 & $p^{(1)}=1$, $0< p^{(2)}<1$ &  1 & 0 & $(2) \le (1)$\\
41 & 14 & $A_q^{(1)} = A_q^{(2)} < A_p^{(1)} = A_p^{(2)}$ & 10 & $0 < p^{(1)} < p^{(2)} < 1$ &  0 & 1 &\\
42 & 14 & $A_q^{(1)} = A_q^{(2)} < A_p^{(1)} = A_p^{(2)}$ & 11 & $0 < p^{(2)} < p^{(1)} < 1$ &  1 & 0 &\\
\end{longtable}
\end{center}

If arbitrary large number of trials costs nothing and possible, then due to decreasing variance of $a_1$ following from Equation \ref{EqVarianceOfSampleMean} the author thinks that
\begin{equation}
\label{EqAxiom1}
\begin{split}
\textrm{if} \; A_p^{(1)} p^{(1)} + A_q^{(1)} q^{(1)} > A_p^{(2)} p^{(2)} + A_q^{(2)} q^{(2)}, \; N^{(1)} \rightarrow \infty, \; N^{(2)} \rightarrow \infty, \; \textrm{then} \\
f(\{A_p^{(1)},A_q^{(1)},p^{(1)},N^{(1)}\}, \{A_p^{(2)},A_q^{(2)}, p^{(2)},N^{(2)}\}) \rightarrow 1,
\end{split}
\end{equation}
\begin{equation}
\label{EqAxiom2}
\begin{split}
\textrm{if} \; A_p^{(1)} p^{(1)} + A_q^{(1)} q^{(1)} < A_p^{(2)} p^{(2)} + A_q^{(2)} q^{(2)}, \; N^{(1)} \rightarrow \infty, \; N^{(2)} \rightarrow \infty, \; \textrm{then} \\
f(\{A_p^{(1)},A_q^{(1)},p^{(1)},N^{(1)}\}, \{A_p^{(2)},A_q^{(2)}, p^{(2)},N^{(2)}\}) \rightarrow 0,
\end{split}
\end{equation}
\begin{equation}
\label{EqAxiom3}
\begin{split}
\textrm{if} \; A_p^{(1)} p^{(1)} + A_q^{(1)} q^{(1)} = A_p^{(2)} p^{(2)} + A_q^{(2)} q^{(2)}, \; N^{(1)} \rightarrow \infty, \; N^{(2)} \rightarrow \infty, \; \textrm{then} \\
f(\{A_p^{(1)},A_q^{(1)},p^{(1)},N^{(1)}\}, \{A_p^{(2)},A_q^{(2)}, p^{(2)},N^{(2)}\}) \rightarrow \textrm{undefined}.
\end{split}
\end{equation}

The author believes that \textit{respondents unanimously} select the \textit{sure} 3,000.01 instead of 3,000.00 and -4,000.00 instead of -4,000.01. For equal sure amounts the choice is undefined. These are the combinations Table \ref{TblTheoryF1}(3) + Table \ref{TblProbabilitiesF1}(7), \ref{TblTheoryF1}(4) + \ref{TblProbabilitiesF1}(7, 8, or 9), \ref{TblTheoryF1}(6) + \ref{TblProbabilitiesF1}(9), \ref{TblTheoryF1}(11) + \ref{TblProbabilitiesF1}(1, 4, or 7), \ref{TblTheoryF1}(12) + \ref{TblProbabilitiesF1}(any), \ref{TblTheoryF1}(13) + \ref{TblProbabilitiesF1}(1), \ref{TblTheoryF1}(14) + \ref{TblProbabilitiesF1}(1, or 9), \ref{TblTheoryF1}(15) + \ref{TblProbabilitiesF1}(1), \ref{TblTheoryF1}(16) + \ref{TblProbabilitiesF1}(1, 2, or 3), \ref{TblTheoryF1}(20) + \ref{TblProbabilitiesF1}(9), \ref{TblTheoryF1}(23) + \ref{TblProbabilitiesF1}(3, 6, or 9), \ref{TblTheoryF1}(24) + \ref{TblProbabilitiesF1}(3). In general, 0, 1, and $\uparrow$ postulated for $f^{(1)}$ and/or $f^{(2)}$ in Tables \ref{TblTheoryF1} and \ref{TblCartesianF1} indicate combinations of arguments and values, which, together with Equations \ref{EqAxiom1} - \ref{EqAxiom3}, are candidates to \textit{axioms}. Kahneman and Tversky present experimental $F^{(1)}$ and $F^{(2)}$ for $N^{(1)} = N^{(2)} = 1$ \cite{kahneman1979}.
\begin{figure}
  \centering
  \includegraphics[width=65mm]{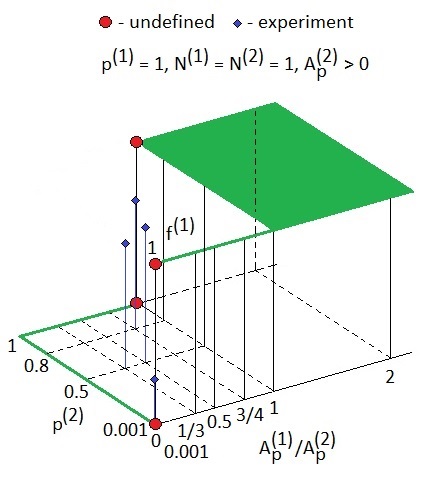}
  \caption[FigF1]
   {A 3D projection of predictable (axiomatic) properties of $f^{(1)}$ together with blue diamonds, the experimental points 2, 10, 12, and 14 from Table \ref{TblF1}. Green lines and surface correspond to the postulates. Red points indicate ambiguity. $A_q^{(1)} = A_q^{(2)} = 0$.}
  \label{FigF1}
\end{figure}
The postulated and experimental points $F^{(1)}$ are applied to sketch $f^{(1)}$. $f^{(1)}$ and $f^{(2)}$ are the functions of at least eight arguments. We can plot 3D projections of the eight dimensional space choosing suitable coordinates, Figure \ref{FigF1}.

Robinson Crusoe took to his desert island from the sunk ship a gun instead of coins. A donation can be decided to be greater. Respondents can be mentally ill. Other reasons can be invented with an intention to invalidate the postulates: Table \ref{TblTheoryF1} (1 - 4, 9 - 12, 16 - 18, 23 - 26); Table \ref{TblCartesianF1} (1, 3, 4, 8, 10 - 13, 15, 19, 22, 26, 29 - 32, 34, 36 - 42); Equations \ref{EqAxiom1} - \ref{EqAxiom3}. The paper omits such exercises.

\section{Available data and reducing dimension eight}

Two-point random variables are the essence of a few prospects reviewed here. This paper suggests an extension, sample means as awards, replacing these variables with their mean sums obtained in sequences of independent trials. The properties of both are well studied theoretically and supported experimentally including computer simulations. While some properties of sequential Bernoulli trials such as the \textit{law of arcsine} are paradoxical and "contradict to common sense" \cite[pp. 91 - 97]{kolmogorov1982}, the variables are fundamental "simple" candidates to investigate how \textit{randomness} of prospects influences on the fractions of respondents selecting between the prospects. With this proposal $f^{(1)}=f(a_1^{(1)}, a_1^{(2)}) = f(\{A_p^{(1)}, A_q^{(1)},p^{(1)},N^{(1)}\}, \{A_p^{(2)}, A_q^{(2)},p^{(2)},N^{(2)}\})$ depends on eight parameters. Fixing two coordinates for each dimension, a rough estimate of $f^{(1)}$, requires $2^8 = 256$ points of an experimental plan. Table \ref{TblF1} collects estimates $F^{(1)}$ and \textit{median cash equivalents} found in literature.
\begin{center}
\begin{longtable}{|r|r|r|r|r|r|r|r|c|c|c|}
\caption[History of Mega Millions]{Empirical estimates of $f^{(1)}=f(a_1^{(1)}, a_1^{(2)}) = f(\{A_p^{(1)}, A_q^{(1)},p^{(1)},N^{(1)}\}, \{A_p^{(2)}, A_q^{(2)},p^{(2)},N^{(2)}\})$; $N^{(1)}=N^{(2)}=1.$} \label{TblF1} \\
 \hline
 \multicolumn{1}{|c|}{\#} &
 \multicolumn{1}{c|}{$A_p^{(1)}$} &
 \multicolumn{1}{c|}{$A_q^{(1)}$} &
 \multicolumn{1}{|c|}{$p^{(1)}$} &
 \multicolumn{1}{c|}{$A_p^{(2)}$} &
 \multicolumn{1}{|c|}{$A_q^{(2)}$} &
 \multicolumn{1}{c|}{$p^{(2)}$} &
 \multicolumn{1}{c|}{$F^{(1)}$} &
 \multicolumn{1}{c|}{Voters} &
 \multicolumn{1}{c|}{Ref} &
 \multicolumn{1}{c|}{Table 2/3/4}\\
 \hline 
 \endfirsthead
 \multicolumn{11}{c}%
 {\tablename\ \thetable{} -- continued from previous page} \\
 \hline
 \multicolumn{1}{|c|}{\#} &
 \multicolumn{1}{c|}{$A_p^{(1)}$} &
 \multicolumn{1}{c|}{$A_q^{(1)}$} &
 \multicolumn{1}{|c|}{$p^{(1)}$} &
 \multicolumn{1}{c|}{$A_p^{(2)}$} &
 \multicolumn{1}{|c|}{$A_q^{(2)}$} &
 \multicolumn{1}{c|}{$p^{(2)}$} &
 \multicolumn{1}{c|}{$F^{(1)}$} &
 \multicolumn{1}{c|}{Voters} &
 \multicolumn{1}{c|}{Ref} &
 \multicolumn{1}{c|}{Table 2/3/4}\\
 \hline 
 \endhead
 \hline \multicolumn{11}{|r|}{{Continued on next page}} \\ \hline
 \endfoot
 \hline
 \endlastfoot
1 & 2500 & 0 & 0.33 & 2400 & 0 & 0.34 &  0.83 & 72 & \cite{kahneman1979} & 15/10/*\\
2 & 4000 & 0 & 0.80 & 3000 & 3000 & 1 &  0.20 & 95 & \cite{kahneman1979} & 8/6/25\\
3 & 4000 & 0 & 0.20 & 3000 & 0 & 0.25 &  0.65 & 95 & \cite{kahneman1979} & 15/10/*\\
4 & 6000 & 0 & 0.45 & 3000 & 0 & 0.90 &  0.14 & 66 & \cite{kahneman1979} & 15/10/*\\
5 & 6000 & 0 & 0.001 & 3000 & 0 & 0.002 &  0.73 & 66 & \cite{kahneman1979} & 15/10/*\\
6 & 0 & -4000 & 0.20 & -3000 & -3000 & 1 &  0.92 & 95 & \cite{kahneman1979} & 8/6/25\\
7 & 0 & -4000  & 0.80 & 0 & -3000 & 0.75 &  0.42 & 95 & \cite{kahneman1979} & 6/11/14\\
8 & 0 & -6000 & 0.55 & 0 & -3000 & 0.10 &  0.92 & 66 & \cite{kahneman1979} & 6/11/14\\
9 & 0 & -6000 & 0.999 & 0 & -3000 & 0.998 &  0.30 & 66 & \cite{kahneman1979} & 6/11/14\\
10 & 1000 & 0 & 0.50 & 500 & 500 & 1 &  0.16 & 70 & \cite{kahneman1979} & 8/6/25\\
11 & 0 & -1000 & 0.50 & -500 & -500 & 1 &  0.69 & 68 & \cite{kahneman1979} & 8/6/25\\
12 & 5000 & 0 & 0.001 & 5 & 5 & 1 &  0.72 & 72 & \cite{kahneman1979} & 8/6/25\\
13 & 0 & -5000 & 0.999 & -5 & -5 & 1 &  0.17 & 72 & \cite{kahneman1979} & 8/6/25\\
14 & 1260 & 1260 & 1 & 3394 & 0 & 0.50 &  0.78 & 72 & \cite{kahneman1979} & 22/8/*\\
15 & 1260 & 0 & 0.10 & 3394 & 0 & 0.05 &  0.33 & 72 & \cite{kahneman1979} & 13/11/35\\
16 & 9 & 9 & 1 & 50 & 0 & 0.10 &  ? & 25 & \cite{tversky1992} & 22/8/*\\
17 & 21 & 21 & 1 & 50 & 0 & 0.50 &  ? & 25 & \cite{tversky1992} & 22/8/*\\
18 & 37 & 37 & 1 & 50 & 0 & 0.90 &  ? & 25 & \cite{tversky1992} & 22/8/*\\
19 & -8 & -8 & 1 & 0 & -50 & 0.90 &  ? & 25 & \cite{tversky1992} & 22/8/*\\
20 & -21 & -21 & 1 & 0 & -50 & 0.50 &  ? & 25 & \cite{tversky1992} & 22/8/*\\
21 & -39 & -39 & 1 & 0 & -50 & 0.10 &  ? & 25 & \cite{tversky1992} & 22/8/*\\
22 & 14 & 14 & 1 & 100 & 0 & 0.05 &  ? & 25 & \cite{tversky1992} & 22/8/*\\
23 & 25 & 25 & 1 & 100 & 0 & 0.25 &  ? & 25 & \cite{tversky1992} & 22/8/*\\
24 & 36 & 36 & 1 & 100 & 0 & 0.50 &  ? & 25 & \cite{tversky1992} & 22/8/*\\
25 & 52 & 52 & 1 & 100 & 0 & 0.75 &  ? & 25 & \cite{tversky1992} & 22/8/*\\
26 & 78 & 78 & 1 & 100 & 0 & 0.95 &  ? & 25 & \cite{tversky1992} & 22/8/*\\
27 & -8 & -8 & 1 & 0 & -100 & 0.95 &  ? & 25 & \cite{tversky1992} & 22/8/*\\
28 & -23.5 & -23.5 & 1 & 0 & -100 & 0.75 &  ? & 25 & \cite{tversky1992} & 22/8/*\\
29 & -42 & -42 & 1 & 0 & -100 & 0.50 &  ? & 25 & \cite{tversky1992} & 22/8/*\\
30 & -63 & -63 & 1 & 0 & -100 & 0.25 &  ? & 25 & \cite{tversky1992} & 22/8/*\\
31 & -84 & -84 & 1 & 0 & -100 & 0.05 &  ? & 25 & \cite{tversky1992} & 22/8/*\\
32 & 10 & 10 & 1 & 200 & 0 & 0.01 &  ? & 25 & \cite{tversky1992} & 22/8/*\\
33 & 20 & 20 & 1 & 200 & 0 & 0.10 &  ? & 25 & \cite{tversky1992} & 22/8/*\\
34 & 76 & 76 & 1 & 200 & 0 & 0.50 &  ? & 25 & \cite{tversky1992} & 22/8/*\\
35 & 131 & 131 & 1 & 200 & 0 & 0.90 &  ? & 25 & \cite{tversky1992} & 22/8/*\\
36 & 188 & 188 & 1 & 200 & 0 & 0.99 &  ? & 25 & \cite{tversky1992} & 22/8/*\\
37 &  -3 & -3 & 1 & 0 & -200 & 0.99 &  ? & 25 & \cite{tversky1992} & 22/8/*\\
38 & -23 & -23 & 1 & 0 & -200 & 0.90 &  ? & 25 & \cite{tversky1992} & 22/8/*\\
39 & -89 & -89 & 1 & 0 & -200 & 0.50 &  ? & 25 & \cite{tversky1992} & 22/8/*\\
40 & -155 & -155 & 1 & 0 & -200 & 0.10 &  ? & 25 & \cite{tversky1992} & 22/8/*\\
41 & -190 & -190 & 1 & 0 & -200 & 0.01 &  ? & 25 & \cite{tversky1992} & 22/8/*\\
42 & 12 & 12 & 1 & 400 & 0 & 0.01 &  ? & 25 & \cite{tversky1992} & 22/8/*\\
43 &  377 & 377 & 1 & 400 & 0 & 0.99 &  ? & 25 & \cite{tversky1992} & 22/8/*\\
44 & -14 & -14 & 1 & 0 & -400 & 0.99 &  ? & 25 & \cite{tversky1992} & 22/8/*\\
45 &  -380 & -380 & 1 & 0 & -400 & 0.01 &  ? & 25 & \cite{tversky1992} & 22/8/*\\
46 &  59 & 59 & 1 & 100 & 50 & 0.10 &  ? & 25 & \cite{tversky1992} & 22/8/*\\
47 &  71 & 71 & 1 & 100 & 50 & 0.50 &  ? & 25 & \cite{tversky1992} &  22/8/*\\
48 &  83 & 83 & 1 & 100 & 50 & 0.90 &  ? & 25 & \cite{tversky1992} &  22/8/*\\
49 &  -59 & -59 & 1 & -50 & -100 & 0.90 &  ? & 25 & \cite{tversky1992} & 22/8/*\\
50 &  -71 & -71 & 1 & -50 & -100 & 0.50 &  ? & 25 & \cite{tversky1992} & 22/8/*\\
51 &  -85 & -85 & 1 & -50 & -100 & 0.10 &  ? & 25 & \cite{tversky1992} & 22/8/*\\
52 & 64 & 64 & 1 & 150 & 50 & 0.05 &  ? & 25 & \cite{tversky1992} & 22/8/*\\
53 & 72.5 & 72.5 & 1 & 150 & 50 & 0.25 &  ? & 25 & \cite{tversky1992} & 22/8/*\\
54 & 86 & 86 & 1 & 150 & 50 & 0.50 &  ? & 25 & \cite{tversky1992} & 22/8/*\\
55 & 102 & 102 & 1 & 150 & 50 & 0.75 &  ? & 25 & \cite{tversky1992} & 22/8/*\\
56 & 128 & 128 & 1 & 150 & 50 & 0.95 &  ? & 25 & \cite{tversky1992} & 22/8/*\\
57 & -60 & -60 & 1 & -50 & -150 & 0.95 &  ? & 25 & \cite{tversky1992} & 22/8/*\\
58 & -71 & -71 & 1 & -50 & -150 & 0.75 &  ? & 25 & \cite{tversky1992} & 22/8/*\\
59 & -92 & -92 & 1 & -50 & -150 & 0.50 &  ? & 25 & \cite{tversky1992} & 22/8/*\\
60 & -113 & -113 & 1 & -50 & -150 & 0.25 &  ? & 25 & \cite{tversky1992} & 22/8/*\\
61 & -132 & -132 & 1 & -50 & -150 & 0.05 &  ? & 25 & \cite{tversky1992} & 22/8/*\\
62 & 118 & 118 & 1 & 200 & 100 & 0.05 &  ? & 25 & \cite{tversky1992} & 22/8/*\\
63 & 130 & 130 & 1 & 200 & 100 & 0.25 &  ? & 25 & \cite{tversky1992} & 22/8/*\\
64 & 141 & 141 & 1 & 200 & 100 & 0.50 &  ? & 25 & \cite{tversky1992} & 22/8/*\\
65 & 162 & 162 & 1 & 200 & 100 & 0.75 &  ? & 25 & \cite{tversky1992} & 22/8/*\\
66 & 178 & 178 & 1 & 200 & 100 & 0.95 &  ? & 25 & \cite{tversky1992} & 22/8/*\\
67 & -112 & -112 & 1 & -100 & -200 & 0.95 &  ? & 25 & \cite{tversky1992} & 22/8/*\\
68 & -121 & -121 & 1 & -100 & -200 & 0.75 &  ? & 25 & \cite{tversky1992} & 22/8/*\\
69 & -142 & -142 & 1 & -100 & -200 & 0.50 &  ? & 25 & \cite{tversky1992} & 22/8/*\\
70 & -158 & -158 & 1 & -100 & -200 & 0.25 &  ? & 25 & \cite{tversky1992} & 22/8/*\\
71 & -179 & -179 & 1 & -100 & -200 & 0.05 &  ? & 25 & \cite{tversky1992} & 22/8/*\\
72 & 100 & -100 & 0.65 & 0 & 0 & 1 &  0.50 & ? & \cite{williams1966} & 8/6/25\\
73 & 0 & -200 & 0.20 & -100 & -100 & 1 &  0.50 & ? & \cite{williams1966} & 8/6/25\\
\end{longtable}
\end{center}
For points 14 and 15 the prospects were formulated with non-monetary outcomes where (1) is a one-week tour of England and (2) is a three-week tour of England, France, and Italy. At the time of this writing the site \url{http://www.budgetyourtrip.com/countrylist.php} presents the following costs per day per person: United Kingdom \pounds 111.67, France \euro 128.58, and Italy \euro 111.65. With the exchange rates 0.6205 \pounds/\$ and 0.788 \euro/\$ we get modern dollar equivalents (1) $\frac{111.67}{0.6205} \times 7 \approx 1260$ and (2) $(\frac{111.67}{0.6205}+\frac{128.58}{0.788} + \frac{111.65}{0.788}) \times 7 \approx 3394$. These modern absolute values with the ratio $\approx \frac{1}{3}$ and the fractions of respondents obtained more than 35 years ago should be used carefully with low weights. The points 72 and 73 are referenced by \cite{kahneman1979}.

For points 1 - 15 the numbers of respondents selecting each prospect were counted \cite{kahneman1979} and being divided by their total numbers give the estimates $F^{(1)}$ and complementary $F^{(2)}$.

\paragraph{Median Cash Equivalents.} In contrast, for points 16 - 71 a \textit{certainty equivalent} was estimated \cite[pp. 305 - 306]{tversky1992}. The latter concept, and method can be illustrated by an example. Errors in the example belong to the author.

Let $a_1^{(2)}$ is $\{A_p^{(2)}=150, A_q^{(2)}=50, p^{(2)}=0.25, N^{(2)}=1\}$. The certainty equivalent is the prospect $a_1^{(1)}=\{A_p^{(1)}=?,A_q^{(1)}=A_p^{(1)}, p^{(1)}=1, N^{(1)}=1\}$. Given $a_1^{(2)}$ and the expected value $75 = 0.75 \times 50 + 0.25 \times 150$, the goal was to determine experimentally $A_p^{(1)}$. The authors \cite{tversky1992} do not speak about sample means. Such an interpretation attempts to match their results with the proposed approach. \textit{"The display also [VS: together with the original prospect and expected value] included a descending series of seven sure outcomes (gains or losses) logarithmically spaced between the extreme outcomes of the prospect. The subject indicated a preference between each of the seven sure outcomes and the risky prospect"} \cite[p. 305]{tversky1992}. It does not say whether the seven plus two extreme values belong to one dependence, where neighboring logarithms are equidistant. In the latter case, the logarithm base is not important and the descending values are members of the geometric progression $A_i = A_0c^{-i}, \; i = 0, ..., n + 1$, where $n = 7$ is the number of sure outcomes. For $A_0 = 150$ and $A_{n+1=8}=50$ we get $c = \sqrt[n+1]{\frac{A_0}{A_{n+1}}} = \sqrt[8]{\frac{150}{50}} \approx 1.147$ and seven sure outcomes 130.75 (+), 113.98 (+), 99.35 (+), 86.60 (+), 75.49 (-), 65.80 (-), 57.36 (-). Accepted (the sure outcome is chosen) and rejected (the prospect is selected) values are marked by + and -. The signs in the example are fictional. \textit{"To obtain a more refined estimate of the certainty equivalent, a new set of seven outcomes was then shown, linearly spaced between a value 25\% higher than the lowest amount accepted in the first set and a value 25\% lower than the highest amount rejected"} \cite[pp. 305 - 306]{tversky1992}. Therefore, the lowest accepted value 86.60 would be increased by 25\% to 108.25 and the highest rejected value 75.49 would be decreased by 25\% to 60.39. The interval $[60.39, 108.25]$ would be added by new seven \textit{linearly spaced} sure outcomes: 102.27 (+), 96.29 (+), 90.30 (+), 84.32 (+), 78.34 (+), 72.36 (-), 66.37 (-). The mid point between the lowest accepted 78.34 and highest rejected 72.36 values $\frac{78.34 + 72.36}{2} = 75.35$ was used as a certainty equivalent. Inconsistent selections from the first and second sets were rejected by a computer. \textit{The median cash equivalents of the 25 subjects} were reported and are cited for points 16 - 71 in columns $A_p^{(1)}, \; A_q^{(1)}$.

Placing seven sure and two extreme outcomes on one logarithmic dependence, representing a geometric progression with $c > 0$, fails, if two extreme outcomes have a different sign or one of them is zero. The author thinks that inserting seven values under such conditions creates bias. For two extreme outcomes 0 and 50, points 16 - 18, one, dividing $\log(50)$ by 8 and gradually decrementing $\log(50)$, reaches 0, where $\log(50) = \log(50) - 0 = \log(50) - \log(1)$ imply the prospect $\{A_p^{(2)}=50,A_q^{(2)}=1,p^{(2)},N^{(2)}=1\}$ but not $\{A_p^{(2)}=50,A_q^{(2)}=0,p^{(2)},N^{(2)}=1\}$. The same problem arises for points 16 - 45, 73. It would be also difficult to apply this method to point 72 $(-50, 100)$ provoking bias.

In order to combine results of \cite{kahneman1979} and \cite{tversky1992}, the method cited above \cite{tversky1992} being applied to prospects \cite{kahneman1979} must reproduce the fractions of respondents. \cite{tversky1992} has a different goal - median cash equivalent. The number of subjects 25 in \cite{tversky1992} is 2 - 4 times less than the numbers of respondents in \cite{kahneman1979}. The author cannot conclude that the median cash equivalent associates with the fractions of respondents $F^{(1)} = F^{(2)} = \frac{1}{2}$ or other obvious fraction. $F^{(1)}$ could be obtained after requesting respondents to select between two prospects, for instance, $\{72.5, 72.5, 1, 1\}$ and $\{150, 50, 0.25, 1\}$, point 53, and counting votes. In axioms, \textit{if two variables are deterministically equivalent, then the fractions of respondents are undefined. Estimates made under such conditions would fluctuate between zero and one}.

\paragraph{Points 1, 3 - 5, and 15} with $A_q^{(1)} = A_q^{(2)} = 0$ are on a six dimensional projection. The same is true for points 7 - 9 with $A_p^{(1)} = A_p^{(2)} = 0$.

\paragraph{Sobol Sequences for Experimental Plans.} Previous experiments, Table \ref{TblF1}, had no a goal to fit $f^{(1)}$ and $f^{(2)}$. They neither fill "uniformly" parts of the eight dimensional space nor the projection on Figure \ref{FigF1}, where a plan should fill the product-square $(0 \le p^{(2)} \le 1)\times(0 \le \frac{A_p^{(1)}}{A_p^{(2)}} \le 1)$. Ilya Meerovich Sobol mentions that if dots of a TV $16 \times 16$ raster are transfered not by rows but \textit{quasi-randomly}, then recognition of large objects and their moves is possible without waiting when the image will fill all cells and independently on the object location \cite[pp. 3 - 4]{sobol1985}. \textit{It is worth to consider Sobol sequences \cite{sobol1976}, \cite{antonov1979}, \cite{sobol1985}, \cite{joe2008} for selecting points in regions, not covered by the axioms, for experimental estimation of $f^{(1)}$ and $f^{(2)}$.}

\paragraph{The ratio} $\frac{A_p^{(1)}}{A_p^{(2)}}$ decreases the number of variables. The reduction can be inadequate, if absolute outcomes matched with income of respondents will result in outliers, where under other equal conditions more than one value $F^{(1)}$ will correspond to a single ratio. Similar notes relate to the ratio $\frac{p^{(1)}}{p^{(2)}}$.

\paragraph{Swapping indexes 1 and 2} in Tables \ref{TblTheoryF1} - \ref{TblCartesianF1} neither eliminates nor adds uncertainty. It swaps $f^{(1)}$ and $f^{(2)}$ and transforms entries marked by '*' in the last column of Table \ref{TblF1} and meaning "absent in Table \ref{TblCartesianF1}" to an entry found in the latter table: $15/10/* \leftrightarrow 13/11/35$, $22/8/* \leftrightarrow 8/6/25$. For instance, experiment 1 from Table \ref{TblF1} corresponds to a case, existing in Table \ref{TblCartesianF1}, and can be applied together with other points for estimating parameters of a model function $f(x,y)$. From 18 unpredictable cases 2, 5 - 7, 9, 14, 16 - 18, 20, 21, 23 - 25, 27, 28, 33, 35 in Table \ref{TblCartesianF1}, only three 14, 25, 35 with the help of swapping are covered by available to the author data. The author does not see how to reuse certainty equivalents for estimating the functions of respondents. \textit{In later sections it will be shown that statistics of lotteries can provide additional points.}

\paragraph{It is natural} to "aggregate" the eight parameters to means, variances, other sample mean moments, entropy. These "secondary" quantities may supply a clearer meaning: variance tells about deviations from mean, entropy is a measure of uncertainty, etc. \textit{The formula for $f^{(1)}$ may get "physical sense".}

Figure \ref{FigF1} and the axioms hint that $f^{(1)}$ can be a combination of continuous and step functions.

We should not forget that $f^{(1)}$ can depend not only on the eight parameters fully characterizing the sample means underlying the prospects but uncovered parameters related to respondents. This can lead to several $F^{(1)}$ values collected from different groups of respondents and corresponding to the same eight values of arguments. The relationships may change in time. Discrepancies can be treated as "experimental errors" using statistics as long as they are "acceptably small". Otherwise, they would promote new dimensions of the problem. It will be beneficial to sketch projections of $f^{(1)}$ dependent on the eight arguments or quantities derived from them, \textit{where the role of time is in the increasing $N^{(1)}$ and $N^{(2)}$ and erasing uncertainty}.

\section{The role of time. Trading, and speculation}

\begin{figure}
  \centering
  \includegraphics[width=65mm]{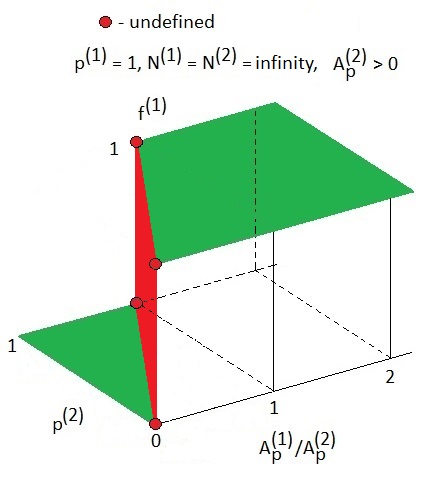}
  \caption[FigF1Inf]
   {3D projection of the theoretical fraction of respondents selecting $a_1^{(1)}$ with infinite numbers of trials applied for both $a_1^{(1)}$ and $a_1^{(2)}$. Green lines and surfaces correspond to the postulates. Red areas indicate ambiguity forming the "vertical wall" of the step function probability slices. $A_q^{(1)} = A_q^{(2)} = 0$.}
  \label{FigF1Inf}
\end{figure}

If an award or loss is a sample mean, then the role of time is in the elimination of uncertainty. With time, a high frequency of games increases the number of trials $N$. This decreases a sample mean variance, if the variance of a single trial is finite \cite{gnedenko1949}. In the limit $N \rightarrow \infty$, random sample means become deterministic mathematical expectations. The latter will dominate in decision making, if large $N$ is reachable and costs nothing, axioms \ref{EqAxiom1} - \ref{EqAxiom3}. Estimates made with the Prokhorov's inequality \ref{EqProkhorov} show that required $N$ can be large. Figure \ref{FigF1} should "converge" to Figure \ref{FigF1Inf}.
By the reasons explained later the following is chosen for $N^{(1)} \rightarrow \infty, \; N^{(2)} \rightarrow \infty, \; f^{(1,\infty)} = 1 - f^{(2,\infty)}$
\begin{equation}
\label{Eqf1inf}
f^{(1,\infty)} = \frac{\max(E(a_1^{(1)}) - E(a_1^{(2)}), 0)}{\max(E(a_1^{(2)}) - E(a_1^{(1)}), 0) + \max(E(a_1^{(1)}) - E(a_1^{(2)}), 0)}.
\end{equation}
It takes 0 or 1, is undefined, if the mathematical expectations of two prospects given by Equation \ref{EqMeanOfSampleMean} are equal each to other, and agrees with the axioms and Equation \ref{EqFunctional}: $f^{(\infty)}(a_1^{(1)},a_1^{(2)}) + f^{(\infty)}(a_1^{(2)},a_1^{(1)}) = 1; \; 0 \le f^{(\infty)}(a_1^{(1)},a_1^{(2)}) \le 1$ .

In contrast, if a respondent feels that a large number of trials almost zeroing variance is not reachable in a reasonable time, then his or her choice will be influenced not only by mathematical expectations. Buyers of Mega Million and Powerball tickets "prove" it, when the jackpots are renewed and the expectation of the game is negative. The case $N = 1$ is  embedded into Kahneman's and Tversky's prospects \cite{kahneman1979}, \cite{tversky1992}.

In trading and market speculation there is a lack of control that a decision to buy or sell a security or contract is made under the same conditions. A participant focusing on a rule or pattern ignores other factors possibly affecting a next price or rate move. The majority feels a uniqueness of trading opportunity and inability to "play" endlessly and deviates from mathematical expectations. This tendency was described by Livermore \cite{livermore1940} prior Kahneman's and Tversky's systematic research, demolishing the utility theory and adding valuable experimental points to the functions of choice.

An apologist of the \textit{Technical Analysis} \cite{murphy1999} can argue that formations of trading price patterns or indicator values themselves point to repetition of underlying conditions. Without presenting evidences of such dependencies a healthy criticism is born because many patterns can appear as a result of \textit{random walk} \cite{malkiel2007}. Less cited but not less important mathematical properties of a \textit{symmetric random walk} are described in \cite{kolmogorov1982}, \cite{feller1964}.The review of the Technical Analysis \cite{neftci1991} leaves to a speculator "hopes on systematic success", following from non-linear properties of the real financial time-series. There is a lack of published results in this area. In \cite{salov2011}, \cite{salov2012}, \cite{salov2013} the author concludes that as long as a market creates frequent opportunities to make large profits, speculation will exist, and suggests measuring the frequencies and potential profits using the \textit{maximum profit strategy}, which is an objective market property.

A point of view distinguishing between economic risk and uncertainty situations is presented in \cite{knight1964}. While the risk assumes probability distributions known at the outset, uncertain situations imply absence of the probability models describing them. Frank Knight talks about \cite[p. 232]{knight1964}: \textit{... that higher form of uncertainty not susceptible to measurement and hence to elimination}. He concludes: \textit{It is this true uncertainty which by presenting the theoretically perfect outworking of the tendencies of competition gives the characteristic form of "enterprise" to economic organizations as a whole and accounts for the peculiar income of the entrepreneur}. The author believes that inability to measure such uncertainty also deals with fundamental limits of the probability theory requiring repetition of events and trials. Trading is full of \textit{individual random objects}. An E-mini S\&P 500 future contract price time series is unique. Ability to check presence of the "same" conditions for making decisions based on the past time series is limited. Classics well understood the limits of the probability theory \cite{kolmogorov1987}. The latter work summarizes three approaches and definitions of randomness of an individual sequence of 0s and 1s based on the theory of algorithms. In contrast with these difficulties, the size and frequency of market offers measured by the maximum profit strategy do not become less intensive \cite{salov2013}. In the sense described here, time decreases the variance of the sample mean awards and "evaporates" their uncertainty.

John Jr. Kelly \cite{kelly1956} introduces the \textit{exponential rate of growth of the gambler's capital} as the limit $G = \lim_{N \rightarrow \infty} \frac{1}{N} \log_2\frac{V_N}{V_0}$. It is possible to show that under his conditions and for his function $V_N = (1 + \ell)^W (1 - \ell)^L V_0$ with the constant bet allocation fraction $\ell$, where $W$ and $L$ are the numbers of wins and losses and $V_0$ and $V_N$ is the initial and after $N$ bets capital, this limit coincides with the mathematical expectation $E(G_N)$ for arbitrary positive integer $N$. This implies that the process $G_N$ is not a \textit{martingale}, see definition in \cite[p. 91]{doob1990}. Formulas suitable for futures trading based on the allocation fraction of the capital for a next trade are considered in \cite{vince1992}, \cite{vince1995}, and \cite[pp. 55 - 79]{salov2007}. Kelly's approach is a bright example demonstrating that people may rely not only on a mathematical expectation of a game. Another famous example is the \textit{Sharpe ratio} \cite{sharpe2007}.

Khinchin \cite{khinchin1925} well understood the influence of the growing number of trials on the properties of the arithmetic and geometric sample means in the St. Petersburg game. However, his two theorems do not allow to estimate a convergence rate of the means. Such estimates based on his theorems have been obtained on a computer \cite{salov2014}. Prokhorov's estimate, when it can be applied, is a key to verification of influence of time linearly affecting the number of trials under a condition of constant frequency of trials.

\section{The functions of choice $f^{(1)}$ and $f^{(2)}$}

\paragraph{The random difference} of the two \textit{independent} random variables $\eta=a_1^{(1)}-a_1^{(2)}$ characterizes absolute advantage, if it is positive, or disadvantage, if it is negative, of the award $a_1^{(1)}$ over $a_1^{(2)}$. Independence simplifies formulas for the moments due to known relationships for $\xi=\xi^{(1)}+\xi^{(2)}$ \cite[p. 70]{gnedenko1949}:
\begin{displaymath}
\alpha_1=\alpha_1^{(1)} + \alpha_2^{(2)}, \; \mu_2=\mu_2^{(1)}+\mu_2^{(2)}, \; \mu_3=\mu_3^{(1)}+\mu_3^{(2)}, 
\end{displaymath}
\begin{displaymath}
\mu_4=\mu_4^{(1)} + 6\mu_2^{(1)}\mu_2^{(2)} + \mu_4^{(2)}.
\end{displaymath}
It can be proved that for the difference $\xi=\xi^{(1)}-\xi^{(2)}$ two of the four formulas have to change to $\alpha_1=\alpha_1^{(1)} - \alpha_2^{(2)}$ and $\mu_3=\mu_3^{(1)}-\mu_3^{(2)}$. Using equations from section "Problem 3 from Kahneman and Tversky" we get for two-point variables
\begin{equation}
\label{EqMeanDiff}
E(\eta) = \alpha_1(\eta) = (A_p^{(1)}-A_q^{(1)})p^{(1)}+A_q^{(1)}-(A_p^{(2)}-A_q^{(2)})p^{(2)}-A_q^{(2)},
\end{equation}
\begin{equation}
\label{EqVarianceDiff}
D(\eta) = \mu_2(\eta) = \frac{(A_p^{(1)}-A_q^{(1)})^2p^{(1)}(1-p^{(1)})}{N^{(1)}}+ \frac{(A_p^{(2)}-A_q^{(2)})^2p^{(2)}(1-p^{(2)})}{N^{(2)}},
\end{equation}
\begin{equation}
\label{EqThirdCentralMomentDiff}
\mu_3(\eta) = \frac{(A_p^{(1)}-A_q^{(1)})^3   (p^{(1)}-3p^{(1)2}+2p^{(1)3})}{N^{(1)2}} - \frac{(A_p^{(2)}-A_q^{(2)})^3   (p^{(2)}-3p^{(2)2}+2p^{(2)3})}{N^{(2)2}},
\end{equation}
\begin{equation}
\label{EqFourthCentralMomentDiff}
\begin{split}
\mu_4(\eta) = \frac{(A_p^{(1)}-A_q^{(1)})^4p^{(1)}(1-p^{(1)})[1 + 3(N^{(1)}-2)p^{(1)}(1-p^{(1)})]}{N^{(1)3}} +\\
+ 6 \frac{(A_p^{(1)}-A_q^{(1)})^2p^{(1)}(1-p^{(1)})  (A_p^{(2)}-A_q^{(2)})^2p^{(2)}(1-p^{(2)})  }{N^{(1)}N^{(2)}} + \\
+ \frac{(A_p^{(2)}-A_q^{(2)})^4p^{(2)}(1-p^{(2)})[1 + 3(N^{(2)}-2)p^{(2)}(1-p^{(2)})]}{N^{(2)3}},
\end{split}
\end{equation}
\begin{equation}
\label{EqSkewnessDiff}
\gamma_1(\eta) = \frac{\mu_3(\eta)}{\sqrt{\mu_2(\eta)^3}},
\end{equation}
\begin{equation}
\label{EqExcessKurtosisDiff}
\gamma_2(\eta) = \frac{\mu_4(\eta)}{\mu_2(\eta)^2} - 3.
\end{equation}
\begin{figure}
  \centering
  \includegraphics[width=100mm]{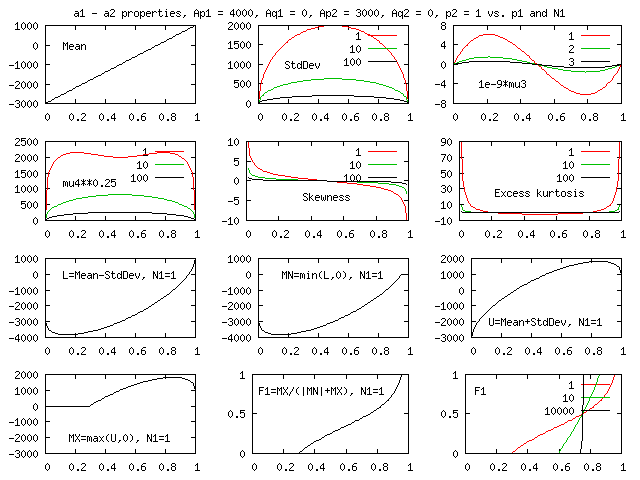}
  \caption[FigA1DiffProblem3]
   {Properties of $\eta=a_1^{(1)}-a_2^{(2)}$ for Problem 3 (certainty effect), where $p_1^{(1)}$ changes from 0 to 1 for different values of $N^{(1)}$. Plots are done using gnuplot \url{http://www.gnuplot.info/}.}
  \label{FigA1DiffProblem3}
\end{figure}
\begin{figure}
  \centering
  \includegraphics[width=100mm]{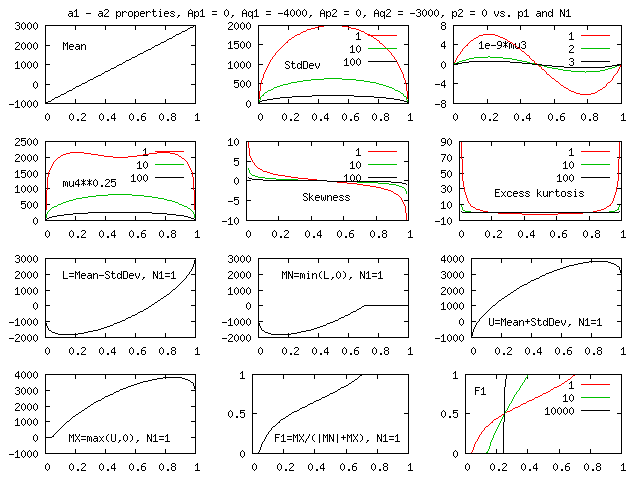}
  \caption[FigA1DiffProblem3b]
   {Properties of $\eta=a_1^{(1)}-a_2^{(2)}$ for Problem 3' (reflection effect), where $p_1^{(1)}$ changes from 0 to 1 for different values of $N^{(1)}$. Plots are done using gnuplot \url{http://www.gnuplot.info/}.}
  \label{FigA1DiffProblem3b}
\end{figure}
For $N^{(1)}=N^{(2)}=1$, the four values of $\eta$ and corresponding probabilities are
\begin{equation}
\label{EqEtaValuesProbabilities}
\begin{split}
\eta_1=A_p^{(1)}-A_p^{(2)}, \; p_1=p^{(1)}p^{(2)};\\
\eta_2=A_p^{(1)}-A_q^{(2)}, \; p_2=p^{(1)}(1-p^{(2)});\\
\eta_3=A_q^{(1)}-A_p^{(2)}, \; p_3=(1-p^{(1)})p^{(2)};\\
\eta_4=A_q^{(1)}-A_q^{(2)}, \; p_4=(1-p^{(1)})(1-p^{(2)});\\
p_1+p_2+p_3+p_4=1, \; \eta_1p_1+\eta_2p_2+\eta_3p_3+\eta_4p_4=E(\eta).
\end{split}
\end{equation}
Dependencies of statistics of the random difference $\eta$ on probability of the outcome $A_p^{(1)}=4000$ and number of trials $N^{(1)}$ for Problem 3 (certainty effect) are plotted on Figure \ref{FigA1DiffProblem3} and illustrate formulas \ref{EqMeanDiff} - \ref{EqExcessKurtosisDiff}. Dependencies of $\eta$ on probability of the outcome $A_q^{(1)}=-4000$ and number of trials $N^{(1)}$ for Problem 3' (reflection effect) are presented on Figure \ref{FigA1DiffProblem3b}.

\paragraph{Ratio of Fractions.} Let $\frac{f(a_1^{(1)},a_1^{(2)})}{f(a_1^{(2)},a_1^{(1)})}=r$. Then, from Equation \ref{EqFunctional} $f(a_1^{(1)}, a_1^{(2)})=\frac{r}{1+r}$, $f(a_1^{(2)}, a_1^{(1)})=\frac{1}{1+r}$. We shall search for an expression for $r$.

How much $a_1^{(1)}$ \textit{can be} greater than $a_1^{(2)}$? If both are deterministic, then our answer is $\max(a_1^{(1)} - a_1^{(2)}, 0)$. The function $\max$ is applied to the difference in order to emphasize the adjective \textit{greater}. If the difference is negative, then $a_1^{(1)}$ is \textit{not greater} than $a_1^{(2)}$: the positive excess is zero. If both are random, then \textit{on average} our answer is $\max(E(a_1^{(1)})-E(a_1^{(2)}),0)=\max(E(\eta),0)$. But not \textit{on average} it can be greater and we express it adding $a(\eta) \ge 0$ standard deviations to the mean $\max(E(\eta)+a(\eta)\sqrt{D(\eta)},0)=\max(\eta_{high},0)$, where $\eta_{high} \le \eta_{max}$.

When one or both sample means are random, $a_1^{(1)}$ can be less than $a_1^{(2)}$. How much less we express as a non-negative quantity $|\min(E(\eta)-b(\eta)\sqrt{D(\eta)},0)|=|\min(\eta_{low},0)|=-\min(\eta_{low},0)$, where $\eta_{min} \le \eta_{low}$. The number $b(\eta) \ge 0$ does not have to be equal to $a(\eta) \ge 0$, if the distribution of $\eta$ is not symmetrical. How much $a_1^{(1)}$ \textit{can be} less than $a_1^{(2)}$ is equal to that how much $a_1^{(2)}$ can be greater than $a_1^{(1)}$: $|\min(E(a_1^{(1)})-E(a_1^{(2)}) - b(\eta)\sqrt{D(a_1^{(1)})+D(a_1^{(2)})},0)|=\max(E(a_1^{(2)})-E(a_1^{(1)}) + b(\eta)\sqrt{D(a_1^{(1)})+D(a_1^{(2)})},0)$. How much $a_1^{(1)}$ can be greater than $a_1^{(2)}$ is equal to that how much $a_1^{(2)}$ can be less than $a_1^{(1)}$: $\max(E(a_1^{(1)})-E(a_1^{(2)}) + a(\eta)\sqrt{D(a_1^{(1)})+D(a_1^{(2)})},0)=|\min(E(a_1^{(2)})-E(a_1^{(1)}) - a(\eta)\sqrt{D(a_1^{(1)})+D(a_1^{(2)})},0)|$.

\paragraph{Hypothesis.} \textit{The ratio of the fractions of respondents is the ratio of how much} $a_1^{(1)}$ \textit{can be greater than} $a_1^{(2)}$ \textit{to how much} $a_1^{(2)}$ \textit{can be greater than} $a_1^{(1)}$
\begin{equation}
\label{EqTheRatio}
r = \frac{f^{(1)}}{f^{(2)}} = \frac{f(a_1^{(1)},a_1^{(2)})}{f(a_1^{(2)},a_1^{(1)})} =\frac{\max(E(\eta)+a(\eta)\sqrt{D(\eta)},0)}{|\min(E(\eta)-b(\eta)\sqrt{D(\eta)},0)|},
\end{equation}
\begin{equation}
\label{EqThef1}
f^{(1)}=\frac{\max(E(\eta)+a(\eta)\sqrt{D(\eta)},0)}{\max(-E(\eta)+b(\eta)\sqrt{D(\eta)},0)+\max(E(\eta)+a(\eta)\sqrt{D(\eta)},0)},
\end{equation}
\begin{equation}
\label{EqThef2}
f^{(2)}=\frac{\max(-E(\eta)+b(\eta)\sqrt{D(\eta)},0)}{\max(-E(\eta)+b(\eta)\sqrt{D(\eta)},0)+\max(E(\eta)+a(\eta)\sqrt{D(\eta)},0)}.
\end{equation}
\textit{In general, the ratio of the fractions of respondents choosing between two random or deterministic prospects is the ratio of how much the first prospect can be "better" than the second to how much the second can be "better" than the first.} Equations \ref{EqTheRatio} - \ref{EqThef2} express what is "better" for two sample means. As long as \textit{common rational sense} underlies "better", fractions depend on prospects only. If "better" is dominated by subjective preferences, then the formulas can stop working. In a private discussion with Pavel Grosul, it was emphasized that the same money means different things for respondents with different wealth. In \cite{kahneman1979} and \cite{tversky1992} a care was taken to choose some outcomes  comparable or intentionally deviating from a monthly income. Accounting such effects requires a distribution of respondents by income and/or wealth. Equations above can hide subjective complexity in $a(\eta, \; subjective \; parameters)$ and $b(\eta, \; subjective \; parameters)$, but do not pretend to become \textit{formula amoris} or \textit{equations of attractiveness ...}

\paragraph{Constraints.}If $D(\eta) = 0$, then Equation \ref{EqThef1} $\equiv$ \ref{Eqf1inf}. Natural constraints are
\begin{equation}
\label{EqConstraintsAB}
\begin{split}
0 \le a \le \frac{\eta_{max} - E(\eta)}{\sqrt{D(\eta)}}; \; 0 \le b \le \frac{E(\eta)-\eta_{min}}{\sqrt{D(\eta)}}; \; 0 \le a+b \le \frac{\eta_{max}-\eta_{min}}{\sqrt{D(\eta)}},\\
\eta_{max}=a_{1,max}^{(1)}-a_{1,min}^{(2)}=\frac{A_{max}^{(1)}N^{(1)}}{N^{(1)}} - \frac{A_{min}^{(2)}N^{(2)}}{N^{(2)}}=A_{max}^{(1)}-A_{min}^{(2)},\\
\eta_{min}=a_{1,min}^{(1)}-a_{1,max}^{(2)}=\frac{A_{min}^{(1)}N^{(1)}}{N^{(1)}} - \frac{A_{max}^{(2)}N^{(2)}}{N^{(2)}}=A_{min}^{(1)}-A_{max}^{(2)},\\
\eta_{max}-\eta_{min}=(A_{max}^{(1)}-A_{min}^{(1)})+(A_{max}^{(2)}-A_{min}^{(2)}).
\end{split}
\end{equation}
For two-point variables $A_{max}^{(1)}=A_p^{(1)}$, $A_{min}^{(1)}=A_q^{(1)}$, $A_{max}^{(2)}=A_p^{(2)}$, $A_{min}^{(2)}=A_q^{(2)}$
\begin{equation}
\label{EqBinaryConstraintsAB}
\begin{split}
0 \le a(\eta) \le \frac{(A_p^{(1)}-A_q^{(1)})(1-p^{(1)})+(A_p^{(2)}-A_q^{(2)})p^{(2)}}{\sqrt{D(\eta)}},\\
0 \le b(\eta) \le \frac{(A_p^{(1)}-A_q^{(1)})p^{(1)}+(A_p^{(2)}-A_q^{(2)})(1-p^{(2)})}{\sqrt{D(\eta)}}.
\end{split}
\end{equation}
The $a_{max}(a_1^{(1)}, a_1^{(2)})$ and $b_{max}(a_1^{(1)}, a_1^{(2)})$ posses the property that swapping indexes 1 and 2 transforms $a$ to $b$ and vice versa. If the same property holds for other $a(\eta)$ and $b(\eta)$, then $f^{(1)}(a_1^{(1)}, a_1^{(2)})$ and $f^{(2)}(a_1^{(1)}, a_1^{(2)})$ will have it too because the remaining $E(\eta)$ and $D(\eta)$ already have it. This would allow to write $f^{(1)}(a_1^{(1)}, a_1^{(2)}) = f(a_1^{(1)}, a_1^{(2)})$ and $f^{(2)}(a_1^{(1)}, a_1^{(2)}) = f(a_1^{(2)}, a_1^{(2)})$ and use one function. If $\sqrt{D(\eta)}=0$, then $a(\eta)$ and $b(\eta)$ can be arbitrary finite numbers.

In Problem 3 $\eta_{max} = 4000 - 3000 = 1000$, $\eta_{min} = 0 - 3000 = -3000$, $E(\eta)=200$, and $\sqrt{D(\eta)}=1600$. Therefore, $0 \le a \le \frac{1}{2}$ and $0 \le b \le 2$. Any pair $(a,b)$ satisfying the equation $b=4a+\frac{5}{8}$ with $0\le a$ being substituted to Equations \ref{EqThef1} and \ref{EqThef2} yields $f^{(1)}=0.2$ and $f^{(2)}=0.8$ experimentally found for Problem 3. Intersection with constrains gives $0 \le a \le \frac{11}{32}$ and $\frac{5}{8} \le b \le 2$.

In Problem 3' $\eta_{max} = 0 - (-3000) = 3000$, $\eta_{min} = -4000 - (-3000) = -1000$, $E(\eta)=-200$, and $\sqrt{D(\eta)}=1600$. Therefore, $0 \le a \le 2$ and $0 \le b \le \frac{1}{2}$.  Any pair $(a,b)$ obeying $b = \frac{2}{23}a - \frac{25}{184}$ with $0 \le b$ yields $f^{(1)}=0.92$ and $f^{(2)}=0.08$ observed for Problem 3'. Intersection with constraints gives $\frac{25}{16} \le a \le 2$ and $0 \le b \le \frac{7}{184}$. The solution lines for Problem 3 and 3' intersect at $(a=-\frac{7}{36},b=-\frac{11}{72})$, which violates constraints: there is no one pair satisfying both problems. The $a$ and $b$ obeying constraints \ref{EqConstraintsAB} ensure predictable properties.

If $f^{(1)}$ or $f^{(2)}$ are not equal to zero or one, then the formulated relationships \ref{EqTheRatio} - \ref{EqConstraintsAB} imply dependence between $a$ and $b$ because from two fractions of respondents only one is independent. If $\eta_{low} < 0 < \eta_{high}$, then
\begin{equation}
\label{EqDependenceAB}
b(\eta) = \frac{E(\eta) + a(\eta) \sqrt{D(\eta)}(1 - f^{(1)})}{f^{(1)}\sqrt{D(\eta)}} = \frac{a(\eta)}{r}+\frac{E(\eta)}{\sqrt{D(\eta)}}\frac{1}{f^{(1)}},
\end{equation}
where $\eta=a_1^{(1)}-a_1^{(2)}$. $\frac{E(\eta)}{\sqrt{D(\eta)}}$ can be interpreted as a Sharpe ratio, where $a_1^{(1)}$ is not a return but a sample mean over or under a benchmark sample mean $a_1^{(2)}$.

\paragraph{Case of i.i.d.} If $a_1^{(1)}$ and $a_1^{(2)}$ are i.i.d., then $E(\eta)=0$, $0 < D(\eta)$ (non-constant variables), and $f^{(1)}=\frac{a(\eta)}{a(\eta) + b(\eta)}$. From Equations \ref{EqVarianceDiff} and \ref{EqBinaryConstraintsAB}, $a_{max} = b_{max} = \frac{N}{2p(1-p)}$. Similar to equal constants, there is no rational sense to select between two identical random quantities and $f^{(1)}$ must be undefined. \textit{We conclude that in this case} $a(\eta)=b(\eta)=0$.

\paragraph{Disjoint and adjacent intervals.} In Equation \ref{EqThef1} the first $\max$ in the denominator is zero, if $-E(\eta) + b(\eta)\sqrt{D(\eta)} \le -E(\eta) + b_{max}(\eta)\sqrt{D(\eta)} = -E(\eta) + E(\eta) - \eta_{min} = -\eta_{min} = A_{max}^{(2)} - A_{min}^{(1)} \le 0$. $f^{(1)}=\frac{\max(E(\eta)+a\sqrt{D(\eta)},0)}{0 + \max(E(\eta)+a\sqrt{D(\eta)},0)}$ for $A_{max}^{(2)} \le A_{min}^{(1)}$. The second max is zero, if $E(\eta) + a(\eta)\sqrt{D(\eta)} \le E(\eta) + a_{max}(\eta)\sqrt{D(\eta)} = E(\eta) + \eta_{max} - E(\eta) = \eta_{max} = A_{max}^{(1)}-A_{min}^{(2)} \le 0$. Thus, $f^{(1)}=\frac{0}{0+0}$ is undefined, if $A_{max}^{(2)} \le A_{min}^{(1)}$ and $A_{max}^{(1)} \le A_{min}^{(2)}$. This implies $A_{min}^{(2)} = A_{max}^{(2)} = A_{min}^{(1)} = A_{max}^{(1)}$. If only one of the inequalities holds, then Equation \ref{EqThef1} correctly returns 1 or 0 corresponding to predictable properties.

\paragraph{Cases $E(\eta) = 0$.} For 4, 5, 8, 9, 10, 11, 12, 13, 23, and 33 in Table \ref{TblF1}, $\sqrt{D(\eta)}$ vanishes from Equations \ref{EqTheRatio} - \ref{EqThef2}, and $r=\frac{a}{b}$, $f^{(1)} = \frac{a}{a + b}, \; f^{(2)}=\frac{b}{a+b}$, Table \ref{Tbla1a2eta}.

\begin{center}
\begin{longtable}{|r|c|r|r|r|r|r|r|r|r|r|c|}
\caption[History of Mega Millions]{Computed properties of $a_1^{(1)}$, $a_2^{(2)}$, and $\eta=a_1^{(1)} - a_2^{(2)}$ for $N^{(1)}=N^{(2)}=1$, $E(\eta)=0$, and experimental $F$ from Table \ref{TblF1}.} \label{Tbla1a2eta} \\
 \hline
 \multicolumn{1}{|c|}{\#} &
 \multicolumn{1}{c|}{$\xi$} &
 \multicolumn{1}{c|}{$A_p$} &
 \multicolumn{1}{c|}{$A_q$} &
 \multicolumn{1}{c|}{$p$} &
 \multicolumn{1}{c|}{$E$} &
 \multicolumn{1}{|c|}{$\sqrt{D}$} &
 \multicolumn{1}{c|}{$\gamma_1$} &
 \multicolumn{1}{|c|}{$\gamma_2$} &
 \multicolumn{1}{c|}{$H$} &
 \multicolumn{1}{c|}{$F$} &
 \multicolumn{1}{c|}{Table \ref{TblF1}}\\
 \hline 
 \endfirsthead
 \multicolumn{9}{c}%
 {\tablename\ \thetable{} -- continued from previous page} \\
 \hline
 \multicolumn{1}{|c|}{\#} &
 \multicolumn{1}{c|}{$\xi$} &
 \multicolumn{1}{c|}{$A_p$} &
 \multicolumn{1}{c|}{$A_q$} &
 \multicolumn{1}{c|}{$p$} &
 \multicolumn{1}{c|}{$E$} &
 \multicolumn{1}{|c|}{$\sqrt{D}$} &
 \multicolumn{1}{c|}{$\gamma_1$} &
 \multicolumn{1}{|c|}{$\gamma_2$} &
 \multicolumn{1}{c|}{$H$} &
 \multicolumn{1}{c|}{$F$} &
 \multicolumn{1}{c|}{Table \ref{TblF1}}\\
 \hline 
 \endhead
 \hline \multicolumn{12}{|r|}{{Continued on next page}} \\ \hline
 \endfoot
 \hline
 \endlastfoot
1 & $a_1^{(1)}$ & 6000 & 0 & 0.45 & 2700 & 2985 & 0.201 & -1.96 & 0.993 &  0.14 & 4\\
   & $a_1^{(2)}$ & 3000 & 0 & 0.9 & 2700 & 900 & -2.67 & 5.11 & 0.469 &  0.86 & 4\\
   & $\eta$ &  &  &  & 0 & 3118 & 0.24 & -1.61 & 1.46 &  1.00 & \\
2 & $a_1^{(1)}$ & 6000 & 0 & 0.001 & 6 & 190 & 31.6 & 995 & 0.011 &  0.73 & 5\\
   & $a_1^{(2)}$ & 3000 & 0 & 0.002 & 6 & 134 & 22.3 & 495 & 0.021 &  0.27 & 5\\
   & $\eta$ & & & & 0 & 232 & 12.9 & 497 & 0.032 &  1.00 & \\
3 & $a_1^{(1)}$ & 0 & -6000 & 0.55 & -2700 & 2985 & -0.201 & -1.96 & 0.993 &  0.92 & 8\\
   & $a_1^{(2)}$ & 0 & -3000 & 0.1 & -2700 & 900 & 2.67 & 5.11 & 0.469 &  0.08 & 8\\
   & $\eta$ & & & & 0 & 3118 & -0.24 & -1.61 & 1.46 &  1.00 & \\
4 & $a_1^{(1)}$ & 0 & -6000 & 0.999 & -6 & 190 & -31.6 & 995 & 0.011 &  0.30 & 9\\
   & $a_1^{(2)}$ & 0 & -3000 & 0.998 & -6 & 134 & -22.3 & 495 & 0.021 &  0.70 & 9\\
   & $\eta$ & & & & 0 & 232 & -12.9 & 497 & 0.032 &  1.00 & \\
5 & $a_1^{(1)}$ & 1000 & 0 & 0.5 & 500 & 500 & 0 & -2 & 1 &  0.16 & 10\\
   & $a_1^{(2)}$ & 500 & 500 & 1 & 500 & 0 & 0 & -3 & 0 &  0.84 & 10\\
   & $\eta$ & & & & 0 & 500 & 0 & -2 & 1 &  1.00 & \\
6 & $a_1^{(1)}$ & 0 & -1000 & 0.5 & -500 & 500 & 0 & -2 & 1 &  0.69 & 11\\
   & $a_1^{(2)}$ & -500 & -500 & 1 & -500 & 0 & 0 & -3 & 0 &  0.31 & 11\\
   & $\eta$ & & & 0 & & 500 & 0 & -2 & 1 &  1.00 & \\
7 & $a_1^{(1)}$ & 5000 & 0 & 0.001 & 5 & 158 & 31.6 & 995 & 0.011 &  0.72 & 12\\
   & $a_1^{(2)}$ & 5 & 5 & 1 & 5 & 0 & 0 & -3 & 0 &  0.28 & 12\\
   & $\eta$ & & & & 0 & 158 & 31.6 & 995 & 0.011 &  1.00 & \\
8 & $a_1^{(1)}$ & 0 & -5000 & 0.999 & -5 & 158 & -31.6 & 995 & 0.011 &  0.17 & 13\\
   & $a_1^{(2)}$ & -5 & -5 & 1 & -5 & 0 & 0 & -3 & 0 &  0.83 & 13\\
   & $\eta$ & & & & 0 & 158 & -31.6 & 995 & 0.011 &  1.00 & \\
9 & $a_1^{(1)}$ & 25 & 25 & 1 & 25 & 0 & 0 & -3 & 0 &  ? & 23\\
   & $a_1^{(2)}$ & 100 & 0 & 0.25 & 25 & 43 & 1.15 & -0.667 & 0.81 &  ? & 23\\
   & $\eta$ & & & & 0 & 43 & -1.15 & -0.667 & 0.81 &  1.00 & \\
10 & $a_1^{(1)}$ & 20 & 20 & 1 & 20 & 0 & 0 & -3 & 0 &  ? & 33\\
   & $a_1^{(2)}$ & 200 & 0 & 0.1 & 20 & 60 & 2.67 & 5.11 & 0.469 &  ? & 33\\
   & $\eta$ & & & & 0 & 60 & -2.67 & 5.11 & 0.469 &  1.00 & \\
\end{longtable}
\end{center}

If profits are comparable Table \ref{Tbla1a2eta}(1, 2, 5), then the greater entropy, the less the fraction of respondents. The entropy, characterizing uncertainty, becomes a measure of the certainty effect. If one profit is significantly greater than other and the latter is not attractive, then uncertainty is less important than a possibility to become rich, Table \ref{Tbla1a2eta}(7). The entropy depends only on probabilities.

If losses are comparable Table \ref{Tbla1a2eta}(3, 4, 6), then the greater entropy, the greater the fraction of respondents. Now, entropy measures reflection effect. If one loss is significantly worse than other and the latter is not critical, then uncertainty is less important comparing with a possibility of financial ruin, Table \ref{Tbla1a2eta}(8).

For profits Table \ref{Tbla1a2eta}(5, 7), $E(\eta) = 0$, $A_q^{(1)} = 0$, $p^{(2)} = 1$. Therefore, $A_p^{(1)}p^{(1)}=A_p^{(2)}$, $\sqrt{D(a_1^{(2)})}=0$, $\sqrt{D(a_1^{(1)})}=  A_p^{(1)}\sqrt{\frac{p^{(1)}(1-p^{(1)})}{N^{(1)}}}=A_p^{(2)}\sqrt{\frac{1-p^{(1)}}{p^{(1)}N^{(1)}}}$, $\lim_{p^{(1)\rightarrow 0}}{\sqrt{D(a_1^{(1)})}=\infty}$. Author's attention is attracted by one \textit{rational} and two experimental points $(\sqrt{D(a_1^{(1)})}, F^{(1)})$: $(0, 1)_{p^{(1)}=1}$, $(158, 0.72)_{12}$, $(500, 0.16)_{10}$, which are close to a straight line.

For losses Table \ref{Tbla1a2eta}(6, 8), $E(\eta) = 0$, $A_p^{(1)} = 0$, $p^{(2)} = 0$. Hence, $A_q^{(1)}(1-p^{(1)})=A_q^{(2)}$, $\sqrt{D(a_1^{(2)})}=0$, $\sqrt{D(a_1^{(1)})}=|A_q^{(1)}|\sqrt{\frac{p^{(1)}(1-p^{(1)})}{N^{(1)}}}=|A_q^{(2)}|\sqrt{\frac{p^{(1)}}{(1-p^{(1)}) N^{(1)}}}$, $\lim_{p^{(1)\rightarrow 1}}{\sqrt{D(a_1^{(1)})}=\infty}$. Author's attention is attracted by one rational and two experimental points $(\sqrt{D(a_1^{(1)})}, F^{(1)})$: $(0, 0)_{p^{(1)}=0}$, $(158, 0.17)_{13}$, $(500, 0.69)_{11}$, which are close to a straight line.

In both cases the standard deviation can approach infinity, while $0 \le F^{(1)} \le 1$. Therefore, lines $F^{(1)}=slope \times \sqrt{D(a_1^{(1)})} + intercept$ cannot be adequate on the beam $[0, \infty)$. The behavior on the right side is not obvious for the author. At first glance, decreasing dependence for profits could asymptotically approach the standard deviation axis $F^{(1)} \rightarrow 0$ being a mirrored S-shaped curve bound to the point (0, 1). However, if $A_p^{(2)}$ is tiny, one cent, comparing with a healthy monthly income, then large but less probable $A_p^{(1)}=\frac{A_p^{(2)}}{p^{(1)}}$ can become attractive. It gives a chance to "become rich" momentarily, while one cent would not make a principal difference anyway. We recollect people playing lotteries.

Similarly, a tiny definite loss can be preferred to a huge, \textit{life}, but probabilistically negligible one, where $|A_q^{(1)}| = \frac{|A_q^{(2)}|}{1-p^{(1)}}$. Such a preference does not look for the author irrational.

\begin{figure}
  \centering
  \includegraphics[width=80mm]{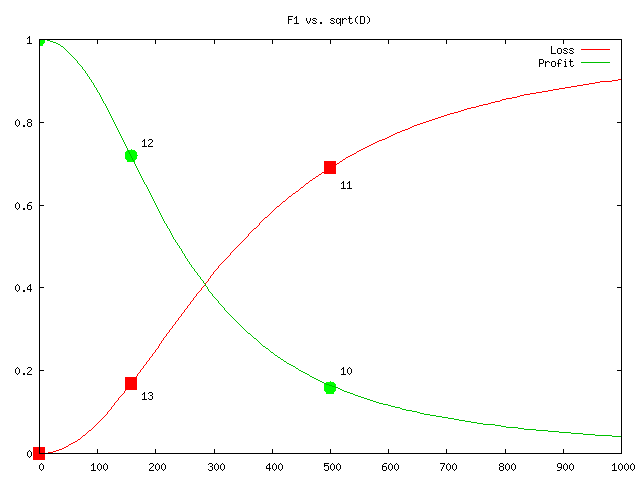}
  \caption[FigF1Inf]
   {Table \ref{TblF1}(10, 12): $E(\eta) = 0$, $A_q^{(1)} = A_q^{(2)} = 0$, $p^{(2)} = 1$, $N^{(1)}=N^{(2)}=1$, $F_{profit}^{(1)}=\frac{238505.5242}{238505.5242+(\sqrt{D(a_1^{(1)})})^{2.259253618}}$. Table \ref{TblF1}(11, 13): $E(\eta) = 0$, $A_p^{(1)} = A_p^{(2)} = 0$, $p^{(2)} = 0$, $N^{(1)}=N^{(2)}=1$, $F_{loss}^{(1)}=\frac{(\sqrt{D(a_1^{(1)})})^{2.071333117}}{174976.0287+(\sqrt{D(a_1^{(1)})})^{2.071333117}}$.}
  \label{FigF1Deta0}
\end{figure}

Two curves of fractions of respondents dependent on variances of random prospects and fitting rational and experimental points are plotted on Figure \ref{FigF1Deta0}. They asymptotically approach zero (profits) or one (losses). The fitting curves are $F_{profit}^{(1)}=\frac{K}{K+(\sqrt{D(a_1^{(1)})})^k}$, and $F_{loss}^{(1)}=\frac{(\sqrt{D(a_1^{(1)})})^m}{M+(\sqrt{D(a_1^{(1)})})^m}$. The latter equation is mathematically equivalent to the \textit{Archibald Hill's equation} describing the sigmoidal oxygen binding curve of hemoglobin \cite{hill1910}. In the particular case $m=2$, the equation transforms to $\frac{D(a_1^{(1)})}{M+D(a_1^{(1)})}$, which is mathematically equivalent to the \textit{Leonor Michaelis - Maud Menten equation} describing enzyme kinetics. Both can be written as straight line equations in bilogarithmic scales $\ln(r_{profit}) = \ln(K)-\frac{k}{2}\ln(D(a_1^{(1)}))$, and $\ln(r_{loss}) = \frac{m}{2}\ln(D(a_1^{(2)}))-\ln(M)$ each fitting two points exactly with $K=238505.5242, k=2.259253618$, and $M=174976.0287, m=2.071333117$. The differences in $K$, $M$, and $k$, $m$ take care about description of certainty and reflection effects. \textit{Absence of data on the right side of dependencies leaves the question about asymptotic properties open}.

\paragraph{Selecting between Constant and Random Two-Point Variables.} Equations \ref{EqThef1} and \ref{EqThef2} correctly describe rational properties where $D(\eta)=0$. Under other conditions, they switch computation of fractions of respondents to evaluation of $a$, $b$, and possible dependences between them.

\begin{figure}
  \centering
  \includegraphics[width=70mm]{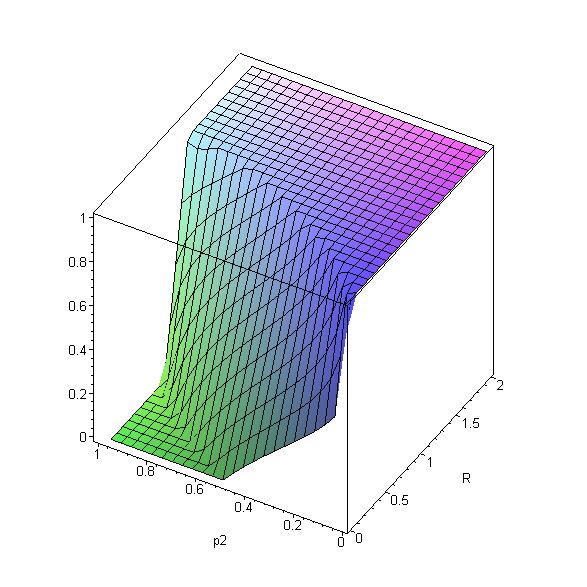}
  \caption[FigF1ConstRand1]
   {$f^{(1)}(\{A_p^{(1)}=3000, A_q^{(1)}=3000, p^{(1)}=1, N^{(1)}=1\}$, $\{A_p^{(1)}/R, A_q^{(2)}=0, p^{(2)}, N^{(2)}=1\})$ vs. $p^{(2)} \in [0, 1]$ and $R \in [0.000001, 2]$, $a=1$, $b=2/3$. Plot is done using Maple 10 from Maplesoft.}
  \label{FigF1ConstRand1}
\end{figure}

\begin{figure}
  \centering
  \includegraphics[width=70mm]{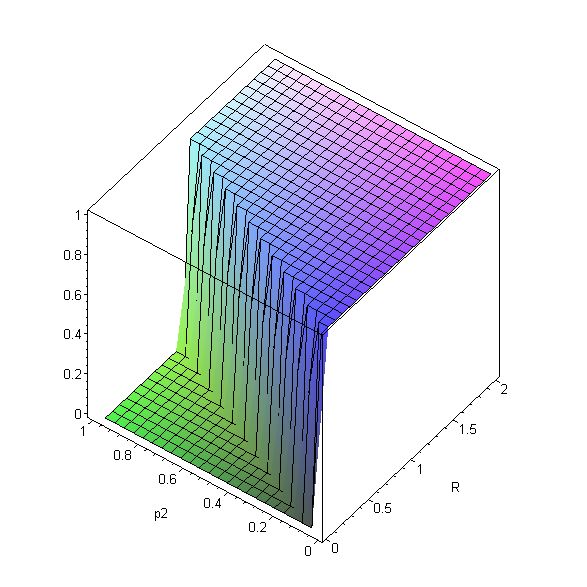}
  \caption[FigF1ConstRand2]
   {$f^{(1)}(\{A_p^{(1)}=3000, A_q^{(1)}=3000, p^{(1)}=1, N^{(1)}=10^9\}$, $\{A_p^{(1)}/R, A_q^{(2)}=0, p^{(2)}, N^{(2)}=10^9\})$ vs. $p^{(2)} \in [0, 1]$ and $R \in [0.000001, 2]$, $a=1$, $b=2/3$. Plot is done using Maple 10 from Maplesoft.}
  \label{FigF1ConstRand2}
\end{figure}

Let the first variable is constant $\{A_p^{(1)}, A_q^{(1)}=A_p^{(1)}, p^{(1)} = 1, 1 \le N^{(1)}=N^{(2)}\}$ and second - random $\{A_p^{(2)}=A_p^{(1)}/R, A_q^{(2)}=0, p^{(2)} \in [0, 1],  1 \le N^{(2)}=N^{(1)}\}$, where $0 < R$. Compare the surface on Figure \ref{FigF1ConstRand1} with experimental and axiomatic points on Figure \ref{FigF1}. For large $N^{(1)}=N^{(2)}=10^9$ Figure \ref{FigF1ConstRand2} is getting closer to Figure \ref{FigF1Inf} following from Equation \ref{Eqf1inf}. The surfaces are computed using Equations \ref{EqMeanDiff}, \ref{EqVarianceDiff}, and \ref{EqThef1}. Fitting depends on $a$, and $b$.

From Equations \ref{EqVarianceDiff} and \ref{EqBinaryConstraintsAB}, if the first variable is constant and the second is random two-point, then $a_{max}=\sqrt{\frac{p^{(2)}N^{(2)}}{1-p^{(2)}}}$, $b_{max}=\sqrt{\frac{(1-p^{(2)})N^{(2)}}{p^{(2)}}}$, $a_{max}b_{max}=N^{(2)}$. Swapping constant and random variables yields $a_{max}=\sqrt{\frac{(1-p^{(1)})N^{(1)}}{p^{(1)}}}$, $b_{max}=\sqrt{\frac{p^{(1)}N^{(1)}}{1-p^{(1)}}}$, $a_{max}b_{max}=N^{(1)}$.

\paragraph{Lottery.} In lotteries, one selects between a random variable with large win, small ticket price, tiny winning probability $\{A_p^{(1)}=large, A_q^{(2)}=-small, p^{(1)}=tiny, N^{(1)}=1\}$ and the constant $\{A_p^{(2)}=0, A_q^{(2)}=0, p^{(2)}=1, N^{(2)}=1\}$.
\begin{figure}
  \centering
  \includegraphics[width=90mm]{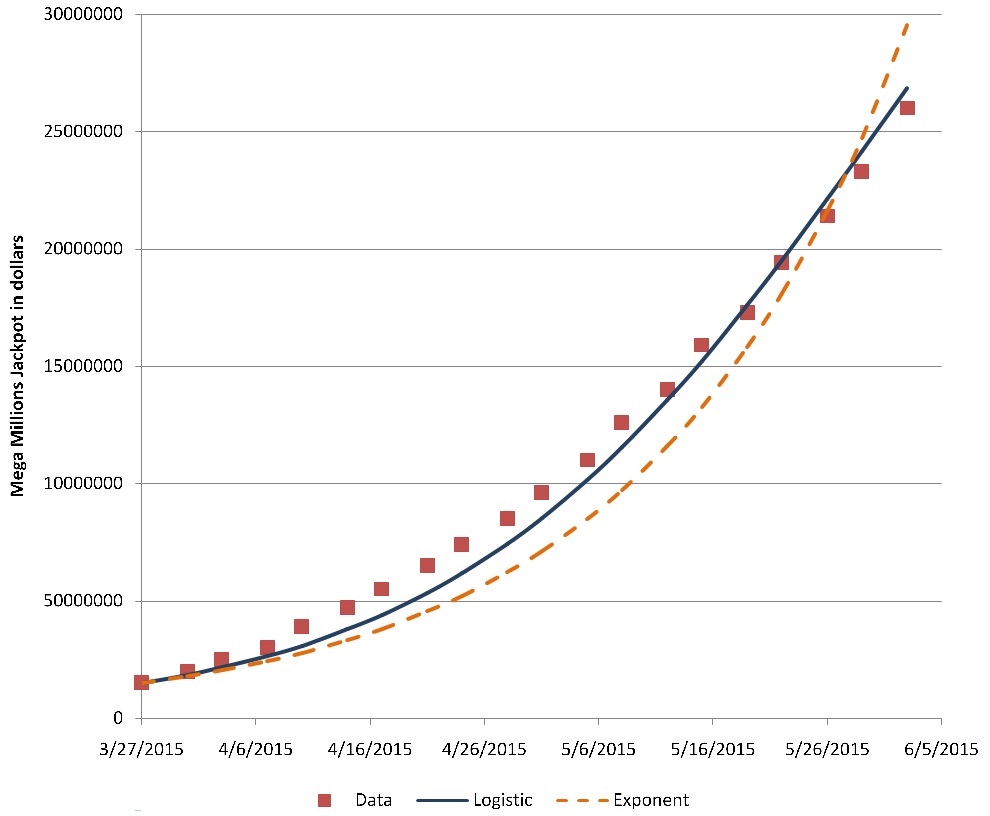}
  \caption[FigMegaMillionsJackpot]
   {A typical Mega Millions jackpot growth curve before dropping to initial jackpot. The data is taken from \url{http://www.lottostrategies.com/script/jackpot_history/draw_date/113}. Plot is done using Microsoft Excel.}
  \label{FigMegaMillionsJackpot}
\end{figure}

Usually, the greater the jackpot, the faster it grows, Figure \ref{FigMegaMillionsJackpot}. If the jackpot is won, then the curve drops to the initial jackpot $J_0$. This creates the sequence of "teeth" with random height and width at the bottom. Dependence after $J_0$ resembles the exponential solution $J_E(t) = J_0 \exp^{k(t-t_0)}$ of the equation $\frac{dJ}{dt} = kJ$ with the initial condition $J_E(t_0)=J_0$. This solution cannot describe random shape of the teeth and sharp dropping at random time. The jackpot, even, without dropping, could not grow to infinity. Hence, the S-shaped \textit{logistic curve} $J_L(t)=\frac{J_{max}J_0\exp(k(t-t_0))}{J_{max}+J_0(\exp(k(t-t_0))-1)}$, solving the equation $\frac{dJ}{dt}=kJ(1-\frac{J}{J_{max}})$ developed with other denominations by Pierre Francois Verhulst and Lamberte Adolphe Jacques Quetelet  for describing population growth, is a better fitting choice. Without data close to $J_{max}$, estimation of this level is inaccurate. The Richards' \textit{generalized logistic function} \cite{richards1959} is more flexible. With credit to \url{http://www.lottostrategies.com/script/jackpot_history/draw_date/113}, the author presents the data in Table \ref{TblJackpot} for those who wants to try other fitting options. The exponential and logistic curves were applied minimizing the sum of square deviations (Microsoft Excel, Solver), which is not a well justified criterion in this case. $A_p^{(1)}=J+A_q^{(1)}$ with $A_q^{(1)} < 0$ does not take into account a possibility of sharing jackpot between winners.  For simplicity, we also ignore multiple winning outcomes and taxes described in Section "Mega Millions".
\begin{center}
\begin{longtable}{|c|c|r|r|r|c|r|c|}
\caption[A growth of Mega Millions Jackpot]{A growth of Mega Millions Jackpot $J$ before dropping to $J_0=15,000,000$, $J_L(t)=\frac{500,000,000\times 15,000,000 \times \exp(19.748t)}{500,000,000 + 15,000,000 \times (\exp(19.748t)-1)}$, $J_E(t)=15,000,000 \times \exp(16.232t)$, $\sum{\Delta J_L^2}$=1.3e+15, $\sum{\Delta J_E^2}$=7.2e+15.} \label{TblJackpot} \\
\hline
\multicolumn{1}{|c|}{Date} &
\multicolumn{1}{c|}{$t$, years} &
\multicolumn{1}{c|}{$J$, real} &
\multicolumn{1}{c|}{$J_{next} - J$} &
\multicolumn{1}{c|}{$J_L(t)$} &
\multicolumn{1}{c|}{$(J-J_L)^2$} &
\multicolumn{1}{c|}{$J_E(t)$} &
\multicolumn{1}{c|}{$(J-J_E)^2$}\\
\hline 
\endfirsthead
\multicolumn{7}{c}%
{\tablename\ \thetable{} -- continued from previous page} \\
\hline
\multicolumn{1}{|c|}{Date} &
\multicolumn{1}{c|}{$t$, years} &
\multicolumn{1}{c|}{$J$, real} &
\multicolumn{1}{c|}{$J{next} - J$} &
\multicolumn{1}{c|}{$J_L(t)$} &
\multicolumn{1}{c|}{$(J-J_L)^2$} &
\multicolumn{1}{c|}{$J_E(t)$} &
\multicolumn{1}{c|}{$(J-J_E)^2$}\\
\hline 
\endhead
\hline \multicolumn{8}{|c|}{{Continued on next page}} \\ \hline
 \endfoot
 \hline
 \endlastfoot
3/27/2015 & 0.0000 & 15000000 & 5000000 & 15000000 & 0.0e+00 & 15000000 & 0.0e+00\\
3/31/2015 & 0.0110 & 20000000 & 5000000 & 18490306 & 2.3e+12 & 17920342 & 4.3e+12\\
4/3/2015 & 0.0192 & 25000000 &  5000000 & 21608026 & 1.2e+13 & 20478002 & 2.0e+13\\
4/7/2015 & 0.0301 & 30000000 &  9000000 & 26551757 & 1.2e+13 & 24464854 & 3.1e+13\\
4/10/2015 & 0.0384 & 39000000 & 8000000 & 30941404 & 6.5e+13 & 27956572 & 1.2e+14\\
4/14/2015 & 0.0493 & 47000000 & 8000000 & 37851574 & 8.4e+13 & 33399424 & 1.8e+14\\
4/17/2015 & 0.0575 & 55000000 & 10000000 &  43935991 & 1.2e+14 & 38166318 & 2.8e+14\\
4/21/2015 & 0.0685 & 65000000 & 9000000 &  53417755 & 1.3e+14 & 45596900 & 3.8e+14\\
4/24/2015 & 0.0767 & 74000000 & 11000000 & 61670446 & 1.5e+14 & 52104664 & 4.8e+14\\
4/28/2015 & 0.0877 & 85000000 & 11000000 & 74355473 & 1.1e+14 & 62248896 & 5.2e+14\\
5/1/2015 & 0.0959 & 96000000 & 14000000 & 85225587 & 1.2e+14 & 71133297 & 6.2e+14\\
5/5/2015 & 0.1068 & 110000000 & 16000000 & 101632292 & 7.0e+13 & 84982204 & 6.3e+14\\
5/8/2015 & 0.1151 & 126000000 & 14000000 & 115408834 & 1.1e+14 & 97111190 & 8.3e+14\\
5/12/2015 & 0.1260 & 140000000 & 19000000 & 135724730 & 1.8e+13 & 116017720 & 5.8e+14\\
5/15/2015 & 0.1342 & 159000000 & 14000000 & 152355179 & 4.4e+13 & 132576214 & 7.0e+14\\
5/19/2015 & 0.1452 & 173000000 & 21000000 & 176195256 & 1.0e+13 & 158387412 & 2.1e+14\\
5/22/2015 & 0.1534 & 194000000 & 20000000 & 195128304 & 1.3e+12 & 180993072 & 1.7e+14\\
5/26/2015 & 0.1644 & 214000000 & 19000000 & 221398922 & 5.5e+13 & 216230524 & 5.0e+12\\
5/29/2015 & 0.1726 & 233000000 & 27000000 & 241565629 & 7.3e+13 & 247091774 & 2.0e+14\\
6/2/2015 & 0.1836 & 260000000 & ? & 268580467 & 7.4e+13 & 295197950 & 1.2e+15\\
\end{longtable}
\end{center}
The choice $\{A_p^{(1)}=232,999,999, A_q^{(1)}=-1, p^{(1)}=\frac{1}{258,890,850}, N^{(1)}=1\}$ vs. $\{A_p^{(2)}=0, A_q^{(2)}=0, p^{(2)}=0, N^{(2)}=1\}$ should be made on 5/29/2015. The jackpot has grown to 260,000,000. We assume that 27,000,000 is one third of the number of purchased tickets 81,000,000. Since quoting is done by annuity, the number of tickets is less. Based on Section "Mega Millions", we estimate the number of purchased tickets as half $\approx$ 40,000,000.

Mega Millions \url{http://www.megamillions.com/where-to-play} is played in 44 states AR 2,966,369, AZ 6,731,484, CA 38,802,500, CO 5,355,866, CT 3,596,677, DE 935,614, FL 19,893,297, GA 10,097,343, IA 3,107,124, ID 1,634,464, IL 12,880,580, IN 6,596,855, KS 2,904,021, KY 4,413,457, LA 4,649,676, MA 6,745,408, MD 5,976,407, ME 1,330,089, MI 9,909,877, MN 5,458,333, MO 6,063,589, MT 1,023,579, NC 9,943,964, ND 739,482, NE 1,881,503, NH 87,137 2013, NJ 8,938,175, NM 2,085,572, NY 19,746,227, OH 11,594,163, OK 3,878,051, OR 3,970,239, PA 12,787,209, RI 1,055,173, SC 4,832,482, SD 853,175, TN 6,549,352, TX 27,695,284, VA 8,326,289, VT 626,562, WA 7,061,530, WI 5,757,564, WV 1,850,326, WY 584,153, the District of Columbia 658,893, and U.S. Virgin islands 106,405 2010 census. The population estimates are from Wikipedia for the year 2014 unless the year is cited. The sum is 302,681,519. One must be 18 years or older to purchase lottery tickets. In accordance with the United States Census Bureau \url{http://www.census.gov/population/age/}, in 2010 the population under 18 years was 24 percent. We estimate the number of potential buyers as $0.76 \times 302,681,519 \approx 230,037,954$.

If the numbers of people from $P_1$ to $P_m$ buy the numbers of tickets from $n_1$ to $n_m$, then the number of buyers is $\sum_{i=1}^{i=m}P_i$, the number of bought tickets is $\sum_{i=1}^{i=m}P_in_i$, the mean number of purchased tickets per buyer is $n_{mean}=\frac{\sum_{i=1}^{i=m}P_in_i}{\sum_{i=1}^{i=m}P_i}$, the number of not playing people is $P - \sum_{i=1}^{i=m}P_i$, and the number of potential tickets, which they did not buy, is $n_{mean}(P - \sum_{i=1}^{i=m}P_i)$. The fraction of purchased tickets to the total of purchased and potentially not purchased ones is $\frac{\sum_{i=1}^{i=m}P_in_i}{\sum_{i=1}^{i=m}P_in_i+n_{mean}(P - \sum_{i=1}^{i=m}P_i)}=\frac{\sum_{i=1}^{i=m}P_i}{P}$. The latter is the fraction $f^{(1)}$ of those, who decided to play. With $n_{mean}=5$, $f^{(1)}=\frac{40,000,000}{5 \times 230,037,954} \approx 0.035$.  \textit{We conclude that lotteries can be used to estimate fractions of respondents selecting between random and constant variables or prospects}.

For this problem, $E(\eta)=-0.10$, $\sqrt{D(\eta)}=14,480.97$, $\eta_{min}=-1$, $\eta_{max}=232,999,999$, $\gamma_1(\eta)=16,090.09$, $\gamma_2(\eta)=258,890,848$, $H(\eta)$=1.15e-7. The big positive skewness $\gamma_1$ overweights the certain loss prediction following from the tiny entropy $H$. As long as $J_L$ is not won, the $A_p^{(1)}=J_L(t) + A_q^{(1)}$, $\eta_{max}$, $E(\eta)$, and $D(\eta)$ increase with $t$. In contrast, in this hypothetical lottery, probabilities and dependent only on them $\gamma_1$, $\gamma_2$, and $H$ remain intact. $J_L(t)$, Figure \ref{FigMegaMillionsJackpot}, grows with acceleration before it reaches the second part of S-shaped curve, where the capital of \textit{dreamers} exhausts. This acceleration is the increasing number of tickets from game to game and, therefore, increasing $f^{(1)}$ and $\frac{a}{b}$. To estimate $f^{(1)}$, multiply the values from column $J(next)-J$ of Table \ref{TblJackpot} by $\frac{3 \times \frac{1}{2}}{5 \times 230,037,954} \approx 1.3\times 10^{-9}$.

\begin{figure}
  \centering
  \includegraphics[width=90mm]{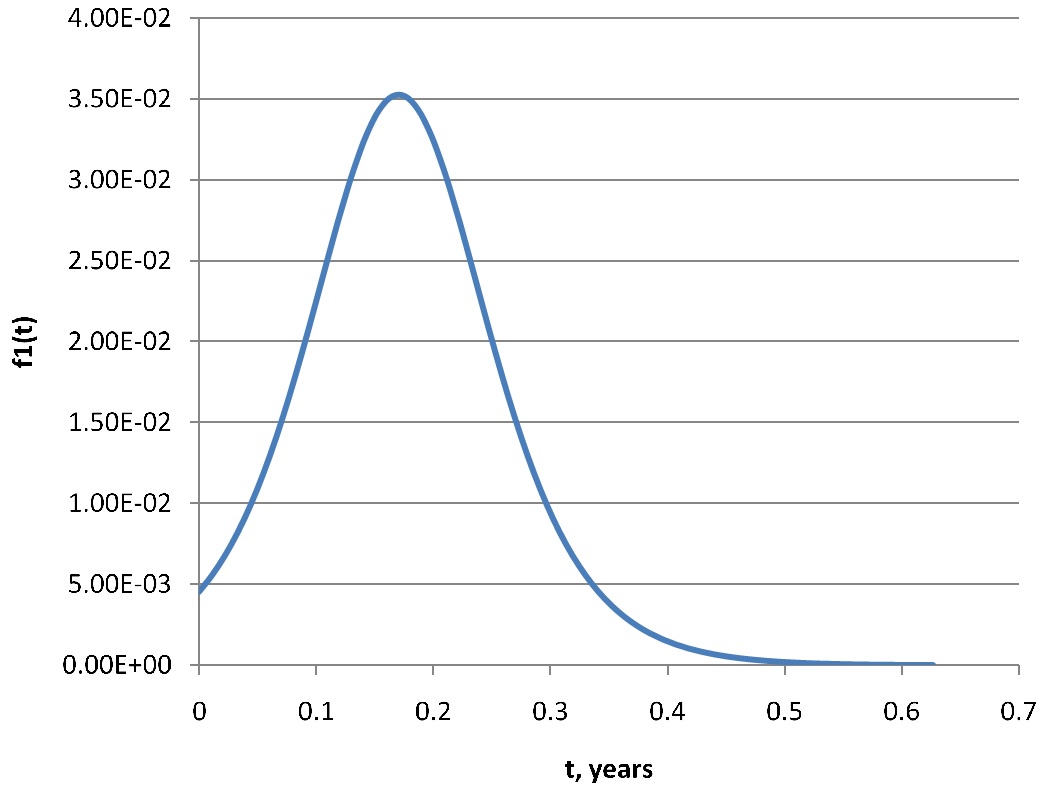}
  \caption[FigtLotteryf1t]
   {Equation \ref{EqLotteryf1}, $C=1.3\times 10^{-9}$, $J_{max}$ = 500,000,000, $J_0$ = 15,000,000, $\Delta t = 0.011$, $k = 19.748$,  Plot is done using Microsoft Excel.}
  \label{FigtLotteryf1t}
\end{figure}

The previous paragraph evaluates $f^{(1)}(t)=1.3 \times 10^{-9} \times (J(t+\Delta t) - J(t))=C \times (J(t+\Delta t) - J(t))$. Replacing $J(t)$ with $J_L(t)$ yields
\begin{equation}
\label{EqLotteryf1}
f^{(1)}(t)=\frac{C \times (J_{max}-J_0)J_{max}J_0e^{kt}(e^{k\Delta t}-1)}{(J_{max}-J_0)^2+(J_{max}-J_0)J_0e^{kt}(e^{k \Delta t}+1)+(J_0e^{kt})^2e^{k \Delta t}}
\end{equation}
This formula implies that, if the jackpot is not won, then $f^{(1)}$ has a maximum: \textit{growing interest during the first phase is suppressed by a lack of the capital later}. Setting $\frac{df^{(1)}}{dt}=0$, we get $t_{max}=\frac{1}{k}\ln(\frac{J_{max}-J_0}{J_0e^{\frac{k \Delta t}{2}}})$. Substituting $t_{max}$ into Equation \ref{EqLotteryf1} leads to $f_{max}^{(1)}=\frac{C \times J_{max}(e^{\frac{k \Delta t}{2}}-1)}{e^{\frac{k \Delta t}{2}}+1}$, Figure \ref{FigtLotteryf1t}. The estimate $J_{max}$ $\approx$ 500,000,000 is inaccurate because of a lack of data at the saturation of $J$. Twice jackpot was greater, Table \ref{TblLottery}.
Due to \ref{EqBinaryConstraintsAB}, $0 \le a(\eta) \le \sqrt{\frac{1-p^{(1)}}{p^{(1)}}}=16,090.09$ and $0 \le b(\eta) \le \sqrt{\frac{p^{(1)}}{1 - p^{(1)}}}=\frac{1}{16,090.09}$. The $a_{max}$ and $b_{max}$ are constant reciprocals of each other but $a$ and $b$ depend on $A_p^{(1)}$ and $J$. 

\paragraph{Relationship between $a(a_1^{(1)},a_1^{(2)})$ and $b(a_1^{(1)},a_1^{(2)})$.} From odd $E(-\eta)=-E(\eta)$ and even $D(-\eta)=D(\eta)$ properties and Equation \ref{EqThef1} it follows that Equation \ref{EqFunctional} holds and one function of choice $f(a_1^{(1)},a_1^{(2)})$ is sufficient, if $a(a_1^{(1)},a_1^{(2)}) = b(a_1^{(2)},a_1^{(1)})$ and $b(a_1^{(1)},a_1^{(2)}) = a(a_1^{(2)},a_1^{(1)})$.

\begin{figure}
  \centering
  \includegraphics[width=75mm]{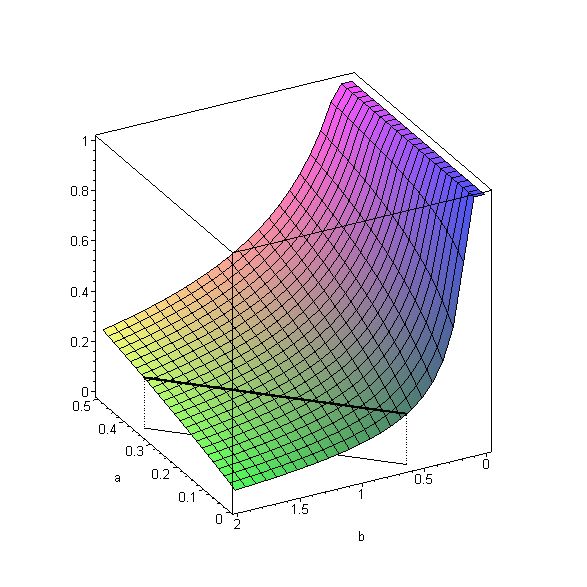}
  \caption[FigF1ABProblem3]
   {Problem 3. Surface $f^{(1)}(a, b)$ for $E(\eta)=200$, $\sqrt{D(\eta)}=1600$. Line $b=4a+\frac{5}{8}$ corresponds to $f^{(1)}=0.20$. Plot is done using Maple 10 from Maplesoft.}
  \label{FigF1ABProblem3}
\end{figure}

\begin{figure}
  \centering
  \includegraphics[width=75mm]{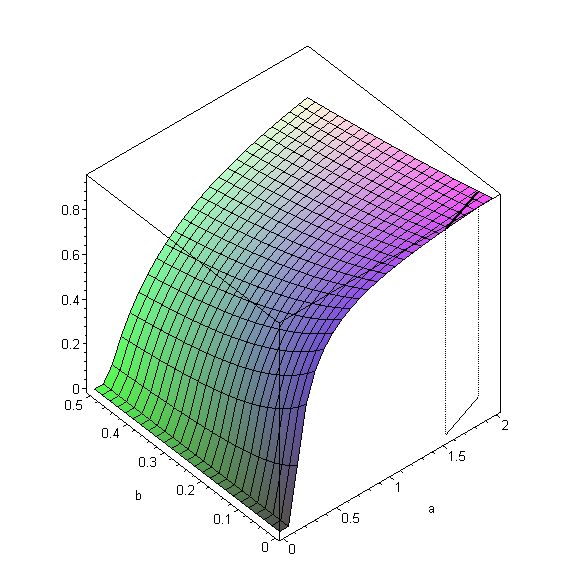}
  \caption[FigF1ABProblem3l]
   {Problem 3'. Surface $f^{(1)}(a, b)$ for $E(\eta)=-200$, $\sqrt{D(\eta)}=1600$. Line $b=\frac{2}{23}a-\frac{25}{184}$ corresponds to $f^{(1)}=0.92$. Plot is done using Maple 10 from Maplesoft.}
  \label{FigF1ABProblem3l}
\end{figure}

Fixing $E(\eta)$ and $\sqrt{D(\eta)}$, Equation \ref{EqThef1} draws a surface above the $ab$-plane. A curve on the $ab$-plane defines a dependence between $a$ and $b$. This dependence is linear, Equation \ref{EqDependenceAB}, if $f^{(1)}$ is also fixed. It is easy to prove that intersection of the surface and the plane, containing this straight line, and orthogonal to the $ab$-plane is also a \textit{horizontal} straight line. Figure \ref{FigF1ABProblem3} illustrates these geometric properties for Problem 3, certainty effect. The surface looking curvy is a set of horizontal straight line segments, each located at its own characteristic height $f^{(1)}$, with slope and intersect given by Equation \ref{EqDependenceAB}. Figure \ref{FigF1ABProblem3l} presents a similar picture for Problem 3', reflection effect.

Dependence $b(a)$ for fixed $E(\eta)$ and $\sqrt{D(\eta)}$ does not have to be linear. If it is not linear or linear but has different slope and intercept than those, implied by Equation \ref{EqDependenceAB}, then $f^{(1)}$ is not constant. Let $E(\eta)=const_1$, and $D(\eta)=const_2 \ge 0$ for $f^{(1)}(\{A_p^{(1)},A_q^{(1)},p^{(1)},N^{(1)}\}, \{A_p^{(2)},A_q^{(2)}=A_p^{(2)},p^{(2)}=1,N^{(2)}\})$, random two-point variable vs. constant. Then, from Equations \ref{EqMeanDiff} - \ref{EqExcessKurtosisDiff}
\begin{equation}
\label{EqPconstED}
p^{(1)}=\frac{1}{1+\frac{const_2N^{(1)}}{(const_1+A_q^{(2)}-A_q^{(1)})^2}},
\end{equation}
\begin{equation}
\label{EqGamma12}
\gamma_1(\eta)=\frac{1-2p^{(1)}}{\sqrt{p^{(1)}(1-p^{(1)})N^{(1)}}}, \; \gamma_2(\eta)=\frac{1}{p^{(1)}(1-p^{(1)})N^{(1)}}-\frac{6}{N^{(1)}}.
\end{equation}
With Equation \ref{EqPconstED} we can design $\eta=a_1^{(1)}-a_1^{(2)}$, change $p^{(1)}$, $N^{(1)}$, and keep $E(\eta)$, $D(\eta)$ constant. From Equations \ref{EqGamma12} this affects $\eta$th skewness and excess kurtosis. \textit{Symmetry of the distribution can influence on respondents preferences. This leaves degrees of freedom for $f^{(1)}$ with constant $E(\eta)$ and $D(\eta)$.}

\begin{figure}
  \centering
  \includegraphics[width=65mm]{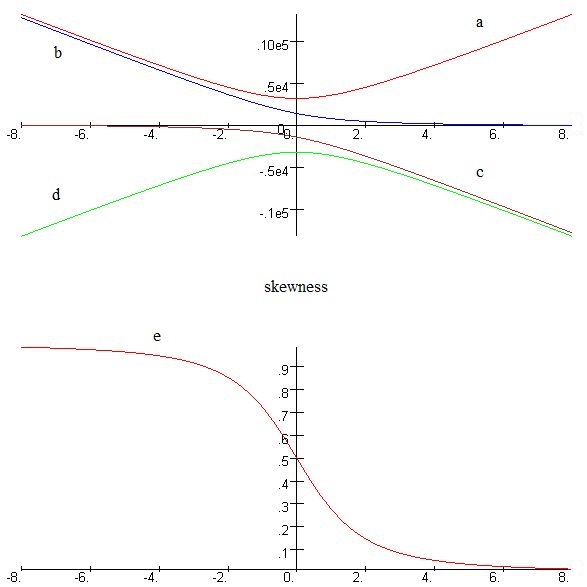}
  \caption[FigEDG]
   {a - $A_p^{(1)}=1600\sqrt{4+\gamma_1^2}$, b - $A_q^{(2)}=-200+1600\sqrt{\frac{\sqrt{4+\gamma_1^2}-\gamma_1}{\sqrt{4+\gamma_1^2}+\gamma_1}}$, c - $A_q^{(1)}=-1600\sqrt{4+\gamma_1^2}$, d - $A_q^{(2)}=-1600\sqrt{4+\gamma_1^2}+200+1600\sqrt{\frac{\sqrt{4+\gamma_1^2}-\gamma_1}{\sqrt{4+\gamma_1^2}+\gamma_1}}$, e - $p^{(1)}=\frac{1}{2}(1-\gamma_1\sqrt{\frac{1}{4+\gamma_1^2}})$. Plot is done using Maple 10 from Maplesoft.}
  \label{FigEDG}
\end{figure}

From Equations \ref{EqGamma12}, $\gamma_1(\eta)>0$ for $p^{(1)}\in (0, \frac{1}{2})$, $\gamma_1(\eta)<0$ for $p^{(1)}\in (\frac{1}{2}, 1)$, and $\gamma_1(\eta)=0$ for $p=\frac{1}{2}$, see also Figures \ref{FigA1DiffProblem3}, \ref{FigA1DiffProblem3b}. Let us rewrite equation for $\gamma_1$ as $(4+N^{(1)}\gamma_1^2)p^{(1)2}-(4+N^{(1)}\gamma_1^2)p^{(1)}+1=0$ and solve the latter $p^{(1)}=\frac{1}{2}\pm \frac{1}{2} \sqrt{\frac{N^{(1)}\gamma_1^2}{4+N^{(1)}\gamma_1^2}}$. Selection of sign and evaluation of root should correspond to the above $\gamma_1$ inequalities and probability intervals. Hence,
\begin{equation}
\label{EqPGamma1}
p^{(1)}=\frac{1}{2} - \frac{\gamma_1}{2} \sqrt{\frac{N^{(1)}}{4+N^{(1)}\gamma_1^2}}, \; \frac{p^{(1)}}{1-p^{(1)}}=\frac{\sqrt{4+N^{(1)}\gamma_1^2}-\gamma_1\sqrt{N^{(1)}}}{\sqrt{4+N^{(1)}\gamma_1^2}+\gamma_1\sqrt{N^{(1)}}}.
\end{equation}
From Equation \ref{EqPconstED} $A_q^{(1)}-A_q^{(2)}=const_1-\sqrt{const_2}\sqrt{\frac{p^{(1)}N^{(1)}}{1-p^{(1)}}}$ and
\begin{equation}
\label{EqAq12Gamma}
A_p^{(2)}=A_q^{(2)}=A_q^{(1)}-const_1+\sqrt{const_2N^{(1)}}\sqrt{\frac{\sqrt{4+N^{(1)}\gamma_1^2}-\gamma_1\sqrt{N^{(1)}}}{\sqrt{4+N^{(1)}\gamma_1^2}+\gamma_1\sqrt{N^{(1)}}}}.
\end{equation}
From Equations \ref{EqVarianceDiff} and \ref{EqPconstED} under the specified conditions
\begin{equation}
\label{EqAp1Gamma}
A_p^{(1)}=A_q^{(1)}+\sqrt{const_2 N^{(1)} (4+N^{(1)}\gamma_1^2)}.
\end{equation}
Two-point variable vs. constant problems matching $E(\eta)=200$, $\sqrt{D(\eta)}=1600$, $A_q^{(1)}=0$ of Problem 3 are $\{A_p^{(1)}(\gamma_1(\eta)),0,p^{(1)}(\gamma_1(\eta)), 1\}$ vs. $\{A_p^{(2)}(\gamma_1(\eta))=A_q^{(2)}(\gamma_1(\eta)),A_q^{(2)}(\gamma_1(\eta)),1,1\}$, where $p^{(1)}(\gamma_1(\eta))$, $A_q^{(2)}(\gamma_1(\eta))$, $A_p^{(1)}(\gamma_1(\eta))$ are given by Equations \ref{EqPGamma1} - \ref{EqAp1Gamma}. For Problem 3' the problems with $E(\eta)=-200$, $\sqrt{D(\eta)}=1600$, $A_p^{(1)}=0$ are $\{0,-\sqrt{const_2 N^{(1)} (4+N^{(1)}\gamma_1^2)},p^{(1)}(\gamma_1(\eta)), 1\}$ vs. $\{A_p^{(2)}=A_q^{(2)},A_q^{(2)}(\gamma_1(\eta)),1,1\}$, Figure \ref{FigEDG}. \textit{In both cases skewness $\gamma_1(\eta)$ can be arbitrary. It adds a degree of freedom to influence on $a(\eta)$ and $b(\eta)$}.

\begin{figure}
  \centering
  \includegraphics[width=130mm]{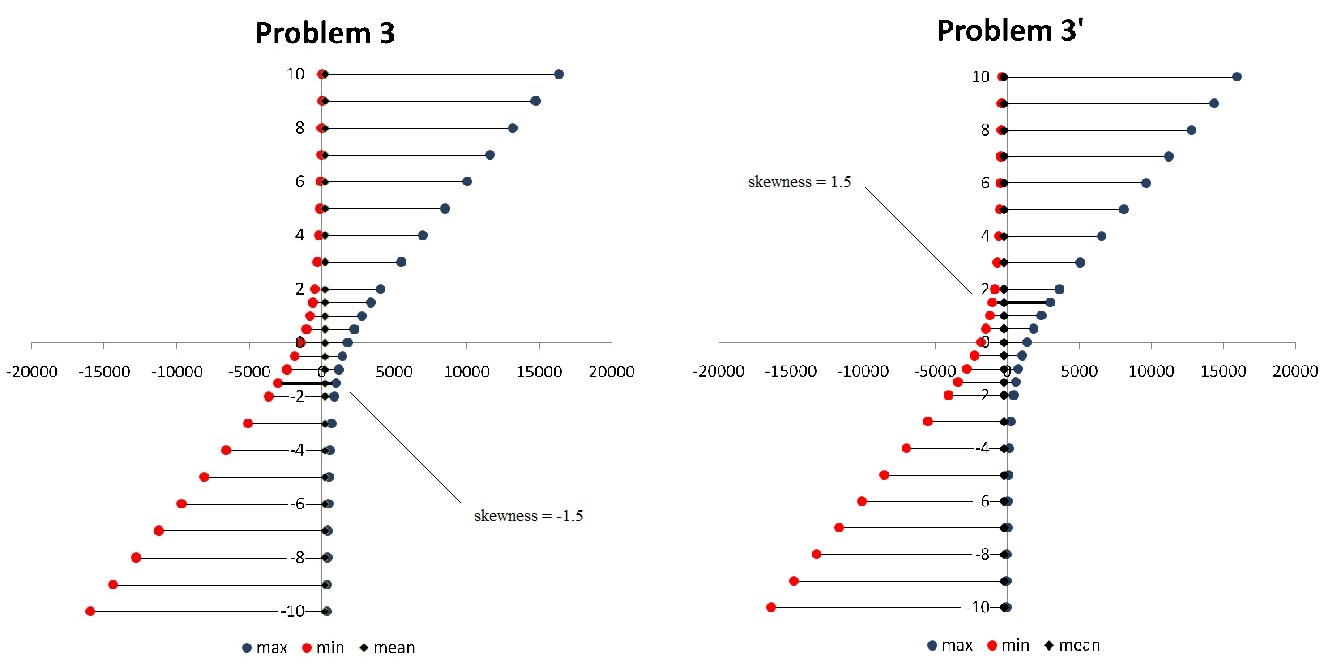}
  \caption[FigMMMGamma]
   {Minimum, maximum, and mean values of variables $\eta$ corresponding to different values of $\gamma_1$ with constant $\sqrt{D(\eta)}=1600$ and $E(\eta)=200$ in Problem 3 and $E(\eta)=-200$ in Problem 3'. Plot is done using Microsoft Excel.}
  \label{FigMMMGamma}
\end{figure}

Keeping $E(\eta)$, $D(\eta)$, $A_q^{(1)}$ intact, variation of skewness $\gamma_1(\eta)$ changes $A_p^{(1)}(\eta)$, $A_p^{(2)}(\eta)=A_q^{(2)}(\eta)$, Equations \ref{EqAp1Gamma}, \ref{EqAq12Gamma}, $\eta_{max}(\eta)$, $\eta_{min}(\eta)$, $a_{max}(\eta)$, $b_{max}(\eta)$, Equations \ref{EqConstraintsAB}, \ref{EqBinaryConstraintsAB}. Under the specified conditions these equations yield
\begin{equation}
\label{EqAmaxBmax}
\begin{split}
a_{max}=\frac{\sqrt{N^{(1)}}}{2}(\sqrt{4+N^{(1)}\gamma_1^2} + \gamma_1\sqrt{N^{(1)}}),\\
b_{max}=\frac{\sqrt{N^{(1)}}}{2}(\sqrt{4+N^{(1)}\gamma_1^2} - \gamma_1\sqrt{N^{(1)}}),\\
a_{max}b_{max}=N^{(1)}.
\end{split}
\end{equation}
A random difference $\eta$ can be depicted as a horizontal line segment $[\eta_{min}, \eta_{max}]$ with an inner point $E(\eta)$. On Figure \ref{FigMMMGamma} such segments are plotted relative each to other for different values of skewness $\gamma_1$ for Problems 3 and 3'. We see that $\eta$ overlaps a greater positive area for $\gamma_1=\frac{3}{2}$ in Problem 3' than for $\gamma_1=-\frac{3}{2}$ in Problem 3. \textit{These diagrams support the reflection effect.}

\begin{figure}
  \centering
  \includegraphics[width=100mm]{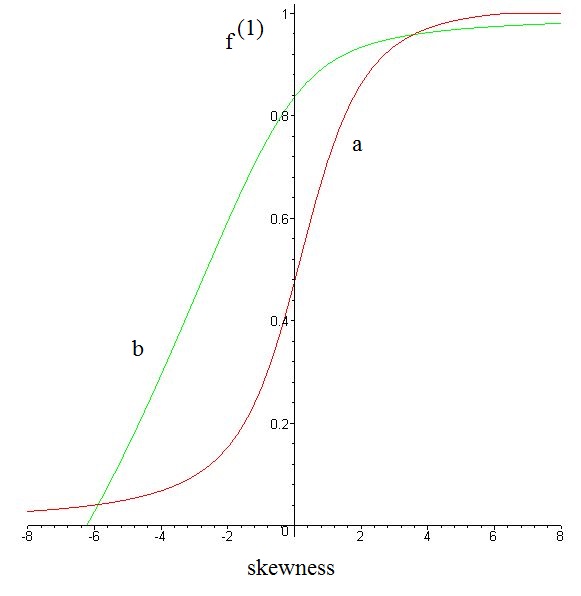}
  \caption[FigF1Gamma]
   {a - $const_a=2$, $const_b=\frac{2}{1\frac{5}{8}}$; b - $const_a=1\frac{1}{4}$, $const_b=153.(3)$. Plot is done using Maple 10 from Maplesoft.}
  \label{FigF1Gamma}
\end{figure}

Let in a hypothetical case $a(\gamma_1)=\frac{a_{max}(\gamma_1)}{const_a}$ and $b(\gamma_1)=\frac{b_{max}(\gamma_1)}{const_b}$. Then, from Equations \ref{EqAmaxBmax} $b(\gamma_1)=\frac{N^{(1)}}{const_a const_b a(\gamma_1)}$ is a hyperbola. This also creates a parametric dependence $a=a(\gamma_1)$ and $b=b(\gamma_1)$ and allows to express $f^{(1)}$ as a function of $\gamma_1$ for constant $E(\eta)$ and $\sqrt{D(\eta)}$. Figure \ref{FigF1Gamma} illustrates fitting $f^{(1)}=0.2$ in Problem 3 with $const_a=2$ and $const_b=\frac{2}{1\frac{5}{8}}\approx 1.230769$ and fitting $f^{(1)}=0.92$ in Problem 3' with $const_a=1\frac{1}{4}$ and $const_b=153.(3)$. More experimental data on fractions of respondents selecting between two-point random variables is needed to clarify further dependencies discussed so far.

\paragraph{Acknowledgments.}  I would like to thank Timur Misirpashaev for the discussion on \cite{salov2014}, the St. Petersburg paradox and its credit component and Pavel Grosul for the discussion on relationships between fractions of respondents and wealth.

\bigskip

\noindent\textbf{Valerii Salov} received his M.S. from the Moscow State University, Department of Chemistry in 1982 and his Ph.D. from the Academy of Sciences of the USSR, Vernadski Institute of Geochemistry and Analytical Chemistry in 1987.  He is the author of the articles on analytical, computational, and physical chemistry, the book Modeling Maximum Trading Profits with C++, \textit{John Wiley and Sons, Inc., Hoboken, New Jersey}, 2007, and papers in \textit{Futures Magazine} and \textit{ArXiv}.

\noindent\textit{v7f5a7@comcast.net}

\end{document}